\documentclass[onecolumn,numberedappendix,nofootinbib,nobalancelastpage]{openjournal}
\usepackage{amsmath} % Equations
\usepackage{amssymb} % Equations
\usepackage{array}
\usepackage{graphicx}
\usepackage{setspace}
\usepackage[colorlinks=true,
            urlcolor=blue,
            citecolor=blue,
            linkcolor=red,
            menucolor=blue,linktocpage=true]{hyperref}
\usepackage[dvipsnames]{xcolor}
\usepackage{amsfonts}
\usepackage{bm}
\usepackage{soul}
\usepackage{color}
\usepackage{mathrsfs}
\usepackage[noabbrev]{cleveref}
\DeclareMathAlphabet{\mathpzc}{OT1}{pzc}{m}{it}

\hypersetup{final,bookmarksdepth=2}

\def\mbf#1{\mathbf{#1}}
\def\mrm#1{\mathrm{#1}}
\def\be{\begin{equation}}\def\ee{\end{equation}}
\def\bea#1\eea{\begin{align}#1\end{align}}
\def\x{{\mathbf{x}}}

\def\r{{\mathbf{r}}}
\def\k{{\mathbf{k}}}
\def\kl{{\mathbf{K}}} %\def\kl{{\bm{k}_l}}
\def\K{{\mathbf{K}}}
\def\q{{\mathbf{q}}}
\def\v{{\mathbf{v}}}

\def\dif{d} %\def\dif{\mathrm{d}}
\def\Mid{\ssp|\ssp}

\newcommand\delD{\delta_\mathrm{D}}

\newcommand\calH{\mathcal{H}}

\newcommand{\DD}{\mathcal{D}}

\newcommand\im{\mrm{i}}

\newcommand\Ptot{P_{\mathrm{tot}}}
\newcommand\Pcross{P_{\mathrm{cross}}}
\newcommand{\wideaverage}[1]{\left\langle #1 \right\rangle}
\newcommand\bard{\bar{\delta}}
\newcommand\barl{\delta^L}
\newcommand\PL{P_{L}}

\def\ssp{\hspace{0.09em}}
\def\sp{\hspace{0.06em}}
\definecolor{myr}{rgb}{0.68, 0.14, 0.14} % {0.65,0,0}

\newcommand{\Geneve}{D\'epartement de Physique Th\'eorique, Universit\'e de Gen\`eve, 24 Quai Ansermet, CH-1211 Gen\`eve 4, Switzerland}

\makeatletter
\def\l@subsection#1#2{}
\def\l@subsubsection#1#2{}
\def\l@paragraph#1#2{}
\def\l@subparagraph#1#2{}
\makeatother

\begin{document}

\title{Density reconstruction from biased tracers:\\[3pt]
Testing the equivalence principle through consistency relations\vspace{-15mm}}
%\shorttitle{\sc{Testing the equivalence principle through consistency relations}}
%\shortauthors{\sc{Dam and Darwish}}

\author{Lawrence Dam\!\!}
\email{lawrence.dam@unige.ch}
\author{\;Omar Darwish}
\email{od261@cantab.ac.uk}

\affiliation{\Geneve}

%\date{\today}

\begin{abstract}
Consistency relations of large-scale structure offer a unique and powerful test of the weak equivalence principle (EP) on cosmological scales.
If the EP is violated, different tracers will undergo different accelerations in response
to a uniform gravitational field.
In the context of large-scale structure,
this loss of universality manifests as a dipole
with a characteristic $1/K$ scale dependence in the squeezed limit of the bispectrum.
In this work we show that such a violation can be identified with a particular anti-symmetric {modulation} in the local cross-power spectrum of distinct tracers.
Based on this observation, we propose to test the EP using quadratic estimators as a more practical alternative to the conventional approach of directly estimating the bispectrum.
We apply our quadratic estimator to a DESI-like survey and forecast constraints on
the overall amplitude of EP violation,
%Including mildly nonlinear scales in our
%reconstruction ($k_\mrm{max}\simeq0.15\,h\,\mrm{Mpc}^{-1}$), 
finding that our estimator
is competitive with the more exhaustive direct bispectrum approach. This shows that surveys like DESI can already benefit from the quadratic estimator approach.
\end{abstract}

\maketitle
% \newpage

% \twocolumngrid
\setstretch{0.1}   % spacing between lines 
\setcounter{tocdepth}{1}
  \tableofcontents
% \onecolumngrid

% \setstretch{1.05}   % spacing between lines 
\setstretch{1.00}   % spacing between lines 

\newpage

% \newpage
\section{Introduction}\label{sec:intro}
Our understanding of gravity across a wide range of scales is founded on the weak equivalence principle (EP)---that 
for the same initial positions and velocities, test bodies (of sufficiently small size)
subject only to gravity will follow the same trajectory, independent of the body's detailed structure. This universality is very well supported by a variety of precision laboratory experiments, satellite measurements, and 
observations within the Solar System~\citep{Will:2014kxa}.
On cosmological scales, however, the validity of the EP is less certain, with
tests few and far between.

Gravity is long range and it is plausible there are additional
long-range interactions (fifth forces) which do not couple universally to matter,
but whose effects only become apparent on large scales due to screening~\citep{Ferreira:2019xrr}.
Nevertheless, in the standard $\Lambda$CDM model it is the working assumption that the EP
remains valid on those cosmic scales.
Although this assumption appears to be consistent with current observations,
it represents an extrapolation of several orders
of magnitude beyond the length scales on which the EP has been stringently tested~\citep{Will:2014kxa}.
With the increasingly large datasets provided by galaxy surveys, it is interesting to ask how
the large-scale structure can be used in a precision test of the EP.

To that end, what are the implications
of EP violation in large-scale structure (LSS)? What
observational signatures can we search for in the clustering of galaxies, such as
in their correlation functions, if there is a violation?

A powerful way to approach these questions is through
{consistency relations}~\citep{Kehagias:2013yd, Peloso:2013zw, Peloso:2013spa, Valageas:2013cma, Creminelli:2013poa, Scoccimarro:1995if}. 
In the context of LSS, these are precise nonperturbative statements,
valid under quite general conditions, about the insensitivity of
local measurements (e.g.\ of small-scale correlations)
{to the presence of uniform, time-dependent} gravitational fields~\citep{Baldauf:2015xfa}. 
In the same way that test objects free falling in a
uniform gravitational field preserve their relative positions,
\`a la Einstein's elevator,
galaxy correlations measured on sufficiently small scales at equal times are
invariant under displacements
induced by long-wavelength fluctuations (long enough that tidal effects can be ignored). 

In Fourier space the consistency relation can be formally stated
as {\it the equal-time bispectrum is absent a $1/K$ scale dependence in the squeezed limit}.
The power of this statement lies in its robustness to the usual complications
of LSS modelling, namely, nonlinear gravitational evolution
(including complex baryonic physics on small scales)~\citep{Horn:2014rta,Esposito:2019jkb}, galaxy biasing~\citep{Kehagias:2013rpa,Kehagias:2015tda}, and
redshift-space distortions~\citep{Creminelli:2013poa}.
Given Gaussian and adiabatic initial conditions, if one detects a
$1/K$ pole in the equal-time
squeezed bispectrum of multiple tracers, such a detection would be
difficult to blame on known physics~\citep{Esposito:2019jkb}.
On the contrary, the detection of such a pole may be taken
as evidence of EP violation~\citep{Creminelli:2013nua,Kehagias:2013rpa,Valageas:2013cma}.

\subsection{The anti-symmetric shift as a signature of EP violation}\label{sec:anti-symmetric-toy}
\begin{figure}[t!]
    \centering
    \includegraphics[scale=0.93]{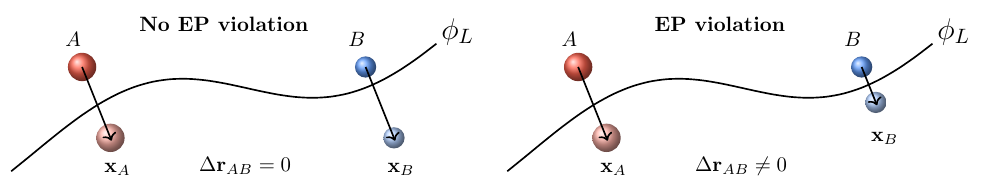}
    % \vspace*{-3mm}
    \caption{The transport of two distinct galaxies when subjected
    to a (near) uniform gravitational field, e.g.\ sourced by
    a long-wavelength potential $\phi_L\propto\delta_L/K^2$.
    If the EP holds, then over a given time interval the gravitational field
    induces {identical} accelerations, resulting in identical displacements
    for both galaxies,
    independent of their mass, type, composition, etc.
    If the EP does {not} hold, each galaxy suffers a {different}
    displacement and a relative shift $\Delta\r_{AB}\neq0$ develops between them,
    directed along the gradient $-\nabla\phi_L$.
    This relative shift---which we call the
    \emph{anti-symmetric shift} due to the dipolar nature of the effect---can be
    targeted as a signature of EP violation.
}
    \label{fig:epschematic}
\end{figure}

We can get a better sense of the consistency relation and its relation to the specific EP-violating effect we are after by studying a simple 
two-particle model (see Figure~\ref{fig:epschematic}).

Consider the evolution in separation $\r$ between
two neighbouring test galaxies,
$A$ and $B$, moving in a  gravitational field with 
static potential $\phi(\x)$. In physical coordinates, each galaxy
moves according to $\ddot\x_A=-\nabla\phi(\x_A)$ and
$\ddot\x_B=-\nabla\phi(\x_B)$, and by taking the difference
for small $\r=\x_A-\x_B$ we have
$\ddot\r+\mbf{T}\ssp\r=0$, where
$\mbf{T}\equiv \nabla_i\nabla_j\phi|_{\x_A}$ is the 
tidal field in the neighbourhood of the pair.
This is the Newtonian analogue of the geodesic deviation equation, 
describing how trajectories of nearby test objects converge or
diverge under tidal forces.
Crucially, note that the deviation is symmetric under interchange of pair positions.

Now consider the effect of an external large-scale potential $\phi_L(\x)=-\x\cdot\mbf{a}_L$
on the separation. Since this potential is a pure gradient,
it sources a uniform acceleration $\mbf{a}_L=-\nabla\phi_L$,
leading to
$\ddot\x_A=-\nabla\phi(\x_A)+c_A\ssp\mbf{a}_L$ and
$\ddot\x_B=-\nabla\phi(\x_B)+c_B\ssp\mbf{a}_L$ {where $c_A$ and $c_B$ are some
tracer-dependent coupling coefficients}.
In this simple model the EP requires $c_A=c_B=1$ (universal coupling)
so that the large-scale potential induces the same acceleration
in both galaxies, resulting in the same displacements over a given time interval. 
See left panel of Figure~\ref{fig:epschematic}.
Importantly, if we change to the accelerated frame, by a (non-Galilean) transformation
$\x\to\x-\frac12t^2\mbf{a}_L$, the gravitational effect of the
large-scale potential disappears for both galaxies and one recovers
the previous equations of motion.%
\footnote{It was shown many years ago~\citep{Trautman_1967,1969ApJ...155..105E,MTW,HeckmannSchuecking1959}
that Newton's equation
$\ddot\x=-\nabla\phi$ is invariant under transformations
$\x'=\x-\mbf{d}(t)$ and $\phi'=\phi+\x\cdot\ddot{\mbf{d}}(t)+f(t)$, for arbitrary
$\mbf{d}(t)$ and $f(t)$. 
This form invariance is in essence the same one underlying the 
Euler--Poisson equations governing the development of
large-scale structure~\citep{Kehagias:2013yd, Peloso:2013zw};
see also the much earlier works by \cite{Rosen_1971,Rosen1972}
who pointed out the same (but in proper coordinates).
As is well known, this transformation is outside the usual group of
Galilean transformations (translations, rotations, boosts) 
and as such has been called 
`extended Galilean invariance' in the LSS literature
(among other names)~\citep{DAmico:2021rdb}. Here we emphasize, in keeping with
the earlier literature, that the transformation
is related to the fact that the
gravitational field can always be removed, at least locally and 
possibly everywhere (if the field is uniform), by
changing to an inertial frame, e.g.\ by choosing $\mbf{d}(t)$ such that
it locally cancels with $-\nabla\phi$ and one has $\ddot\x'=0$
in that frame. That is, the EP allows us to transform away uniform
gravitational fields by a suitable change of (physical) coordinates.
Although the notions of cosmic expansion and comoving observers add a
conceptual twist to the meaning and application of inertial frames to cosmology~\citep{Dai:2015rda,Dai:2015jaa,Inomata:2023faq},
the existence of these privileged frames is the bedrock physics underlying
the LSS consistency relation.
}

Suppose the EP is now violated ($c_A\neq c_B$)
so that
$\ddot\r+\mbf{T}\ssp\r=(c_A-c_B)\sp\mbf{a}_L$. If both
galaxies are initially at rest relative to each other,
then over a small time interval and for small separations
we have approximately
\be
\r_{}(t)
% =\r_0-\frac{1}{2}\sp t^2\sp\mbf{K}\ssp\r_0
% +\frac{1}{2}\sp t^2\epsilon\ssp\mbf{f}
=\r_0-\frac{1}{2}\sp t^2\sp \mbf{T}\ssp\r_0
    +\frac{1}{2}\sp t^2\sp (c_A-c_B)\ssp \mbf{a}_L\,,
\ee
where $\r_0=\x_A(0)$ is the initial separation (assuming
particle $B$ is initially at the origin, $\x_B(0)=0$).
The last term, which acts as a small perturbation to the usual
tidal deviation, is independent of $\r_0$ and is anti-symmetric
in nature.
This means that if we interchange the locations of the 
particles (at any point in time), then we have a sign flip
in the last term, or $c_A-c_B\to -(c_A-c_B)$. In other words,
the separation evolves differently depending on whether we put galaxy $A$ at the end of the vector $\r$ or galaxy $B$.
If we denote by $\r_{AB}$ and $\r_{BA}$ the separation vectors for these interchanged
initial positions (with $\r_0$ pointing in the same direction in both cases), we get the difference
\be\label{eq:Delta-rAB}
\Delta\r_{AB}(t)
\equiv
\r_{AB}(t) - \r_{BA}(t) = t^2(c_A-c_B)\ssp\mbf{a}_L\,,
\ee
where the contribution from tidal deviation has cancelled out
since it 
is symmetric under interchange of initial positions. Clearly
$\Delta\r_{AB}=0$ if the EP holds (or trivially if the galaxies are
identical, $A=B$). And if it does not hold
$\Delta\r_{AB}\neq0$, which implies that we cannot remove the effect
of the large-scale potential by changing frame as we could before;
see right panel of Figure~\ref{fig:epschematic}.
Therefore, if there is an EP violation the \emph{anti-symmetric shift} 
$\Delta\r_{AB}\neq0$  represents a genuine physical effect which cannot
be transformed away by a change of coordinates.

\subsection{Probing the anti-symmetric shift}
In the clustering of galaxies, the anti-symmetric shift can be probed in the spatial correlations of galaxies.
As the simple model above illustrates, for this it is necessary to consider
two distinct populations $A$ and $B$. The aim is then to
look for the effect in the
anti-symmetric cross-correlations between $A$ and $B$.

The key idea is that (cross) correlations on small scales are
generally modulated by large-scale gravitational fields,
sourced by large-scale matter fluctuations $\delta_L(\K)\propto K^2\phi_L(\K)$ (which we can reconstruct).
In particular, the presence of $\delta_L(\K)$
induces a position-dependent cross-spectrum {$P_{AB}(\k;\delta_L)$}
between galaxies $A$ and $B$, {i.e.\ the local small-scale power is modulated by $\delta_L$.}
Importantly, since the cross-spectrum is itself correlated with $\delta_L$, the
form of this modulation is determined by squeezed configurations $K\ll k$
of the bispectrum
{$B_{ABL}(\k,\K)=\langle P_{AB}(\k;\delta_L)\sp\delta_L^*(\K)\rangle'$}~\citep{Chiang:2014oga, Chiang:2017jnm, dePutter:2018jqk}.
{The local cross-spectrum thus gives us access to the squeezed bispectrum through the modulation.}

But as mentioned, the EP protects $B_{ABL}$ from acquiring a $1/K$ pole in the
squeezed limit.
Therefore, a violation of the EP implies a violation of the consistency relation, which manifests as a characteristic $1/K$ scale dependence in the 
equal-time squeezed bispectrum of the form~\citep{Creminelli:2013nua, Peloso:2013spa,Marinucci:2019wdb}
% \be
% f_{AB}(\k, \kl) 
% \sim P_{AB}(\k)\sp \frac{\k\cdot\K}{|\K|^2}\ssp \epsilon \,, \nonumber
% \ee
\be\label{eq:CR}
B_{ABL}(\k, \kl) 
\sim \epsilon\, \frac{\k\cdot\K}{K^2}\ssp P_{AB}(\k) \ssp P_L(\K) \,,
\ee
where $P_{AB}(\k)$ is the usual cross-spectrum
{(in the absence of the modulation $\delta_L$)},
$P_L(\K)$ is the (linear) power spectrum of $\delta_L$, and $\epsilon$
is an EP violation parameter corresponding to $c_A-c_B$ in the
{anti-symmetric shift}~\eqref{eq:Delta-rAB},
with $\epsilon=0$ if the EP holds.
{We can thus view Eq.~\eqref{eq:CR} as a null test in that
the right-hand side should tend to zero if the EP holds.}
Note that a key advantage of this test is that there is no
requirement that the tracer modes measured on small scales live
in the perturbative regime; we only require that $\delta_L(\K)\ll1$
is in the linear regime.
Provided $k\gg K$, the small-scale modes can be as deep in the nonlinear regime
as we choose~\citep{Horn:2014rta,Esposito:2019jkb}.

The most straightforward way to constrain EP violations is through direct measurements of the bispectrum~\citep{Creminelli:2013poa, Lewandowski:2019txi, Sugiyama:2023zvd, Bottaro:2023wkd}. But bispectrum analysis presents significant practical challenges: extracting information efficiently, constructing accurate covariance matrices, and mitigating large-scale systematics---these are all more difficult than in standard power spectrum analysis.
Moreover, combining bispectrum measurements with higher-order $n$-point functions, such as the trispectrum, is not straightforward. Given the potential of these higher-order statistics to probe new physics, {overcoming} these challenges will be important for upcoming surveys like DESI \citep{desicollaboration2016desiexperimentisciencetargeting}, Euclid \citep{Euclid2018}, SPHEREx \citep{dore2015cosmologyspherexallskyspectral}, and future Stage V galaxy spectroscopic surveys \citep{besuner2025spectroscopicstage5experiment}.

Alternatively, as we show in this work, EP violation can be probed using quadratic estimators (QE).
These estimators have the advantage that they are practical and
provide information beyond the power spectrum,
without having to directly estimate the bispectrum or trispectrum.
Moreover, they are straightforwardly integrated into traditional power spectrum analysis.
Indeed, there is a well-developed formalism on QEs by the CMB community, which is now routinely applied to a range of problems including reconstruction of (projected) matter distribution~\citep{Hu:2001kj, LEWIS_2006}, cosmic rotation~\citep{Kamionkowski_2009}, and the kinetic Sunyaev--Zel'dovich effect~\citep{Dore:2003ex, DeDeo:2005yr}. Such probes can easily be combined with external tracers \citep[e.g.][]{Dore:2003ex, Hirata:2004rp, Smith:2007rg, Hirata:2008cb}. Recently, this formalism has  found application
in the context of LSS, including to reconstruct the large-scale density field~\citep{Li:2020uug, Darwish:2020prn, Zhu:2021qmm, Zang:2022qoj}; constrain nonlinear biases and primordial non-Gaussianity \citep[e.g.][]{Schmittfull_2015,Smith:2018bpn, MoradinezhadDizgah:2019xun,Darwish:2020prn};
and to recover modes lost to large-scale foregrounds~\citep{Foreman:2018gnv, Karacayli:2019iyd, Darwish:2020prn, Zang:2022qoj, Qin:2025lyy}. Quadratic estimators have also been proposed in the context of general anti-symmetric galaxy correlations~\citep{Dai:2015wla}, though they have yet to be applied in a test of the EP.

In recent work, EP violations have been explored through models of long-range dark forces that enhance the growth of structure and modify the background expansion (among other effects)~\citep{Bottaro:2023wkd}. Based on an analysis of \emph{Planck} CMB data and BAO data from DESI DR1, \cite{Bottaro:2024pcb} found hints of EP-violating dark forces. These results are suggestive and motivate further study, since if there is such a dark force it should lead to a violation of the
consistency relation which can be verified in a bispectrum analysis of current surveys, independent of other probes.
% \st{The galaxy (multi-tracer) bispectrum offers an independent way to verify these findings, as it uniquely probes a signature not produced by standard gravity.}

\subsection{This work}\label{sec:thiswork}
To avoid the practical challenges to do with direct bispectrum estimation, in this work we will explore quadratic estimators as a simple alternative to constraining EP violation. 
The idea of this approach is to exploit the response
of short-wavelength observables to a long-wavelength matter mode. 
The (linear) response function, describing the modulation of small-scale correlations by the
long mode, encodes key information about
the squeezed bispectrum. Importantly, we will see that the EP fixes
the types of 
responses allowed, or equivalently the analytic structure of the squeezed bispectrum, which must
remain finite as one approaches the limit $K\to0$.

We will show that at leading order a general linear response associated with
two distinct tracers can be decomposed into a set of independent
response functions, corresponding to growth, shift, and tidal effects. Each of these three effects carries both a symmetric and anti-symmetric part,
and hence there are six possible responses. But of these six responses one turns out to be `switched off' by virtue of the EP.
This protected response corresponds
precisely to the anti-symmetric shift described in the previous section. If there is a violation, we expect to see a response with the characteristic
$1/K$ scale dependence, and whose amplitude can be used as a measure of EP violation.

Based on this anti-symmetric shift response function, we construct a quadratic estimator sensitive to EP violations using two different galaxy populations. 
We apply our quadratic estimator to a DESI-like
survey, and forecast constraints on an EP violation amplitude $\tilde{\epsilon}$. Depending on the number of modes used in the quadratic estimator reconstruction, we find $\sigma({\tilde{\epsilon}}) \sim 2\times10^{-3}$ to $10^{-2}$. This is consistent with the independent forecast of \cite{Graham:2025fdt}
on the same DESI-like configuration, but using the direct bispectrum approach.

This paper is organized as follows. In Section \ref{sec:linearresp} we review the effect
of large-scale fluctuations on observables from the point of view of linear response theory, 
discuss the possible types
of responses, and identify the particular response type responsible for a violation of the 
consistency relation.
Given this response, in Section \ref{sec:QEs} we construct a quadratic estimator
sensitive to EP violation. Fisher forecasts for constraints on $\epsilon$ are presented in Section \ref{sec:forecasts} and
discussed in Section~\ref{sec:discussion}. Our main findings are summarized in Section~\ref{sec:conclusions}.
Supporting results and calculation details can be found in a number of appendices.
For numerical work we adopt the {\it Planck} 2018 spatially flat $\Lambda$CDM cosmology~\citep{Planck:2018vyg}
used in the \textsc{AbacusSummit} simulations~\citep{Garrison},
against which we validate our estimator.
Our code is publicly available at~\url{https://github.com/Saladino93/qeep}.

\vspace{1mm}

\paragraph{Notation and conventions.}
We use the Fourier convention
\bea
\tilde{f}(\k)=\mathcal{F}[f]\equiv \int\dif^3\mbf{x}\,e^{\im\k\cdot\mbf{x}}{f}(\mbf{x})\,,
\qquad
f(\mbf{x})=\mathcal{F}^{-1}[\tilde{f}]\equiv \int_\k\,e^{-\im\k\cdot\mbf{x}}\tilde{f}(\k)\,,
\eea
where $\int_\k$ is a shorthand for $\int{\dif^3\k}/(2\pi)^3$. We use
uppercase $\kl$ to denote wavevectors of the long-wavelength modes and
lowercase $\k$ for the  short-wavelength modes.
For any two-point correlator
$P_{AB}(\k_1,\k_2)$, e.g.\ the cross-power spectrum,
we define the symmetric and anti-symmetric parts, respectively, as
\be\label{eq:sym-antisym-defn}
P_{(AB)}=(P_{AB}+P_{BA})/2\,,
\qquad
P_{[AB]}=(P_{AB}-P_{BA})/2\,,
\ee
where the arguments are kept fixed but the labels are swapped.
Functional derivatives are denoted $\partial/\partial\delta(\k)$ and we use the
convention
$\partial \delta(\k)/\partial\delta(\k')=(2\pi)^3\delD(\k-\k')$.
We denote by $\langle\ldots\rangle'$ the ensemble average with the
momentum-conserving delta function stripped away, e.g.\
$\langle\delta(\k_1)\sp\delta(\k_2)\rangle'=P(\k_1)$ with $\k_1+\k_2=0$ understood.

\section{Gravitational responses}\label{sec:linearresp}
The effect of large-scale fluctuations on small-scale observables is well known
in the context of CMB lensing~\citep{Hu:2001kj}, and in this section we begin by reviewing the analogous effect in LSS~\citep{Lewis:2011,Li:2020uug, Darwish:2020prn}.
Here we follow the approach of \cite{Lewis:2011,Lewis:2011fk} (see also \citealt{Chiang:2014oga,Barreira:2017sqa}), building on the idea that
long-wavelength matter fluctuations can be viewed as
weak external sources whose effect on small-scale clustering can be described
using linear response theory.

Without resorting to specific models and assuming only the basic
analytic structure of standard perturbation theory, we 
then present a general framework for characterizing
the different types of responses allowed by the EP, 
showing that violations thereof give rise to a particular response
associated with the {anti-symmetric shift}.
In the limit where the long mode is infinitely long ($K\to0$), we will see
how this response is connected with the consistency relation.

{While the EP can also be tested by measuring unequal-time correlations \citep{Peloso:2013zw, Kehagias:2013yd}, in this work
we will focus on the more practical and interesting case of equal-time correlations.%
\footnote{The unequal-time case, analogous to observing objects free fall in a mass-drop experiment,
generically leads to a $1/K$ pole, whether or not there is EP violation. The origin of 
this pole is due to ordinary relative growth of the long mode
(at different times), rather than a more fundamental 
relative shift~\citep{Esposito:2019jkb}.}
We will therefore suppress the time dependence in the expressions below. 
}

\subsection{Effect of a long mode on two short modes: general analysis}\label{sec:functional-expansion}
Here we consider the general effect of a long-wavelength matter fluctuation $\delta_L(\K)$
on the statistics of two tracers $\delta_A(\k_1)$ and $\delta_B(\k_2)$, or collectively
$\delta_X(\k)$ with $X\in\{A,B\}$.
Assuming $K\ll k_1,k_2$, we will refer to $\delta_L$ as the `long mode' and the tracer modes
as the `short modes', i.e.\ those which we can directly measure.

How do the two-point statistics of the short modes change when measured in 
the presence of the long mode?
Provided $\delta_L(\K)\ll1$, the effect of the long mode
on the two-point function between $A$ and $B$ can be expressed as a
functional Taylor expansion, taking the general form of
a linear response~\citep{Lewis:2011, Horn_2014, Chiang:2014oga}:
\bea
\big\langle\delta_A({\k_1})\sp\delta_B({\k_2})\big\rangle_{\delta_L}
&= \big\langle\delta_A({\k_1})\sp\delta_B({\k_2})\big\rangle_{\delta_L}\big|_{\delta_L=0}
+\int_\kl
        \frac{\partial \big\langle\delta_A(\k_1)\sp\delta_B(\k_2)\big\rangle_{\delta_L}}{\partial\delta_L(\K)}\,\Bigg|_{\delta_L=0}\ssp
        \delta_L(\kl)
    +\mathcal{O}(\delta_L^2)\,,
        \label{eq:2pcf-all}
\eea
where $\langle\cdots\rangle_{\delta_L}$ denotes an average over
the short modes but with fixed long mode $\delta_L(\K)$ for $\K=\k_1+\k_2$.
(We define these conditional averages more rigorously in Appendix~\ref{app:conditional-averages}.)
Note that only the long mode is assumed to be in the linear regime, with no requirement
on $\delta_X(\k)$, so that $\k$ can be deep in the nonlinear regime. Thus the effect of the long mode
is well described by the linear response and we can ignore the quadratic terms.

In the absence of the long mode across a given patch in which the short modes are measured, Eq.~\eqref{eq:2pcf-all} reduces to the usual relation $\langle\delta_A({\k_1})\sp\delta_B({\k_2})\rangle = (2\pi)^3\delD(\k_1+\k_2)P_{AB}(\k_1)$, i.e.\ modes with $\k_1\neq-\k_2$
are statistically independent.
By contrast, the presence of a long mode induces correlations between
modes $\k_1\neq-\k_2$, described by the linear response function
\be\label{eq:fAB-as-deriv}
f_{AB}(\k_1,\k_2)
% \equiv
% \bigg\langle\frac{\partial[\delta_A({\k_1})\sp\delta_B({\k_2})]}{\partial \delta_L(\kl)}\bigg\rangle_0^{\!\!\prime}
\equiv\frac{\partial\big\langle\delta_A(\k_1)\sp\delta_B(\k_2)\big\rangle'_{\delta_L}}{\partial\delta_L(\K)}\,\Bigg|_{\delta_L=0}\;,
\ee
where the prime indicates the triangle equality is understood, $\k_1+\k_2=\K$.
Substituting this into Eq.~\eqref{eq:2pcf-all} yields
\be\label{eq:off-diag}
\big\langle\delta_A({\k_1})\sp\delta_B(\k_2)\big\rangle_{\delta_L}
\simeq f_{AB}(\k_1,\k_2)\sp\delta_L(\kl)\,,
\qquad
\k_1+\k_2=\kl \neq 0\,.
\ee
This key relation tells us that the small-scale power traces large-scale
fluctuations; it is analogous to the coupling between different harmonics in the
CMB induced by a large-scale lensing potential~\citep{Hu:2001kj, LEWIS_2006},
a fact which we will use in Section~\ref{sec:QEs} to construct an estimator for $\delta_L(\K)$.

An explicit expression for the linear response function $f_{AB}$ can be
obtained by
multiplying both sides of Eq.~\eqref{eq:off-diag} by
$\delta_L^{*}(\kl)$, then taking the ensemble average over all
realizations of these long modes.
We thus have
$\langle\langle\delta_A({\k_1})\delta_B({\k_2})\rangle_{\delta_L}\,\delta_L^*(\kl)
\rangle
=f_{AB}(\k_1,\k_2)\ssp\langle \delta_L(\kl)\delta_L^*(\kl)\rangle$,
where the triangle equality $\k_1+\k_2=\kl$ is understood.
Then by the chain rule,
$\langle\langle\delta_A({\k_1})\sp\delta_B({\k_2})\rangle_{\delta_L}\delta_L^*(\kl)\rangle
=\langle\delta_A({\k_1})\sp\delta_B({\k_2})\sp\delta_L^*(\kl)\rangle$,
we have
\be\label{eq:response-bispec}
f_{AB}(\k_1,\k_2)
\simeq\frac{B_{ABL}(\k_1,\k_2,-\K)}{\PL(\K)},
\qquad K\ll k_1,k_2.
\ee
Here the bispectrum is defined as
$\langle\delta_A({\k_1})\sp\delta_B({\k_2})\sp\delta_L^*(\kl)\rangle
=(2\pi)^3\delD(\k_1+\k_2-\K)\sp B_{ABL}(\k_1,\k_2,{-}\K)$.
The power spectrum of long modes, safely in the linear regime,
is $\langle\delta_L(\kl)\sp\delta_L(\kl')\rangle=(2\pi)^3\delD(\kl+\kl')\PL(\K)$.
This relation tells us that $f_{AB}$ is determined by squeezed configurations of the
bispectrum, and vice-versa~\citep{Takada:2013wfa, Sherwin_2012, Wagner_2015, Barreira:2017sqa}.

As in the consistency relations, we emphasize that Eqs.~\eqref{eq:2pcf-all} and
\eqref{eq:response-bispec} only assume the long mode
is in the perturbative regime, described by linear theory.
The short modes $\delta_A(\k_1),\delta_B(\k_2)$ in these
expressions can be deep in the nonlinear regime (and 
may even be nonperturbative).
However, there is a modelling cost to performing the
test with highly nonlinear modes. This is because
the linear response function~\eqref{eq:fAB-as-deriv} needs
to be computed to high enough order in perturbation theory, 
as appropriate for those $\k_1$ and $\k_2$. To keep the
discussion simple, we will now limit our
short modes to scales described by second-order perturbation 
theory, the lowest order at which the anti-symmetric
effect appears.
The quadratic estimator to follow  (Section~\ref{sec:QEs})
is based on filtering the measured modes according to some
maximum wavenumber $k_{\max}$, below which our perturbative
treatment is valid (we test this on simulations).

\subsection{Decomposition and characterization of the linear response function}
We obtained the response~\eqref{eq:response-bispec} without assuming any
particular gravitational evolution or bias model; now we will
compute the response expected from 
standard perturbation theory (SPT)~\citep{Bernardeau:2001qr} together
with the usual treatment of galaxy bias~\citep{bias_review}.
There are two ways the response can be computed: directly from
the definition~\eqref{eq:fAB-as-deriv} or through the
squeezed bispectrum~\eqref{eq:response-bispec}.
Here we compute it using the former approach (with many of the
details relegated to Appendix~\ref{app:response-calculation}).
We assume Gaussian initial 
conditions and postpone a discussion of the non-Gaussian case to
Section~\ref{sec:discussion}.

Working to second order in SPT, as necessary to obtain the leading-order
effects in Eq.~\eqref{eq:response-bispec}, there are three 
ways a density mode can be affected by gravity,
namely, through
isotropic expansion or contraction (growth; G),
displacements due to advection (shift; S), and
anisotropic shear due to tides (tidal; T).
See, e.g.~\cite{1992ApJ...394L...5B, Sherwin_2012}.
These effects are plain to see in position space where
the second-order matter perturbation reads
\be\label{eq:delta_m-2nd}
\delta_m^{(2)}(\x)
\;=\;\frac{17}{21}\sp\underbrace{[\delta_m^{(1)}(\x)]^2\phantom{\Big|\!\!}}_{\text{{\bf G}rowth}}
    \;-\;\underbrace{\bm\psi^{(1)}(\x)\cdot\bm\nabla\delta_m^{(1)}(\x)\phantom{\Big|\!\!}}_{\text{{\bf S}hift}}
    \;+\;\frac27\sp \underbrace{s^{(1)}_{ij}(\x)\sp s^{(1)}_{ij}(\x)\phantom{\Big|\!\!}}_{\text{{\bf T}idal}}\,,
\ee
in which $\bm\psi^{(1)}=-\bm\nabla\nabla^{-2}\delta_m^{(1)}$ is the displacement field and
$s_{ij}=(\nabla_i\nabla_j-\frac13\delta_{ij}\sp\nabla^2)\sp\nabla^{-2}\delta_m$
the tidal field. Although here we assumed SPT and $\Lambda$CDM,
the form of Eq.~\eqref{eq:delta_m-2nd} is quite general.
Indeed, for many alternative theories of gravity the second-order matter perturbation can
be decomposed according to growth, shift and tidal effects~\citep{Fujita:2020xtd,DAmico:2021rdb}; only the
coefficients in Eq.~\eqref{eq:delta_m-2nd} differ. 

In this work we will take for our baseline response the 
standard coefficients $17/21$ and $2/7$, which we recall are
specific to an Einstein-de
Sitter cosmology~\citep{Bernardeau:2001qr} and, to a very good 
approximation, $\Lambda$CDM as well~\citep{Takahashi:2008yk}.
{That the coefficient of the shift is
unity is because this contribution is related more fundamentally to the
symmetries of the gravitational theory 
and not to the background cosmology~\citep{Dai:2015jaa,Scoccimarro:1995if}
nor the particular details of spherical collapse~\citep{Bernardeau:2001qr}.}
For example, Horndeski theories (which include $f(R)$ gravity, 
quintessence, k-essence models) modify the coefficients of the growth and tidal
contributions, while leaving the shift unchanged~\citep{Crisostomi:2019vhj,Lewandowski:2019txi}.

\subsubsection{Response of small-scale matter correlations to a large-scale matter mode}
The simplest response that we can consider is that of matter only, and here
we compute the response $f(\k_1,\k_2)$  of two short matter modes to a long matter mode.
For this we need the matter fluctuation up to second order. We have
$\delta_m(\k)\simeq\delta_m^{(1)}(\k)+\delta_m^{(2)}(\k)$ with
$\delta_L(\k)=\delta_m^{(1)}(\k)$ the linear mode and
\be\label{eq:delta_m-2nd-Fourier}
\delta_m^{(2)}(\k)
=\int_\q F_2(\q,\k-\q)\sp\delta_m^{(1)}(\q)\sp\delta_m^{(1)}(\k-\q)\,,
\ee
where $F_2(\q_1,\q_2)$ is the usual second-order
mode-coupling kernel for matter, encoding the three effects in 
Eq.~\eqref{eq:delta_m-2nd}. It is then convenient to decompose
$F_2$ into three mode-coupling kernels corresponding to
the three effects in Eq.~\eqref{eq:delta_m-2nd}:
$F_2(\k_1,\k_2) =\sum_\alpha F_\alpha(\k_1,\k_2)$, $\alpha\in\{\mrm{G,S,T}\}$, 
where $F_\alpha$ is given in Table~\ref{tab:modecouplings}.

Starting from the response definition~\eqref{eq:fAB-as-deriv}, and inserting into it
Eq.~\eqref{eq:delta_m-2nd-Fourier}, the leading-order response is found to be
(see Appendix~\ref{app:f-matter} for derivation; see also 
\citealt{Jeong:2012df,Li:2020uug,Darwish:2020prn})
\be\label{eq:standard-response}
f(\k_1,\k_2)
\simeq\bigg\langle
    \frac{\partial\delta_m^{(2)}(\k_1)}{\partial\delta_m^{(1)}(\K)}
    \delta_m^{(1)}(\k_2)
\bigg\rangle^{\!\!\prime}
+(\k_1\leftrightarrow\k_2)
=2\sp\big[F_2(\k_1+\k_2,-\k_2)\sp \PL(k_2)+(\k_1\leftrightarrow\k_2)\big]\,,
\ee
where the two terms follow from the product rule,
$\PL$ is the linear matter power spectrum, and in
the last equality $\k_1+\k_2=\K$.
Note that the response can be computed directly from the 
definition~\eqref{eq:fAB-as-deriv}, as we have done here,
or alternatively using Eq.~\eqref{eq:response-bispec} by taking the squeezed limit
of the bispectrum. The latter is the more common approach
(see, e.g.\ \citealt{Darwish:2020prn}), while we find
the former approach more convenient for isolating the effects we are after.

\renewcommand{\arraystretch}{1.45}
\renewcommand{\tabcolsep}{6pt}
\begin{table}[t!]
\centering
\vspace{2mm}
\begin{tabular}{ l l r r }
\toprule
$\alpha$ & {\bf Kernel} $\,F_{\alpha}(\k_1, \k_2)$ & {\bf Kernel coefficient} $\,c^\alpha_X$ & {\bf Response coefficient} $C^{\alpha}_{XY}\equiv c^\alpha_X\sp b_{1Y}$ 
 \\
\cline{1-4}
G & ${17}/{21}$ &  $b_{1X}+\frac{21}{17}\sp b_{2X} + \color{myr}{\epsilon\sp b_{\epsilon,\mrm{G}X}}$ &  $(b_{1X}+\frac{21}{17}\sp b_{2X} + {\color{myr}\epsilon b_{\epsilon,\mrm{G}X}})\sp b_{1Y}$
\\[5pt]
S & $\frac{1}{2}\big(\frac{1}{k_1^2}+\frac{1}{k_2^2}\big)(\k_1\cdot\k_2)$ & $b_{1X}+\color{myr}{\epsilon\sp b_{\epsilon,\mrm{S}X}}$ & $(b_{1X}+{\color{myr}\epsilon\sp b_{\epsilon,\mrm{S}X}})\sp b_{1Y}$ 
\\[5pt]  
T & $\frac{2}{7} \big[ \big(\frac{\k_1\cdot \k_2}{k_1 k_2}\big)^2 -\frac{1}{3} \big]$ & $b_{1X} + \frac{7}{2}\sp b_{s^2X}+\color{myr}{\epsilon\sp b_{\epsilon,\mrm{T}X}}$ &  $(b_{1X} + \frac{7}{2}\sp b_{s^2X}+{\color{myr}\epsilon\sp b_{\epsilon,\mrm{T}X}})\sp b_{1Y}$ \\[3pt]
\botrule
\end{tabular}
\caption{
Mode-coupling types $\alpha$, mode-coupling kernels $F_\alpha$, mode-coupling coefficients $c^\alpha_X$, 
and response coefficients $C^{\alpha}_{XY}\equiv c^{\alpha}_{X}b_{1Y}$
of the quadratic bias model.
Here $b_{1X}$ is the linear bias, $b_{2X}$ is the
quadratic local bias, and $b_{s^2X}$ is the tidal bias
of tracer $X$. Note that the EP protects the shift (S) from acquiring a second-order bias. But when the EP is violated, additional terms (in {\color{myr}red}) arise as `perturbations' to the usual total biases $c_{X}^{\alpha}$ (in black).}
\label{tab:modecouplings}
\end{table}

\subsubsection{Response of small-scale tracer correlations to a large-scale matter mode}
The structure of the response for two distinct
tracers $A$ and $B$ is more complex, giving rise to response types
precluded by matter. Regardless, the general response can be
decomposed into the G, S, T basis:
\be
f_{AB}(\k_1,\k_2)
\equiv\sum_{\alpha \in \{\mathrm{G,S,T}\}} 
f^\alpha_{AB}(\k_1,\k_2)\,. \label{eq:full_response_AB}
\ee
To obtain the individual responses $f^\alpha_{AB}$ we begin as before
with the definition
of the linear response~\eqref{eq:fAB-as-deriv} and evaluate
it using second-order SPT and now galaxy bias
(see Appendix~\ref{app:response-calculation} for explicit calculation).
In terms of the $F_\alpha$ kernels given in Table~\ref{tab:modecouplings}, 
$f^\alpha_{AB}$ is found to have the basic form
\be\label{eq:f-alpha-AB}
f^\alpha_{AB}(\k_1,\k_2)
=2\sp\Big[C^\alpha_{AB}\sp{F}_{\alpha}(\k_1+\k_2,-\k_2)\sp {\PL}(k_2)
+C^\alpha_{BA}\sp{F}_{\alpha}(\k_1+\k_2,-\k_1)\sp {\PL}(k_1)\Big]
\equiv f^\alpha_{(AB)}+f^\alpha_{[AB]}\,.
\ee
Here $C^\alpha_{AB}$ and $C^\alpha_{BA}$ are
tracer-dependent coefficients
given by quadratic combinations of the bias parameters; in general,
they consist of symmetric and anti-symmetric parts (under exchange of $A$ and $B$).%
\footnote{Note that the form of Eq.~\eqref{eq:f-alpha-AB}
is generally
valid even if $\alpha$ runs over more than just G,S,T; it may
include other types of mode coupling, as allowed by
primordial non-Gaussianity. See 
\cite{Darwish:2020prn} for examples.}
Thus, let us decompose each response into symmetric and anti-symmetric
parts
as $f^\alpha_{AB}=f^\alpha_{(AB)}+f^\alpha_{[AB]}$, where each
term is defined as in Eq.~\eqref{eq:sym-antisym-defn}.
The response coefficients $C^\alpha_{(AB)}$ and $C^\alpha_{[AB]}$ are defined
similarly.
With these decompositions, we can extract from
Eq.~\eqref{eq:f-alpha-AB} the symmetric and anti-symmetric parts,
\be\nonumber
f^\alpha_{(AB)}(\k_1,\k_2)
= {C}^\alpha_{(AB)}\, f^{(+)}_\alpha(\k_1,\k_2)\,,
\qquad
f^\alpha_{[AB]}(\k_1,\k_2)
={C}^\alpha_{[AB]}\, f^{(-)}_\alpha(\k_1,\k_2) \,,
% \label{eq:fAB-alpha-anti}
\ee
where $f^{(\pm)}_\alpha$ are the
basic symmetric ($+$) and anti-symmetric ($-$) responses, given by
\be\label{eq:f_alpha-pm}
f^{(\pm)}_\alpha(\k_1,\k_2)
\equiv 2\sp
\Big[{F}_{\alpha}(\k_1+\k_2,-\k_2)\sp {\PL}(k_2)
\pm (\k_1\leftrightarrow\k_2)\Big]\,,
\ee
which are tracer independent. Note that the
sum of the symmetric responses is equal to the {standard} matter
response~\eqref{eq:standard-response}:
$\sum_\alpha f^{(+)}_\alpha(\k_1,\k_2)\equiv f(\k_1,\k_2)$.
Assuming Gaussian initial conditions, we have six distinct
response types, three symmetric
$\{f^{(+)}_\mrm{G},f^{(+)}_\mrm{S},f^{(+)}_\mrm{T}\}$ and
three anti-symmetric $\{f^{(-)}_\mrm{G},f^{(-)}_\mrm{S},f^{(-)}_\mrm{T}\}$.
These six responses define a basis in which the
total linear response~\eqref{eq:full_response_AB} reads
\be\label{eq:fAB-general}
\boxed{
f_{AB}(\k_1,\k_2)
=\sum_{\alpha \in \{\mathrm{G,S,T}\}}^{\phantom{a}}
    \Big[{C}^\alpha_{(AB)}\ssp f^{(+)}_\alpha(\k_1,\k_2)
        +{C}^\alpha_{[AB]}\ssp f^{(-)}_\alpha(\k_1,\k_2)\Big]\,.
}
\ee
Written in this way, the response coefficients ${C}^\alpha_{(AB)}$ and
${C}^\alpha_{[AB]}$ capture the tracer dependence.
In this second-order calculation the response coefficients are made
up of quadratic combinations of the bias coefficients (see Table~\ref{tab:modecouplings_effective_coefficients}).

Gravitational dynamics and the symmetries of galaxy formation (encoded in the galaxy bias expansion)
together determine the types of responses allowed. For the standard galaxy bias expansion,
both symmetric and anti-symmetric parts of the growth and tidal responses are nonzero
($C^\mrm{G}_{AB}\neq0$ and $C^\mrm{T}_{AB}\neq0$, for $A \neq B$) due to nonlinear clustering of biased tracers, as also
pointed out by \cite{Dai:2015wla}.
% as these terms depend on local gravitational effects that cannot be absorbed by a coordinate transformation. 
Furthermore, the symmetric shift response can be nonzero.
On the other hand, for the anti-symmetric shift response the EP requires 
\be\label{eq:EP-condition}
C^\mrm{S}_{[AB]}=0\,.
\ee
It is straightforward to show that the standard galaxy bias relation respects this condition.
As such, this response coefficient encodes EP violation via $C^\mrm{S}_{[AB]}\neq 0$.
We note that \cite{Fujita:2020xtd} obtained a similar relation but
using a different approach.

\subsection{Application to standard galaxy biasing}\label{sec:bias-application}
We now illustrate the formalism by applying the symmetric--anti-symmetric decomposition to
the standard galaxy
bias expansion, showing the structure of the responses in terms of
$f^{(\pm)}_\alpha$. 

To express the auto and cross responses~\eqref{eq:f-alpha-AB} in terms
of $f^{(\pm)}_\alpha$ we will need the quadratic bias expansion~\citep{bias_review}
\be\label{eq:delta_g-second}
\delta_X
=b_{1X}\sp\delta_m + b_{2X}\sp\delta_m^2
+ b_{s^2X}\ssp s^2 ,
    % +b_{\nabla\delta}\ssp v_r^i\nabla_i\sp\delta_L
\ee
where $b_{1X}$ and $b_{2X}$ are the linear and quadratic local biases, 
$b_{s^2 X}$ the tidal bias, and $s^2\equiv s_{ij}s^{ij}$ with
$s_{ij}=(\nabla_i\nabla_j-\frac13\delta_{ij}\sp\nabla^2)\sp\nabla^{-2}\delta_m$ the
tidal field.

\renewcommand{\arraystretch}{1.45}
\renewcommand{\tabcolsep}{6pt}
\begin{table}[t!]
\centering
\vspace{1mm}
\begin{tabular}{lrr}
\toprule
$\alpha$ & {\bf Symmetric response coeff.}\: $C^{\alpha}_{(XY)}=c^\alpha_{(X}\sp b_{1Y)}^{\phantom{\alpha}}$ & {\bf Anti-symmetric response coeff.}\: $C^{\alpha}_{[XY]}=c^\alpha_{[X}\sp b_{1Y]}^{\phantom{\alpha}}$   \\
\cline{1-3}
G & $\textcolor{black}{b_{1X} b_{1Y} + \frac{21}{17}\sp b_{2(X} b_{1Y)}} + \color{myr}{ \epsilon\sp b_{\epsilon,\mrm{G}(X} b_{1Y)}} $ & $\textcolor{black}{\frac{21}{17}\sp  b_{2[X} b_{1Y]}} + \color{myr}{ \epsilon\sp b_{\epsilon,\mrm{G}[X} b_{1Y]}}$
\\[5pt]
S & $b_{1X}b_{1Y} + \color{myr}{\epsilon\sp b_{\epsilon,\mrm{S}(X} b_{1Y)}}$ & $\color{myr}{\epsilon\sp b_{\epsilon,\mrm{S}[X} b_{1Y]}}$
\\[5pt]  
T & $b_{1X} b_{1Y} + \frac{7}{2}\sp b_{s^2(X} b_{1Y)}  + \color{myr}{\epsilon\sp b_{\epsilon,\mrm{T}(X} b_{1Y)}}$  &  $\frac{7}{2}\sp b_{s^2[X} b_{1Y]} + \color{myr}{\epsilon \sp b_{\epsilon,\mrm{T}[X} b_{1Y]}}$ \\[3pt]
\botrule
\end{tabular}
\caption{
Symmetric and anti-symmetric bias coefficients appearing in the quadratic estimator, for different mode couplings. We highlight in red terms that can generally arise in scenarios violating the EP. Notably, the anti-symmetric shift term is the only one serving as a smoking gun of EP violation.
}
\label{tab:modecouplings_effective_coefficients}
\end{table}

Substituting Eq.~\eqref{eq:delta_g-second} into the right-hand side of
Eq.~\eqref{eq:fAB-as-deriv} and using Eq.~\eqref{eq:delta_m-2nd}, we find
\be
\begin{array}{l@{\qquad\quad}l}
\textbf{\textit{Auto response}} &
\textbf{\textit{Cross response}}  \\[3pt]
\begin{aligned}
f_{AA}(\k_1,\k_2)
&= {C}^\mrm{G}_{(AA)}\ssp f^{(+)}_\mrm{G}(\k_1,\k_2) + 0 \\
& + {C}^\mrm{S}_{(AA)}\ssp f^{(+)}_\mrm{S}(\k_1,\k_2) + 0 \\
& + {C}^\mrm{T}_{(AA)}\ssp f^{(+)}_\mrm{T}(\k_1,\k_2) + 0\,,
\end{aligned}
&
\begin{aligned}
f_{AB}(\k_1,\k_2)
&= {C}^\mrm{G}_{(AB)}\ssp f^{(+)}_\mrm{G}(\k_1,\k_2) 
+ {C}^\mrm{G}_{[AB]}\ssp f^{(-)}_\mrm{G}(\k_1,\k_2) \\
& + {C}^\mrm{S}_{(AB)}\ssp f^{(+)}_\mrm{S}(\k_1,\k_2) + 0 \\
& + {C}^\mrm{T}_{(AB)}\ssp f^{(+)}_\mrm{T}(\k_1,\k_2) 
+ {C}^\mrm{T}_{[AB]}\ssp f^{(-)}_\mrm{T}(\k_1,\k_2)\,,
\end{aligned}
\end{array}
\ee
where $C^\alpha_{AB}=c^{\alpha}_A\sp b_{1B}$ with $c^\alpha_{A}$  given in
Table~\ref{tab:modecouplings}.
Here we see that $f_{AA}$ has a similar structure to the 
matter response~\eqref{eq:f_alpha-pm}; however,
each $f_\alpha^{(+)}$ may be enhanced (or suppressed) relative
to matter, depending on the bias parameters.
Note that in $f_{AA}$ the anti-symmetric responses necessarily vanish
in auto correlation, but this is not the case for the cross-response $f_{AB}$.

The tableau of $f_{AB}$ shows that the
long mode excites responses of all types, {\it except the anti-symmetric shift $f_\mrm{S}^{(-)}$}. We already anticipated this result above in Eq.~\eqref{eq:EP-condition}:
${C}^\mrm{S}_{[AB]}=0$ which we can state more explicitly in terms
of the bias coefficients as
$b_{1A}b_{1B}-b_{1B}b_{1A}=0$ for $A\neq B$.
The underlying assumption of the bias expansion~\eqref{eq:delta_g-second}
which leads to this is that the EP holds.
{If it does not hold ${C}^\mrm{S}_{[AB]}\neq0$, leading to
a nonzero anti-symmetric shift.} Since we expect any departures from standard bias coefficients to be small we will parametrize modifications by a dimensionless
parameter $\epsilon$ around standard biases (in general one for each G, S, and T).

\subsection{EP-violating response function }\label{sec:epviolationmodel}

Consider now the simplified second-order bias expansion \citep{Bottaro:2023wkd}:
\begin{equation}\label{eq:ep-violation-bias-model}
    \delta_X = b_{1X}\delta_m+b_{2X}\delta_m^2+b_{s^2X}s^2+b_{rX}\delta_{r} \,,
\end{equation}
where $\delta_{r}=\delta_{\mrm{cdm}}-\delta_\mrm{b}$ is the relative density contrast between dark matter $\delta_{\mrm{cdm}}$ and baryonic matter $\delta_{\mrm{b}}$, and $b_{rX}$ is a relative density parameter. A dark force leads to a different evolution of dark matter with respect to baryons. One can show that this leads to a new EP-violating term of the form $\epsilon b_{\epsilon,S} \delta_m(\mathbf{x})$, where $\epsilon$ is a small parameter with associated bias coefficient $b_{\epsilon,S}$, that leads to the $1/K$ dipole in the squeezed bispectrum \citep{Bottaro:2023wkd}.

Using Table~\ref{tab:modecouplings} it is straightforward to show  that such an operator leads to a shift-type mode coupling, $F_\mrm{S}(\k_1,\k_2)$ (upon symmetrization). This implies
${C}^\mrm{S}_{[AB]}\neq0$, in violation of the condition~\eqref{eq:EP-condition}. An example of a model giving rise to such a violation is a long-range force coupled to dark matter (dark fifth force) that can induce composition-dependent accelerations, manifesting as a velocity-dependent bias term~\citep{Carroll_2009, Bottaro:2023wkd}.

To expose the pole in the bispectrum in Eq.~\eqref{eq:response-bispec} we take the limit $K\to0$ of $f^{(-)}_\mrm{S}$.
Since $K\ll k_1,k_2$ we Taylor expand the 
responses~\eqref{eq:f_alpha-pm} in the small quantity $K/k$.
At leading order we have
%\be
%f^{(-)}_\mrm{S}(\k,\kl-\k)
%=2\sp P_\mrm{lin}(k)\left[\frac{\kl\cdot\k}{K^2}
%    +\mathcal{O}((K/k)^0)\right],
%    \label{eq:responseofinterest}
%\ee
\be
f^{(-)}_\mrm{S}(\k,\kl-\k)
=2\sp \PL(k)
    \Bigg[
        % \underbrace{\frac{\kl\cdot\k}{K^2}}_{{\text{Shift dipole}}}
        \underset{\phantom{\Big|}{\substack{\text{Shift} \\ \text{dipole}}}}{\frac{\kl\cdot\k}{K^2}}
        +\mathcal{O}((K/k)^0)
    \Bigg],
    \label{eq:responseofinterest}
\ee
where $\k_1\equiv\k$ and $\k_2=\kl-\k$
and the order $(K/k)^0$ terms have slight scale dependence in $k$
(through the spectral slope) but remain order unity.
We therefore see that the expected $1/K$ pole, present
when the consistency relations are violated, is contained in
this particular response.
In particular, we emphasize the dipolar nature of this response.

Returning to the case when the EP holds, this response can be contrasted with the
total matter response
$f(\k,\K-\k)=\sum_\alpha f_\alpha^{(+)}(\k,\K-\k)$.
Expressed in terms of $k$, $\mu=\k\cdot \kl/(kK)$ and the small quantity
$K/k\ll1$, this response reads at leading order
\be\label{eq:fullrespgravsqueezed}
f(k,K/k,\mu)
=\PL(k) \Bigg[\underset{\phantom{\Big|}{\substack{\text{Growth} \\ \text{monopole}}}}{\frac{68}{21}}\!\!\!\!\!-\;\underset{\phantom{\Big|}{\substack{\text{Shift} \\ \text{monopole}}}}{\Big(1+\frac{1}{3}\frac{\dif\ln \PL}{\dif\ln k}\Big)}-\underset{\phantom{\Big|}{\substack{\text{Shift} \\ \text{quadrupole}}}}{\frac{2}{3}\frac{\dif\ln \PL}{\dif\ln k}\mathcal{L}_2(\mu)}
    +\underset{\phantom{\Big|}{\substack{\text{Tidal} \\ \text{quadrupole}}}}{\frac{8}{7}\frac{2}{3}\frac{\dif\ln \PL}{\dif\ln k}\mathcal{L}_2(\mu)}
    +\mathcal{O}\Big(\frac{K}{k}\Big)
    \Bigg] \,,
\ee
where $68/21$ and $8/7$ are model-dependent coefficients, and 
$\mathcal{L}_2$ is the Legendre polynomial of second degree.
(See Appendix~\ref{app:response-expansions} for
expansions of $f_\alpha^{(\pm)}$ when $K/k\ll1$.)
This form highlights
the multipole structure of the standard gravitational 
effects~\citep{Akitsu:2016leq, Barreira:2017sqa, Li:2017qgh, Zhu:2021qmm}. Note that
the monopole and quadrupole terms are analogous to the magnification and shear effects of
CMB lensing~\citep{Bucher:2010iv, Prince:2017sms, Schaan:2018tup, Carron:2024mki}.
At leading order the symmetric growth and tidal effects are associated with
a monopole and quadrupole, respectively
(although note that at higher order in $K/k$ there are contributions from
higher-order multipoles).
In contrast, notice that the shift,
nominally associated with the dipole and present in Eq.~\eqref{eq:responseofinterest}, 
does not appear by virtue of the EP.

\section{Quadratic estimators for distinct tracers}\label{sec:QEs}
Having characterized the different response functions in the previous section, and identified the 
anti-symmetric shift~\eqref{eq:responseofinterest} as the response of interest,
we now present the quadratic estimator (QE) framework to probe for EP violation.
Our approach is to reconstruct the long mode for a given input response and then cross-correlate the long mode
with an external tracer (which can be the short modes used in the reconstruction). If the EP is violated,
we will show that it manifests in our estimator as a characteristic flat bias.

In this section we present the main equations used in the forecasts to follow.
Additional plots and supporting material can be found in Appendix~\ref{app:qeestimators}.

\subsection{Formalism}\label{sec:formalism}

Based on Eq.~\eqref{eq:off-diag}, we can write down a quadratic estimator
$\widehat{h}_{XY}^\alpha$ for the long mode $\delta_L(\K)$ by
taking a suitably weighted combination of the short modes $\delta_{X}(\k_1)$ and $\delta_{Y}(\k_2)$
with $X,Y\in\{A,B\}$:
\be
\widehat{h}^{\alpha}_{XY}(\kl)
\equiv \widehat{\delta}_L(\K) = \int_{\k} w_{XY}^{\alpha}(\k, \kl-\k)\ssp\delta_{X}(\k)\ssp\delta_{Y}(\kl-\k) , \label{eq:estimatorgeneral}
\ee
where $w_{XY}^{\alpha}(\k, \kl-\k)$ is some arbitrary weight to be specified
shortly. The integral over small-scales $\k$ runs over wavenumbers in the range $[k_{\mrm{min,rec}},k_{\mrm{max,rec}}]$, where the maximum scale of reconstruction $k_{\mrm{max,rec}}$ is generally imposed by our theoretical control, e.g.\ how well our perturbation theory works. On the other hand, the reconstructed long modes have wavenumbers ranging over $[K_{\mrm{min}}, K_{\mrm{max}}]$, where $K_{\mrm{min}}$ is the lowest wavenumber we want to reconstruct, and $K_{\mrm{max}} < k_{\mrm{min,rec}}$ (to ensure reconstructed modes and modes of reconstruction do not overlap).

The form of the weights $w_{XY}^{\alpha}$ is typically fixed
by requiring that the estimator~\eqref{eq:estimatorgeneral} is both minimum variance (in Gaussian noise) and unbiased, i.e.\ that the true long mode is recovered upon taking the expectation value, $\langle \widehat{h}^\alpha_{XY}\rangle_{\delta_L}=\delta_L$.
However, as we need two different matter tracers, in our case this leads to non-separable expressions that are difficult to evaluate in practice, see Appendix \ref{app:qeestimators}.
Hence, we will relax the requirement of minimum variance at the cost of increasing the reconstruction noise, instead defining the weights as
\begin{equation}\label{eq:w-alpha-AB}
    w_{XY}^{\alpha}(\k, \kl-\k)
    \equiv N_{XY}^{\alpha}(\K)\,
    \frac{f^{\alpha}_{XY}(\k, \kl-\k)}{2 \Ptot^{XX}(\k)\Ptot^{YY}(\kl-\k)}\ ,
\end{equation}
where $P_\mrm{tot}^{XX}(\k)=\langle \delta_X(\k)\delta_X(\k')\rangle'$ is the total tracer power spectrum (which may include shot noise), $N_{AB}^{\alpha}$ is some normalization factor to guarantee an unbiased estimate of the long mode, and $f^{\alpha}_{XY}$ is the relevant response from the bispectrum. For the experimental configurations we explore in this work, this results in little signal-to-noise loss, as we numerically verify in Appendix \ref{app:qeestimators}.

\subsubsection{Displacement estimator}

As we saw above, the anti-symmetric shift response
$f^{(-)}_\mrm{S}$ uniquely probes an EP violation.
For simplicity we take for $f^\alpha_{XY}$ the
response function
\begin{equation}
f^{\mathcal{D}}(\mathbf{k}, \mathbf{K}-\mathbf{k})
\equiv 2\sp \PL(\mathbf{k})\sp\frac{\kl\cdot\k}{K^2}\,,
% \sim g_{\mrm{S}}(\mathbf{k}, \mathbf{K}-\mathbf{k})\ ,
\end{equation}
such that $f^{(-)}_\mrm{S}\sim f^\DD$ on large-scales ($K\to0$), see Eq.~\eqref{eq:responseofinterest}. For brevity, we call this the \emph{displacement response} and label it $\DD$.
This approximate form captures the scale dependence of
$f^{(-)}_\mrm{S}(\k, \kl-\k)$ in the regime when $K/k$ is small and thus when the pole is large and dominant.

The normalization is constrained by requiring that Eq.~\eqref{eq:estimatorgeneral} be unbiased, $\langle \widehat{h}^{\DD}_{XY}\rangle_{\delta_L} = \delta_L(\K)$,
in the event that $f^{\mathcal{D}}$ is the only mode coupling present.
This yields the constraint
$\int_{\,\k} w_{XY}^{\DD}(\k, \kl-\k) f^{\mathcal{D}}(\k, \kl-\k) = 1$.
Inserting this into Eq.~\eqref{eq:w-alpha-AB} with $f^\alpha_{XY}\to f^\DD$ and rearranging for the normalization factor, we obtain
\be\label{eq:ND}
N_{XY}^{\DD}(\K) 
= \bigg(\ssp\int_{\k} \frac{f^{\mathcal{D}}(\k, \kl-\k)^2}{2\sp P_\mrm{tot}^{XX}(\k)P_\mrm{tot}^{YY}(\kl-\k)}\bigg)^{-1}.
\ee
Our estimator, with the normalization~\eqref{eq:ND}, is sub-optimal. This implies that the variance $V_{XY}^{\DD\DD}$ induced by Gaussian chance fluctuations does not equal the normalization, as it does in the case of typical QEs. Nevertheless, it has a practical form which allows for fast evaluation,
and is sufficient for the forecasts to follow (see Appendix~\ref{app:qeestimators} where we 
compare our estimator with an optimal one). 
We leave practical reconstruction improvements for future work.

\begin{figure}
    \centering
\includegraphics[width=\linewidth]{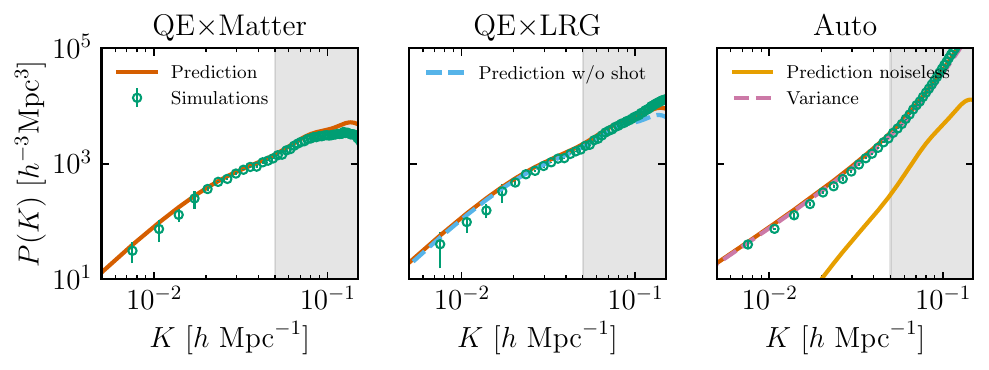}
    \caption{Application of displacement estimator $\widehat{h}^\DD_{XY}$ [Eq.~\eqref{eq:estimatorfourier}] to DESI LRG- and ELG-like mocks from the \textsc{AbacusSummit} simulations ($X=\mrm{LRG}$, $Y=\mrm{ELG}$), for redshift $z = 0.5$. Predictions are given by Eqs.~\eqref{eq:forecast_autorspec} and \eqref{eq:forecast_crossspec}, using rough approximate galaxy bias fits (see Appendix \ref{sec:sims} for details). 
    Error bars are given by the standard deviation of the mean.
    {\textit{Left panel}}: The cross correlation between the reconstruction and input matter field.  {\textit{Center panel}}: The cross-correlation between the reconstruction and
the LRG simulation. {\textit{Right panel}}: The auto correlation of the reconstruction. This is mainly explained by Gaussian variance contributions (dashed pink). As the cross- and auto- correlations of the reconstruction effectively measure three- and four-point functions, respectively, we also include the effect of shot noise from higher-order components (mixed bispectrum and trispectrum). The modes used in the reconstruction range from $k_{\mrm{min,rec}}=0.05h\,\mathrm{Mpc}^{-1}$ to $k_{\mrm{max,rec}}=0.2h\,\mathrm{Mpc}^{-1}$, as indicated by the grey bands.}
    \label{fig:simsgalaxydouble}
\end{figure}

Note that the estimator can also be written in a way that resembles the 
standard CMB lensing estimator~\citep{LEWIS_2006} (indeed, in Appendix~\ref{app:cmblensingder} we show how the
displacement estimator can be derived \`a la CMB lensing). Defining the
Wiener-filtered field
$\barl_X(\k) \equiv \bard_X(\k)\sp \PL(\k)$ and the inverse-variance-filtered field
$\bard_X(\k) \equiv {\delta_X(\k)}/{P^{XX}_{\mathrm{tot}}(\k)}$, the estimator reads
\begin{equation}
   \widehat{h}^\DD_{XY}(\kl) =  N_{XY}^{\DD}(\kl)\int_\k \frac{\K\cdot\k}{K^2} \ssp
   \barl_X(\k)\ssp\bard_{Y}(\kl-\k)\,.
   \label{eq:estimatorfourier}
\end{equation}
Figure \ref{fig:simsgalaxydouble} shows this estimator applied to four simulations from the \textsc{AbacusSummit} suite \citep{Maksimova, Garrison}. 
We use $k_{\mrm{max,rec}}=0.2 h\,\mathrm{Mpc}^{-1}$ rather than the baseline of $k_{\mrm{max,rec}}=0.15h\,\mathrm{Mpc}^{-1}$. Simple theory predictions explain the measured power spectra using approximate fits (see Appendix \ref{sec:sims}), though we see that the auto-spectrum is noise dominated (dashed pink, right panel). 

It is interesting to compare the estimator~\eqref{eq:estimatorfourier} with previous work. In particular, notice that on very large scales ($K \rightarrow 0$) the reconstruction noise goes as $K^2$. This matches the scaling found in the matter estimate from kSZ velocity reconstruction~\citep{Smith:2018bpn}. Both methods use dipole responses that track object shifts (compare Eq.~\eqref{eq:responseofinterest} with equation 24 in \citealt{Smith:2018bpn}). The key difference is that our signature vanishes in $\Lambda$CDM, while the momentum field of \cite{Smith:2018bpn} has a nonzero $1/K$ consistency relation (see also \citealt{Rizzo:2017zow}).

\subsubsection{Growth estimator}

In addition to the displacement estimator, one can also explore other estimators. For example, the (symmetric) growth estimator,
\begin{equation}
   \widehat{h}^{\mrm{G}_+}_{XY}(\kl) =  \int_\k w^{\mrm{G}_+}_{XY}(\k,\K-\k) \ssp 
   \delta_X(\k)\ssp\delta_{Y}(\kl-\k)=N_{XY}^{\mrm{G}_+}(\kl)\int_\k \frac{f^{(+)}_\mrm{G}(\k,\K-\k)}{2\Ptot^{XX}(\k)\Ptot^{YY}(\kl-\k)} \ssp 
   \delta_X(\k)\ssp\delta_{Y}(\kl-\k)
   \label{eq:growthestimatorfourier} \,,
\end{equation}
where the second equality follows from inserting the symmetric growth response $f_{\mrm{G}}^{(+)}$ [Eq.\eqref{eq:f_alpha-pm}] into the weight~\eqref{eq:w-alpha-AB},
and $N^{\mrm{G}_+}_{XY}$ is defined in a similar way to Eq.~\eqref{eq:ND}. This estimator reconstructs matter modes using the 
isotropic monopole moment of the total response. While this response is not protected by the EP, it has the benefit of allowing us to place constraints on higher-order bias parameters such as $b_2$, which in turn helps to break degeneracies
with the parameter of interest, $\epsilon$.
One can of course also build other estimators, using the symmetric shift and tidal effects.
But in this work we focus only on the displacement and the growth
estimators.

\subsection{QE bias for EP violation}
The main advantage of the QE approach is its ability to extract the anti-symmetric component of the response function, which directly probes EP violations. In reality, our estimator~\eqref{eq:estimatorfourier} will respond to various other physical effects besides the anti-symmetric shift term we seek. These effects include the growth and tidal terms, as well as the standard symmetric shift component.

To quantify the sensitivity of the displacement estimator~\eqref{eq:estimatorfourier} to each contribution $f^{(+)}_\alpha$ and $f^{(-)}_\alpha$, we define the QE biased response $B^\DD_{\alpha,XY}(\K)$ and the unnormalized QE response $R^\DD_{\pm\alpha}(\K)$
as follows:%
\footnote{Not to be confused with the theoretical response $f_{AB}$ discussed earlier.}
\bea\label{eq:RD}
B^{\mathcal{D}}_{\alpha, XY}(\kl) 
\equiv\int_{\mathbf{k}} w_{XY}^{\mathcal{D}}(\mathbf{k}, \mathbf{K} - \mathbf{k})\sp f^{\alpha}_{XY}(\mathbf{k}, \mathbf{K} - \mathbf{k})&\,,\\[3pt]
R^{\DD}_{\pm\alpha}(\kl)
\equiv N^{\mathcal{D}}_{XY}(\mathbf{K})^{-1}
    \int_{\mathbf{k}} w^\DD_{XY}(\k,\kl-\k)\sp f^{(\pm)}_{\alpha}(\mathbf{k}, \mathbf{K} - \mathbf{k})&\,. \label{eq:QE-response}
\eea
The former quantifies the tracer-dependent sensitivity
of the $\DD$ estimator to each $\alpha\in\{\mrm{G_+,S_+,T_+,G_-,S_-,T_-}\}$, while the latter, in 
addition to being unnormalized, gives a more tracer-independent measure (but note there is
still dependence on tracer through the weight).
Note that in terms of $R^\DD_{\pm\alpha}$ the QE biased response~\eqref{eq:RD} reads
\begin{align}\label{eq:calR-2}
B^{\mathcal{D}}_{\alpha, XY}(\kl) 
&= N^{\mathcal{D}}_{XY}(\mathbf{K}) \big[ C^{\alpha}_{(XY)} R^{\mathcal{D}}_{+\alpha}(\mathbf{K}) + C^{\alpha}_{[XY]}R^{\mathcal{D}}_{-\alpha}(\mathbf{K})\big].
\end{align}

\begin{figure*}[t!]
\centering
\includegraphics[width=.515\textwidth]{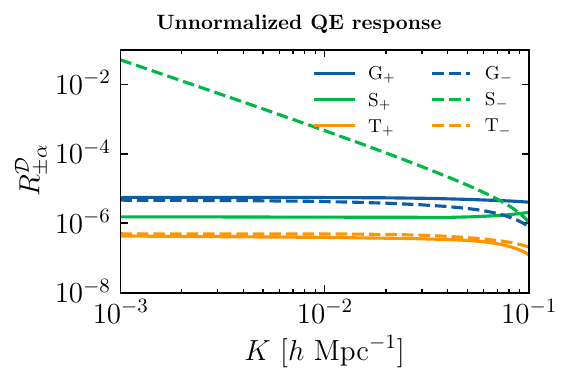}%
\includegraphics[width=.48\textwidth]{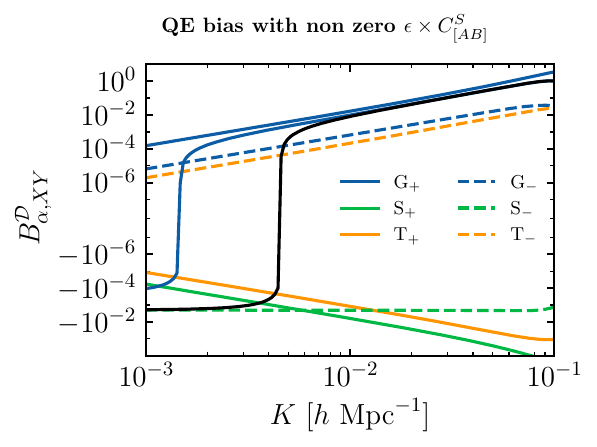}
\caption{{\textit{Left panel}}: Example of unnormalized symmetric and anti-symmetric QE responses $R^{\DD}_{\pm\alpha}$ [Eq.~\eqref{eq:QE-response}] for the growth, shift, and tidal terms. We clearly see a strong scale dependence in the displacement estimator due to the anti-symmetric shift response (dashed green).
{\textit{Right panel}}: Biased response $B^\DD_{\alpha,XY}$ [Eq.~\eqref{eq:RD}] of the displacement estimator. By construction, the anti-symmetric shift component is flat on large scales (dashed green). This is also reflected in the effective bias $b_\DD=\sum_\alpha B^\DD_{\alpha,XY}$ (solid black). For reference, we also show in solid grey $b_\DD$ when the $\epsilon$ amplitude is a factor of ten smaller. In this example we set $\epsilon=10^{-2}$, 
$b_{1X}=1.6,b_{1Y}=1.2$, $b_{2X}=b_{2Y}=-0.3$, and
$b_{\epsilon,\alpha X}=b_{\epsilon,\alpha Y}= 1$, for each $\alpha \in\{\mrm{G,S,T}\}$.}
\label{fig:bias_trend}
\vspace{2\baselineskip}
\end{figure*}

The left panel of Figure \ref{fig:bias_trend} shows these unnormalized QE responses for our displacement estimator. Importantly, we see that the unnormalized response for the anti-symmetric shift term (dashed orange line) exhibits a stronger scale-dependence compared with the other couplings, increasing on larger scales. This makes it an ideal probe for EP violations.\footnote{This was expected by comparing the response in Eq.~\eqref{eq:responseofinterest} with the rest of the responses: the symmetric responses ($f_\mrm{G}^{(+)}$, $f_\mrm{S}^{(+)}$, and $f_\mrm{T}^{(+)}$)
begin at order $(K/k)^0$, while the other anti-symmetric
responses ($f_\mrm{G}^{(-)}$ and $f_\mrm{T}^{(-)}$) begin at order $K/k$, as we show in Appendix \ref{app:response-expansions}.
Assuming ${C}^\mrm{S}_{[AB]}\neq0$, this tells us that as
$K\to0$ the anti-symmetric shift $f_\mrm{S}^{(-)}$ dominates the
total response $f_{AB}(\k,\kl-\k)$.}

We can get a better understanding of our displacement estimator~\eqref{eq:estimatorgeneral} by 
computing the expectation for given $\delta_L(\kl)$. A straightforward 
calculation using Eq.~\eqref{eq:calR-2} shows that
\bea
\big\langle\widehat{h}^\DD_{XY}(\kl)\big\rangle_{\delta_{L}(\K)}
% &= N_{XY}^{\alpha}(\kl)  \int_\k w^{\alpha}_{XY}(\k, \kl-\k)\,
%     \sum_{\beta}f_{XY}^{\beta}(\k, \kl-\k) \delta(\kl)  
%  \nonumber\\
%&\quad=N_{XY}^{\DD}(\kl) \sum_{\beta} \int_\k w^{\DD}_{XY}(\k, \kl-\k) \nonumber\\
%&\qquad\quad
%\times
% \Big[C^{\beta}_{(XY)} f_{\beta}(\k, \kl-\k)+ C^{\beta}_{[XY]} g_{\beta}(\k, \kl-\k)\Big] \delta(\kl) \nonumber\\
% &= \sum_{\beta}N_{XY}^{\DD}(\kl) \Big( C^{\beta}_{(XY)} R^{\DD}_{+\beta}(\kl) + C^{\beta}_{[XY]}R^{\DD}_{-\beta}(\kl)\Big) \delta_{m}(\kl) \nonumber\\
=\Big(\sum_{\alpha} {B}^\DD_{\alpha,XY}(\kl)\Big) \delta_{L}(\kl) 
\equiv b_{\DD}(\kl)\ssp \delta_{L}(\kl)\,, \label{eq:biasedexpectation} \\[-4pt]
\text{with}\quad
b_{\DD}(\kl)\equiv\sum_{\alpha\in\{\mrm{G_\pm,S_\pm,T_\pm}\}} B^\DD_{\alpha,XY}(\kl)\,, \label{eq:effective-bias}
\eea
where we have defined the effective bias $b_\DD$
(note that we have suppressed the labels $X$ and $Y$ for brevity).

This demonstrates that our estimator generally yields a biased estimate of the linear density field $\delta_L(\K)$, with a tracer-dependent bias $b_{\DD}$. In standard gravity, the shift term is protected from anti-symmetric components, i.e. $C^\mrm{S}_{[AB]}=0$ (see Table \ref{tab:modecouplings_effective_coefficients}). But when the EP is violated ($\epsilon \neq 0$), this term becomes nonzero, and gets captured by the effective bias $b_{\DD}$. Figure \ref{fig:bias_trend} illustrates this (using exaggerated values of $\epsilon$). In yellow and black we see how the presence of a non-zero $\epsilon$ leads to a flat behaviour on large-scales in the total bias. This clear separation allows us in principle to use the quadratic estimator as a direct probe of EP violations, though in real cases the value of $\epsilon$ can be much smaller, with current limits at $\mathcal{O}(10^{-3})$ \citep{Bottaro:2023wkd}.

\subsubsection{EP violation effect from other QE estimators}
Why is the displacement estimator best for detecting the $1/K$ anti-symmetric shift?
Consider the general estimator~\eqref{eq:estimatorgeneral}, with weights given by Eq.~\eqref{eq:w-alpha-AB}. The squeezed limit bias $B^\alpha_{\mrm{S},XY}(\K)$ for a general estimator $\alpha$ from the anti-symmetric shift term is
\bea
 \lim_{\kl \rightarrow 0} B^{\alpha}_{\mrm{S},XY}(\kl) 
 &= C_{[AB]}^\mrm{S} \lim_{\kl \rightarrow 0} \int_{\k}w^{\alpha}_{XY}(\k, \kl-\k) f^{(-)}_\mrm{S}(\k, \kl-\k) \nonumber\\
 &\propto \lim_{\kl \rightarrow 0} N_{XY}^{\DD}(\kl) \int_\k f^{\alpha}_{XY}(\k, \kl-\k) \frac{\mu k}{K} \frac{P_L(\k)}{\Ptot^{XX}(\k)\Ptot^{YY}(\kl-\k)}  \, .
\eea
This will vanish if the response function $f^{\alpha}_{XY}$ lacks an odd component. As an example, consider a monopole response like the growth estimator; it captures isotropic effects but remains insensitive to dipoles.

But there is a trade-off. Generally, the displacement estimator poorly traces matter despite its $K^2$ (Gaussian) noise scaling. This is because it fails to robustly extract other effects in the response function that are not the anti-symmetric shift (that if nonzero will also scale as $K^2$, see right panel of Figure \ref{fig:bias_trend}).
Does this also mean that we won't be able to detect EP violations with other estimators? Not necessarily. While this applies to anti-symmetric shifts, EP violations can appear in anti-symmetric growth and tidal coefficients (Table \ref{tab:modecouplings_effective_coefficients}).

\section{Forecast on EP violations}\label{sec:forecasts}
Using the QE formalism above, we now investigate how well a DESI-like survey can constrain EP violation. While an EP violation might affect the growth of perturbations, we focus on the unique information coming from the bispectrum pole.
Our main focus is the potential of the displacement estimator $\widehat{h}^\DD_{AB}$ combined with additional tracers like in a power spectrum analysis. Details on the formalism and additional results can be found in Appendix \ref{app:forecasts}.

\subsection{Data}
Our measured data consist of matter modes obtained from our reconstructions $\widehat{h}_{AB}^{\DD}(\K)$ [Eq.~\eqref{eq:estimatorfourier}] and $\widehat{h}_{AB}^{\mrm{G}_+}(\K)$ [Eq.~\eqref{eq:growthestimatorfourier}], in addition to the galaxy fields,
or `external tracers', $\delta_A(\K)$ and $\delta_B(\K)$, which themselves are used
in the reconstruction. 
From these we construct the following statistics: 
the cross-power spectra $\Pcross^{X\alpha}(\K)=\langle \widehat{h}_{AB}^{\alpha}(\K)\sp \delta_X(\K)^*\rangle'$, for $X\in\{A,B\}$;
the reconstruction auto-spectrum $P_{\mrm{tot}}^{\alpha\alpha}(\K)=\langle \widehat{h}_{AB}^{\alpha}(\K)\sp \widehat{h}_{AB}^{\alpha}(\K)^*\rangle'$; and
the cross-spectrum $\Pcross^{AB}(\K)=\langle \delta_A(\K)\sp \delta_B(\K)^*\rangle'$ on large scales. Additional details can be found in Appendix \ref{sec:powerspectradef}. For our main result we focus on combining cross-correlations between tracers of matter $\Pcross^{XY}$ for $X,Y \in \{A,B,\mathcal{D}\}$, $X\neq Y$. We label this combination $\mathcal{D}\otimes \mrm{Galaxies}$. In Appendix \ref{app:forecasts}, Table \ref{tab:datasetsused} summarizes additional data combinations that we use to make further explorations. Note that we do not consider the total  auto-spectra
$P_{\mrm{tot}}^{XX}$ (signal plus shot noise) to avoid potential contamination from large-scale systematics. We found that including this has little impact on our main result. 

\subsection{Setup}
We base our Fisher forecast on a DESI-like survey configuration. We {adopt DESI-like specifications} for the redshift bins, volumes and number densities~\cite[appendix A.2]{Green:2023uyz}. The total volume is $V \sim 58 h^{-3}\mrm{Gpc}^3$. In particular, for each redshift bin we split the total number density $\bar{n}$ into $\bar{n}_A=\bar{n}/3$ and $\bar{n}_B=\bar{n}/4$, for tracers $A$ and $B$. (The lower values are because,
depending on the nature of EP violation, only a subsample of objects may be suitable for use.) Our fiducial bias parameters over which we marginalize are: $b_{1X}, b_{2X}, b_{s^2X}$, with $X \in \{A,B\}$, as in Eq.~\eqref{eq:delta_g-second}. For the fiducial linear bias parameters, we choose $b_{1A}$ as the DESI linear bias \cite[appendix A.2]{Green:2023uyz}, and set $b_{1B}=0.8 b_{1A}$. We use the fitting formula of \cite{Lazeyras:2015lgp} for $b_{2A}$ and $b_{2B}$; for the tidal biases we use the co-evolution prediction, $b_{s^2A}=-2/7(b_{1A}-1)$ and $b_{s^2B}=-2/7(b_{1B}-1)$, assuming zero tidal bias in Lagrangian space~\citep{Abidi_2018}. {Unless otherwise specified}, the modes {used in the reconstruction} are within $k_{\mathrm{min,rec}}=0.051h\,\mrm{Mpc}^{-1}$ and $k_{\mathrm{max,rec}}=0.15h\, \mrm{Mpc}^{-1}$. Plots involve constraints obtained from the Fisher matrix, integrated over modes from $K_{\mrm{min}}= 2\pi/V^{1/3}$ to $K_{\mrm{max}}=0.05h\,\mrm{Mpc}^{-1}$; see Appendix \ref{app:forecasts}.

\subsection{Results}

\begin{figure}[t!]
    \centering
\includegraphics[width=0.7\linewidth]{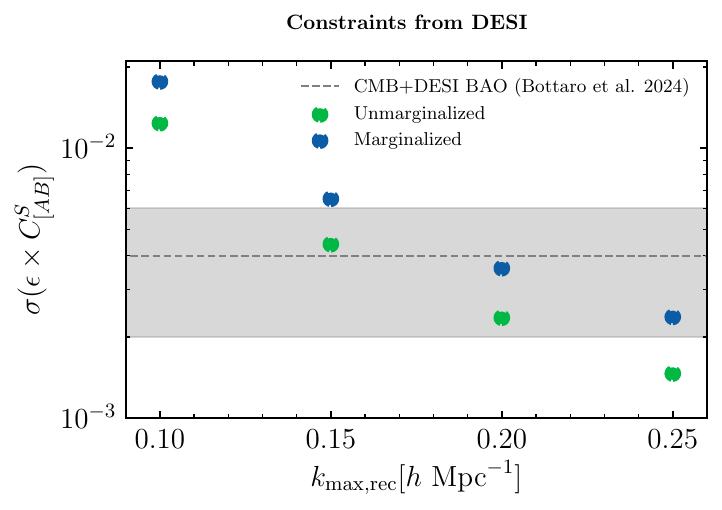}
    \caption{\textbf{Forecasted constraints on EP violation expected from DESI.} Unmarginalized (green) and marginalized (over $b_{1X}, b_{2X}, b_{s^2X}$, blue) constraints on the combination $\epsilon \times C^\mathrm{S}_{[AB]}$ as a function of the maximum wavenumber $k_\mrm{max,rec}$ used in the QE reconstruction. These are the constraints expected from a combined analysis of all cross-correlations
    $\Pcross^{XY}$ for $X,Y \in \{A,B,\mathcal{D}\}$, $X\neq Y$, i.e.\ between
    external tracers $A$, $B$ and the matter modes reconstructed from the
    displacement estimator (or $\mathcal{D}\otimes \mrm{Galaxies}$ for short).
    For comparison, the grey band shows the $1\sigma$ bounds from \emph{Planck} CMB+DESI BAO reported by \cite{Bottaro:2024pcb}. To make this comparison, we assume that both $C^\mathrm{S}_{[AB]}$ and the fraction of fifth-force interacting dark matter are of
    order unity.}
\label{fig:CSAB_forecast}
\end{figure}

Ideally, we would directly constrain the EP violation parameter $\epsilon$. {But this requires determining} the parameter combination $C_{[AB]}^\mrm{S} = (b_{\epsilon,SA}b_{1B}-b_{\epsilon,SB}b_{1A})/2$. Unfortunately, we do not know the EP-violating bias parameters $b_{\epsilon,SA}$ and $b_{\epsilon,SB}$. We will instead constrain the combination $\epsilon \times C_{[AB]}^\mathrm{S}$ as a single quantity, encapsulating both new physics ($\epsilon$) and unknown astrophysical biases ($b_{\epsilon,SA}$ and $b_{\epsilon,SB}$). This also enables comparison with existing bispectrum-based forecasts \citep{Creminelli:2013nua, Bottaro:2023wkd, Graham:2025fdt}.

Figure \ref{fig:CSAB_forecast} is our main result. We constrain the combination $\epsilon \times C_{[AB]}^\mathrm{S}$ using the $\mathcal{D}\otimes \mrm{Galaxies}$ data combination. We show the unmarginalized error bars (green) as a function of the maximum wavenumber $k_\mrm{max,rec}$ used in the displacement estimator. Also shown are the error bars after marginalization over $\epsilon \times C_{[AB]}^\mathrm{S}$ and the standard bias parameters $b_{1X},b_{2X}$, and $b_{s^2X}$, where $\epsilon \times C_{[AB]}^\mathrm{S}$ is treated as an independent parameter. 

For reference, we also show the current best constraints, which derive from the modified growth rate as determined by \emph{Planck} CMB+DESI BAO~\citep{Bottaro:2024pcb}.  These assume a model with dark matter self-interactions mediated by ultralight scalars. Assuming $C_{[AB]}^\mathrm{S}\simeq1$ and that all dark matter interacts, they report (in our notation) $\epsilon = 0.004 \pm 0.002$ (more precisely, they constrain some parameter $\beta = 0.004 \pm 0.002$ that is related to $\epsilon\sim \beta f_{\mrm{dm}}$, where $f_{\mrm{dm}}$ is the fraction of interacting dark matter).

By contrast, we find that going to {mildly nonlinear scales} ($k_{\mrm{max,rec}} \simeq 0.15h\, \mrm{Mpc}^{-1}$) allows us to reach competitive constraints from just the QE cross-spectrum alone. Since the displacement estimator {is by definition quadratic} (consists of two tracer modes), this cross-spectrum contains information about the squeezed-bispectrum.

We compare our results with the recent full-bispectrum forecast for DESI by \cite{Graham:2025fdt}. We use the same DESI configuration \cite[appendix A.2]{Green:2023uyz}, albeit with some differences: we set $b_{1B}(z) = 0.8 b_{1A}(z)$ for our fiducial linear bias parameters at redshift $z$ (over which we marginalize), use a fraction of the total survey objects for $A$ and $B$, and use different long-mode scale cuts. Nevertheless, we get constraints of similar order, showing that the information from the bispectrum is competitive.

%(Note that these independent forecasts use different fiducial bias parameters, long-mode scale cuts, but the survey configuration is the same as used here, see Appendix A2 of Ref.~\cite{Green:2023uyz}.)

Our constraints, while encouraging, are nevertheless based on a number of assumptions. The most crucial one is treating $\epsilon\times C_{[AB]}^\mathrm{S}$ as a single parameter of EP violation. To isolate $\epsilon$ one needs to determine $C_{[AB]}^\mathrm{S}=(b_{\epsilon A}b_{1B}-b_{\epsilon B}b_{1A})/2$, requiring $b_{\epsilon X}$ and $b_{1X}$ for both tracers. While the linear bias parameter $b_{1X}$ is well estimated from the galaxy power spectrum, the bigger issue concerns the parameters $b_{\epsilon X}$; without a good estimate of these parameters, marginalization over $b_{\epsilon X}$ will lead to weaker constraints on $\epsilon$.

Appendix~\ref{app:forecasts} presents forecasts in the case when one assumes a model for $b_{\epsilon X}$, thereby enabling direct constraints on $\epsilon$. We find that if we assume $b_{\epsilon X} \simeq b_{rX}$ (given by the simple model discussed in Section~\ref{sec:epviolationmodel}) our bounds on $\epsilon$ are washed out once we marginalize over the unknown $b_{rX}$.

This makes the bispectrum generally less competitive with the CMB+BAO. This is similar to the conclusion reached by \cite{Bottaro:2023wkd}. However, we underscore that a bispectrum-based measurement contains more robust information owing to the unique $1/K$ dipole signature. Nevertheless, provided one finds suitable tracers (that maximize $C_{[AB]}^\mathrm{S}$) and understands additional biases, 
a detection of a nonzero amplitude of the anti-symmetric-shift ($\epsilon\times C_{[AB]}^\mathrm{S}\neq0$) would constitute direct evidence for EP violation.

\section{Discussion }\label{sec:discussion}
Towards a test of the equivalence principle using large-scale structure, quadratic estimators offer a practical alternative to direct bispectrum methods. In terms of the displacement estimator~\eqref{eq:estimatorfourier}, we saw that the EP-violating response is captured through the effective bias $b_{\DD}$ [Eq.~\eqref{eq:effective-bias}] arising when reconstructing the modulating long mode $\delta_L$ [Eq.~\eqref{eq:estimatorgeneral}]. Our results (Figure \ref{fig:CSAB_forecast})
show that the QE achieves constraining power comparable to the full-bispectrum analysis of \cite{Graham:2025fdt}. This agreement is expected: the consistency relation tells us that the bulk of the information associated with EP violation originates from squeezed-limit configurations of the bispectrum. The QE approach can optimally extract information from these configurations
and constrain amplitude parameters at relatively low computational cost \citep{Schmittfull_2015}. 

The principal limitation of our analysis concerns uncertain EP-violating bias parameters like $b_{\epsilon,\mrm{S}X}$. For this reason we have not marginalized over them when reporting our main results. Instead, we constrain the degenerate combination $\epsilon\times C_{[AB]}^\mrm{S} = \epsilon \times (b_{\epsilon, \mrm{S}A} b_{1B}-b_{\epsilon, \mrm{S}B} b_{1A})/2$. But unlike the very well-determined linear bias  $b_{1X}$, the non-standard parameter $b_{\epsilon,\mrm{S}X}$ is not well understood. This limitation is significant because interpreting any sign of EP violation from the effective bias $b_{\DD}$ requires understanding of this degeneracy. The specific value of $\epsilon$ depends on the model adopted for $b_{\epsilon,\mrm{S}X}$, and in Appendix~\ref{app:additionalforecasts} we explore a simple model. Without knowledge of $b_{\epsilon,\mrm{S}X}$, strong constraints on $\epsilon$ will not be possible, and direct comparison with CMB results may not be possible. This situation is directly analogous to detecting local-type primordial non-Gaussianity (PNG) through scale-dependent large-scale galaxy bias, where constraints depend on the product $b_{\phi X}f_\mrm{NL}$, with $b_{\phi X}$ an astrophysical parameter related to the properties of the tracer~\citep{Barreira:2022sey, Coulton:2022rir}. Just as the bispectrum alone cannot break the bias--$f_\mrm{NL}$ degeneracy, our method suffers a similar bias--$\epsilon$
degeneracy that must be addressed through additional information or modelling assumptions. We note that preliminary studies of the relative density bias $b_{rX}$ of the model in Eq.~\eqref{eq:ep-violation-bias-model} have been carried out \citep{Schmidt:2016coo, bias_review, Chen:2019cfu, Barreira_2020, Khoraminezhad:2020zqe}. Additionally, with improved modelling of the galaxy density, one could also construct quadratic estimators that simultaneously constrain these additional bias parameters.

Previous work has also used quadratic estimators as a means to probe anti-symmetric signals \citep{Jeong:2012df, Vanzan:2024tiq}. While the underlying concepts are similar, our approach differs in two key aspects: we classify the response function using a different basis decomposition ($\mrm{G}_{\pm}, \mrm{S}_{\pm}, \mrm{T}_{\pm}$), and we construct a mixed quadratic estimator designed for practical application to current surveys. Although our estimator is formally sub-optimal, we demonstrate numerically in Appendix \ref{app:qeestimators} that its performance is sufficient for both current and proposed Stage-V surveys, making it a viable tool for near-term EP violation searches.

Other work has also shown the effectiveness of quadratic (and cubic) methods in constraining amplitude bias parameters in various contexts, such as in nonlinear biasing~\citep{Schmittfull_2015} and primordial non-Gaussianity~\citep{Darwish:2020prn,MoradinezhadDizgah:2019xun}; see also Figure \ref{fig:distributions_plot} in Appendix \ref{app:additionalforecasts}. Other applications of the QE include probing the equality scale between matter and radiation, or studying the effect of massive neutrinos in the response function. This broad range of applications demonstrates the versatility of quadratic methods for determining both bias and cosmological parameters. Given these demonstrations and the computational challenges associated with full bispectrum analysis, current surveys like DESI would benefit from incorporating quadratic estimator techniques into their analysis pipelines.

Aside from the characteristic $1/K$ scaling in the squeezed bispectrum, EP violation can also be observed as an enhanced growth of structure~\citep{Bottaro:2023wkd, Bottaro:2024pcb}. While this yields stronger constraints it comes at the cost of detailed modelling of structure formation and the linear growth factor. In contrast, the $1/K$ scaling provides a more universal signature that is robust to complications such as nonlinear evolution, baryonic physics, and redshift-space distortions. This robustness makes the
$1/K$ scaling approach particularly attractive for extracting signatures of EP violation from galaxy survey data, as it does not require the kind of detailed modelling needed for
growth-based methods.

In this work we attempted to incorporate trispectrum information through the auto-spectrum of the displacement estimator, but found it provides minimal additional constraining power for an EP violation detection through the pole. This is noteworthy because on large scales our estimator exhibits a noise scaling similar to that of matter estimation derived from kSZ tomography~\citep{Smith:2018bpn}; see Figure \ref{fig:qevariance}. This can be understood from the similar form of their squeezed-limit responses~\eqref{eq:responseofinterest}~\citep{Rizzo:2017zow, Smith:2018bpn}. However, whereas matter estimation from kSZ tomography is unbiased (assuming perfect knowledge of the filtering power spectra), our displacement estimator is biased, with the bias and variance sharing the same scale dependence.
%Consequently, our displacement estimator is only weakly correlated with the matter field, which limits its sensitivity to the small signal generated by the anti-symmetric shift.
 This unfavourable scaling explains why the variance reduction at large scales does not translate into improved constraints, unlike in kSZ applications where the estimator remains unbiased. Nevertheless, trispectrum information could prove useful in other contexts, for example when including third-order bias parameters \citep[e.g.][]{Darwish:2020prn, Baumann:2021ykm}, which we neglect here.

Finally, throughout our analysis we assumed that the initial fluctuations are Gaussian (and adiabatic), but in principle one should consider the possibility that both EP violation and primordial non-Gaussianity (PNG) are present. In the case of local-type PNG, the squeezed bispectrum acquires a $1/K^2$ pole. 
The question is, what is its multipole structure? Unlike EP violation, 
there is no preferred direction associated with local 
PNG~\citep{Creminelli:2013nua} since it does not permit
gradient-type  operators in the galaxy bias expansion~\citep{bias_review}.
And although local PNG can also induce anti-symmetric components, the 
scale dependence  is not pathological, going as $K$ instead of $1/K$
at leading order~\citep{Dai:2015wla}.
This implies that in order to obtain a $1/K$ scale dependence from PNG
one needs to consider cross terms,
i.e.\ operators mixing EP violation and PNG. But given the smallness
of each sector taken individually, such cross terms are expected to be
of order $\epsilon \cdot f_{NL}$. If $f_{NL}<1$, this is subdominant to
the order $\epsilon$ effect from a pure EP violation.
These arguments suggest that local PNG is not a major concern
for the $1/K$ test.

However, at a practical level one should consider how the cross terms 
impact the QE, which has a more complicated structure.
For this we performed a simple test:
we assumed that when writing down the galaxy bias expansion we could
ignore the cross terms, i.e.\ we considered the pure operators of
EP violation and local PNG.
We carried out a preliminary analysis and found that some level of contamination is present at the QE level (although we note that 
QE bias-hardening techniques may help with this; see \citealt{Namikawa_2013, Darwish:2020prn}).
Detailed calculations will be needed to understand how the
multipole structure changes, how symmetric and anti-symmetric
components couple in the QE, etc.

But we should not lose perspective.
Even if an observed anti-symmetric shift could not be uniquely attributed to PNG or EP violation, detecting a pole in the squeezed bispectrum would represent
a significant finding on its own, since EP violation is competing with PNG, not with
complex but known physics.
Still, a combined treatment including both EP violation and PNG
(and non-adiabatic fluctuations) will eventually be necessary to realise the full
potential of the consistency relation as a test for new physics.

\section{Conclusions}\label{sec:conclusions}

Consistency relations of large-scale structure represent a non-trivial check on the weak equivalence principle (EP) in a regime seldom tested.
By exploiting the specific
coupling between long and short modes 
arising from a violation of these relations,
we have shown that such a check is naturally performed using quadratic estimators.
Our main findings are as follows.
\begin{itemize}
\item A violation of the EP implies the existence of a dipole with $1/K$ scale dependence in the equal-time squeezed bispectrum. In this work we showed that this dipole
is associated with a particular linear response function~\eqref{eq:responseofinterest}.
Physically, this response describes the relative large-scale displacement (or {anti-symmetric shift}, $\mrm{S}_-$) that forms
between distinct tracers when subjected to a near uniform 
gravitational field, e.g.\ sourced by a long-wavelength matter mode. 
If these large-scale displacements are universal (independent of tracer), this
{anti-symmetric shift} vanishes, i.e.\ the EP holds.
\item Based on the characteristic response~\eqref{eq:responseofinterest},
we constructed a quadratic estimator that isolates the signal due to EP violation (if any).
Although our estimator~\eqref{eq:estimatorfourier} is sub-optimal (not minimum variance), it has a simple and
practical form, which allowed us to exploit existing quadratic estimator pipelines. As a basic check, we applied our estimator
to the Abacus simulations by reconstructing the large-scale modes, finding good agreement with basic theoretical predictions. 
\item We forecasted constraints on the EP violation bias $\tilde\epsilon=\epsilon\times C_{[AB]}^\mrm{S}$ (associated with the $\mrm{S}_-$ response)
expected from a DESI-like survey in which we assumed that roughly half of the total number
of objects can be reliably used for QE reconstruction. By going to mildly nonlinear scales ($k_{\mrm{max,rec}} \simeq 0.15h\, \mrm{Mpc}^{-1}$) and after marginalization over the
standard galaxy bias parameters, we forecasted an uncertainty of
$\sigma(\tilde\epsilon) \simeq 6 \times 10^{-3}$. 
The lack of knowledge of the relative bias parameters is a major limiting factor
in one's ability to place stringent constraints on $\epsilon$ alone, whether using the direct bispectrum or the QE approach.
In any case, we expect (naively) \emph{Euclid} to have slightly greater constraining power compared to DESI (mainly due to volume).
Proposed surveys such as MegaMapper \citep{Schlegel:2019eqc} and PUMA \citep{PUMA:2019jwd} will no doubt improve on these numbers.
\end{itemize}

% \vspace{0.5mm}

Future work should extend the present treatment to account for photometric errors, shot-noise marginalization, and redshift-space distortions.
Consistency relations also hold in redshift space~\citep{Kehagias:2013yd, Peloso:2013zw, Creminelli:2013poa}, but for this
one will need to consider additional (line-of-sight) mode couplings and response types.
Furthermore, our work relied on a rather crude estimation of the bias parameters from simulations, and ideally one should aim to self-consistently infer these parameters from the simulations themselves.
Finally, we would like to better understand the extent to which primordial non-Gaussianity complicates the use of the
anti-symmetric shift as a genuine signature of EP violation. A combined treatment of EP violation and
non-Gaussian initial conditions will be presented in a forthcoming work.

\begin{acknowledgments}
We thank
Camille Bonvin for a helpful suggestion on population splitting; 
Simon Foreman for feedback on an early draft;
Lehman Garrison for support with the Abacus simulations; and
Azadeh Moradinezhad Dizgah for interesting discussions.
We also thank Anton Chudaykin, Daniel Green, Colin Hill, Matthew Lewandowski, 
and Joel Meyer for useful exchanges on a range of issues.
OD was supported by an SNSF Eccellenza Professorial Fellowship (No.\ 186879). LD was supported by the European Research Council under the European Union's Horizon 2020 Research and Innovation Programme (grant agreement no.\ 863929). Computations were performed on the Alps-Daint
CSCS cluster of the Swiss National Supercomputing
Center. First data transfers were carried out through the National Energy Research Scientific Computing Center (NERSC).
\end{acknowledgments}

%~~~~~~~~~~~~~~~~~~~~~~~~~~~~~~~~~~~~~~~~~~~~~~~~~~~~~~~~~~~~~

\setstretch{0.98}
\bibliography{main}
% \clearpage

% ~~~~~~~~~~~~~~~~~~~~~~~~~~~~~~~~~~~~~~~~~~~~~~~~~~~~~~~
\setstretch{1.05}
\appendix
\onecolumngrid

\section{Conditional averages and linear response}\label{app:conditional-averages}
In this appendix we clarify the meaning of
$\langle\cdots\rangle_{\delta_L}$ by defining it explicitly in terms of the probability distribution of the linear modes.
Though we have said that this is an `ensemble average with the long mode fixed',
there is some ambiguity here: do 
we fix one long mode and leave the rest to fluctuate (what about neighbouring long modes)? Or do we fix all
modes larger than some scale, because all of these modes are long and could also affect local averages?
For a given realisation the response derives from all long modes.
But statistically, under translation invariance, there is only one mode
that contributes to the \emph{expected} response, namely that for which
$\K=\k_1+\k_2$.

These issues are clarified by effecting a peak--background split
on the linear modes \citep[e.g.][]{dePutter:2018jqk}.
We split the Gaussian density field $\delta_0$ (i.e.\ the linear field)
into long and short modes according to a given cutoff scale $\Lambda$:
$\delta_0(\x)=\delta_S(\x)+\delta_L(\x)$.%
\footnote{In practice one would use a window function instead of a sharp cutoff in $k$-space.}
In Fourier space, using
that $\int_\k=\int_{|\k|>\Lambda}+\int_{|\k|<\Lambda}$, we have
\bea
\delta_0(\k)=\delta_S(\k)+\delta_L(\k)\,,
\quad\text{where}\quad
\delta_S(\k)&=\delta_0(\k)\sp\Theta(|\k|-\Lambda)\,,\nonumber\\
\delta_L(\k)&=\delta_0(\k)\sp\Theta(\Lambda-|\k|)\,,
\eea
with $\Theta$ the Heaviside step function.
The {Gaussian} pdf of $\delta_0$ is $p[\delta_0]=e^{-S[\delta_0]}$ with
$S[\delta_0]=\int_\k {|\delta_0(\k)|^2}/{2P_\mrm{lin}(\k)}$, where we assume statistical homogeneity for the field.
Since $\langle\delta_S\delta_L\rangle=0$ we have that
$p[\delta_0]=p_\Lambda[\delta_S]\sp p_{\bar\Lambda}[\delta_L]=e^{-S_\Lambda[\delta_S]-S_{\bar\Lambda}[\delta_L]}$, where $p_{\Lambda}$ and $p_{\bar\Lambda}$ are the pdfs for the short and long modes respectively.%
\footnote{In the language of field theory, $\delta_0$ is a free field and the
property that $\delta_S$ and $\delta_L$ are independent translates to the 
absence of an interaction term in the action, so
$S[\delta_S,\delta_L]=S_\Lambda[\delta_S]+S_{\bar\Lambda}[\delta_L]$.}
In practice, $\Lambda$
will be set by the survey size and the long modes we will consider
have $|\K|\ll\Lambda$ so that $\delta_L\ll\delta_S$ in a cosmological setting.
The long modes sitting just below $\Lambda$ do not
affect the three-point configurations (squeezed triangles)
considered below.

The idea of the following analysis is quite simple, if obscured
by formalism. The essence of the problem is that we have two
{independent} random variables $x$ and $y$, and we wish to
obtain the statistics of $z=x+y$ for a given $y=y_*$, i.e.\ 
we want the pdf $p(z\Mid y_*)$. 
Because of statistical independence, we have that the probability
of obtaining $z$ given
$y=y_*$ is just the probability of finding $x$
(value of $y$ has no bearing on $x$). Quantitatively, by
statistical independence the joint pdf is separable,
$p(x,y\Mid y_*)=p(x)\sp p(y\Mid y_*)$, so that by the chain rule
we have
\be\label{eq:simple-model}
p(z\Mid y_*)=\int\dif x\int\dif y\,p(z\Mid x,y)\sp p(x)\sp p(y\Mid y_*)
%=\int\dif X\,\delD(Z-X-Y_*)\sp p(X)
=p(x)\ssp\big|_{x=z-y_*}\,,
\ee
where the last equality follows because
$p(z\Mid x,y)=\delD(z-x-y)$ and
$p(y\Mid y_*)=\delD(y-y_*)$.
This is the idea of the following calculation, but
with $x\to\delta_S$, $y\to\delta_L$, and
$z\to\delta_0$. However, note here that
$x,y,z$ are arbitrary random variables;
below we will assume that $\delta_S,\delta_L$, and $\delta_0$
obey Gaussian statistics.
Note also that the split is performed on the linear field,
rather than the fully-evolved, non-Gaussian field.

Return now to the case of fields.
To understand what is meant by $\langle\cdots\rangle_{\delta_L}$, 
it helps to write out the expectations in terms of the pdf. 
Now, modes with distinct wavevectors are statistically
independent under translation invariance and the Fourier support of
$\delta_S(\x)$ and $\delta_L(\x)$
are disjoint and therefore uncorrelated. This means that
$p[\delta_0]=p[\delta_S,\delta_L]=p_\Lambda[\delta_S]\sp p_{\bar\Lambda}[\delta_L]$,
so that
$p[\delta_0\Mid\delta_L]=p_\Lambda[\delta_S]|_{\delta_S=\delta_0-\delta_L}$, as in Eq.~\eqref{eq:simple-model}. Under the assumption that
$\delta_0$, and hence $\delta_S$ and $\delta_L$, are Gaussian
random fields, we have
\be
% S[\delta_0-\delta_L]=\int_{\k}\frac{|\delta_0(\k)-\delta_L(\k)|^2}{2\sp P_\mrm{lin}(\k)}=
p_\Lambda[\delta_S]=e^{-S_\Lambda[\delta_S]}\,,
\quad
S_\Lambda[\delta_S]
\equiv\int_{|\k|>\Lambda}\frac{|\delta_S(\k)|^2}{2\sp P_\mrm{lin}(\k)}
=\int_\k\frac{|\delta_S(\k)|^2}{2\sp P_\mrm{lin}(\k)}
=S[\delta_S]\,,
\ee
where in the second integral we have extended the integration
over all $\k$-space since for $|\k|<\Lambda$ we have 
$\delta_S(\k)=0$ by definition.
This expression shows that the covariance of short linear modes are
`diagonal' in $\k$-space, whether or not we condition on the long modes.
In other words, the effect of conditioning on $\delta_L(\x)$ is to
strike out the rows and columns of the modes corresponding to
$\delta_L(\x)$, as expected by statistical homogeneity.
The expression for $p_{\bar\Lambda}[\delta_L]$ is defined 
analogously to $p_{\Lambda}[\delta_S]$ but with the integration
over wavenumbers $|\k|<\Lambda$.

Now because $p[\delta_0]=p_\Lambda[\delta_S]\sp p_{\bar\Lambda}[\delta_L]$
we have that expectations of quantities like
$F[\delta_S]\sp G[\delta_L]$ are separable:
\be
\langle F[\delta_S]\sp G[\delta_L]\rangle
=\left(\int\mathcal{D}\delta_S\,e^{-S_{\Lambda}[\delta_S]} F[\delta_S]\right)\left(
\int\mathcal{D}\delta_L\,e^{-S_\mrm{\bar\Lambda}[\delta_L]} G[\delta_L]\right)
\equiv
\langle F[\delta_S]\rangle_{p[\delta_S]}\sp
\langle G[\delta_L]\rangle_{p[\delta_L]}
\ee
From this it is not difficult to see that when we condition on the
long modes we are just restricting the measure to the short modes
so that, e.g.\
$
\langle F[\delta_S]\sp G[\delta_L]\rangle_{p_\Lambda[\delta_S]}
% =\langle F[\delta_S]\rangle_{p[\delta_S]}\,
% \langle G[\delta_L]\rangle_{p[\delta_L\Mid\delta_{L*}]}
\equiv\langle F[\delta_S]\rangle_{p_\Lambda[\delta_S]}\ssp
G[\delta_{L}]
$.
Note also that since $\delta_S(\k)=0$ for $|\k|<\Lambda$ we can in
fact just use the full measure, $e^{-S[\delta_S]}$ since the cutoff is by definition built into $\delta_S$
(no cutoff in $k$-space required).

For brevity, in the following we will use the notation
\bea
\langle\cdots\rangle_{\delta_L}
\equiv\langle\cdots\rangle_{p_\Lambda[\delta_S|\delta_L]}
=\langle\cdots\rangle_{p_\Lambda[\delta_S]}\,,
\qquad
\langle\cdots\rangle
\equiv
\langle\cdots\rangle_{p[\delta_S,\delta_L]}
=\langle
    \langle\cdots\rangle_{\delta_L}
    \rangle_{p_{\bar\Lambda}[\delta_L]}\,,
\eea
where the last equality follows by the chain rule.
If $F=F[\delta_S]$ is solely a function of the short modes
then we can simply write
$\langle F[\delta_S]\rangle_{\delta_L}
=\langle F[\delta_S]\rangle$ since
$\langle F[\delta_S]\rangle_{p_{\bar\Lambda}[\delta_L]}=F[\delta_S]$,
i.e.\ the ensemble average is taken over all realizations
of long and short modes, as normal.
Note that the average over long modes with short modes fixed,
$\langle\cdots\rangle_{p[\delta_L|\delta_S]}$, is not needed.
% \bea
% \langle\cdots\rangle_{\delta_L}
% &\equiv
% \int\mathcal{D}\delta_S\ssp e^{-S_\Lambda[\delta_S]}(\cdots) \,,\\
% \langle\cdots\rangle
% &=\int\mathcal{D}(\delta_S,\delta_L)\ssp e^{-S_\Lambda[\delta_S]} e^{-S_{\bar\Lambda}[\delta_L]}\,(\cdots)
% \equiv\int\mathcal{D}\delta_L\ssp e^{-S_{\bar\Lambda}[\delta_L]}
%     \langle\cdots\rangle_{\delta_L}\,.
% \eea

\subsection{Linear response at leading order}
We now compute the tree-level linear response function $f(\k_1,\k_2)$
using the long--short split.
Consider the fully-evolved nonlinear matter fluctuation
$\delta_m=\delta_m[\delta_0]=\delta_m[\delta_S,\delta_L]$.
The two-point function for this non-Gaussian field is
$\langle\delta_m(\k_1)\sp\delta_m(\k_2)\rangle_{\delta_L}$,
where the scales are chosen so that $k_1,k_2>\Lambda\gg K$.
Note that all modes in the
Fourier support of $\delta_L(\x)$ are fixed (not
averaged over). However,
the assumption of statistical homogeneity means that we only need
to fix a single mode, $\delta_L(\K)$ with $\K=\k_1+\k_2$; the other long modes have no impact on the two-point function
and we thus have
$\langle\delta_m(\k_1)\sp\delta_m(\k_2)\rangle_{\delta_L}
\equiv
\langle\delta_m(\k_1)\sp\delta_m(\k_2)\rangle_{\delta_L(\K)}$.

Since there is no mode coupling between long and short modes at
first order, for this calculation it will be necessary to go to
second order in $\delta_0$.
Recall in perturbation theory
\be
\delta_m(\k)=\delta_m^{(1)}(\k)+\delta_m^{(2)}(\k)+\cdots\,,
\qquad
\delta_m^{(1)}(\k)=\delta_0(\k)\,,
\quad
\delta_m^{(2)}(\k)
=\int_\q F_2(\q,\k-\q)\ssp\delta_0(\q)\ssp\delta_0(\k-\q)\,.
\ee
where $F_2$ is the standard mode-coupling kernel.
Inserting $\delta_0(\k)=\delta_S(\k)+\delta_L(\k)$ into
$\delta_m^{(2)}(\k)$ yields
\bea
\delta_m^{(2)}(\k)
% &\equiv\int_\q F_2(\q,\k-\q)\ssp\delta_0(\q)\ssp\delta_0(\k-\q) \nonumber\\
&=\int_\q F_2(\q,\k-\q)\ssp\delta_S(\q)\ssp\delta_S(\k-\q)
+2\int_\q F_2(\q,\k-\q)\ssp\delta_L(\q)\ssp\delta_S(\k-\q)
    +\mathcal{O}(\delta_L^2)\,.
\eea
Here the first term gives the short--short
coupling where the mode coupling is between modes with $|\q|>\Lambda$.
The second term
gives the long--short coupling and
can be restricted to $|\q|<\Lambda$ since if $|\q|>\Lambda$
then $\delta_L(\q)=0$ and there is no contribution. The long--long
coupling contributes to the second-order response and is
neglected here.

With these expressions we now compute the conditional
two-point function
% \be
% \langle\delta_m(\k_1)\sp\delta_m(\k_2)\rangle_{\delta_L}
% =\int\mathcal{D}\delta_S\ssp e^{-S[\delta_S]}\sp
%     \delta_m^{(1)}(\k_1)\sp\delta_m^{(1)}(\k_2)
% +\left(\int\mathcal{D}\delta_S\ssp e^{-S[\delta_S]}\sp
%     \delta_m^{(2)}(\k_1)\sp\delta_m^{(1)}(\k_2)
% +(\k_1\leftrightarrow\k_2)\right)
% +\mathcal{O}(\delta_L^2)
% \ee
\be\label{eq:delta_m-series-LO}
\langle\delta_m(\k_1)\sp\delta_m(\k_2)\rangle_{\delta_L}
=\big\langle
    \delta_m^{(1)}(\k_1)\sp\delta_m^{(1)}(\k_2)\big\rangle_{\delta_L}
+\big[\big\langle
    \delta_m^{(2)}(\k_1)\sp\delta_m^{(1)}(\k_2)\big\rangle_{\delta_L}
+(\k_1\leftrightarrow\k_2)\big]
+\mathcal{O}(\delta_L^2)\,.
\ee
For the first term we can put $\delta_m^{(1)}(\k_1)=\delta_S(\k_1)+\delta_L(\k_1)=\delta_S(\k_1)$
since $\delta_L(\k_1)=0$ for $|\k_1|>\Lambda$; this term is then just
$\langle\delta_S(\k_1)\sp\delta_S(\k_2)\rangle_{\delta_L}=\langle\delta_S(\k_1)\sp\delta_S(\k_2)\rangle$.
We have similarly for $\delta_m^{(1)}(\k_2)$.
For the second term (which is an odd moment and so normally vanishes)
we have mode coupling between
$\delta_L$ and $\delta_S$:
% \bea
% \int\mathcal{D}\delta_S\,e^{-S[\delta_S]}\sp
%     \delta_m^{(2)}(\k_1)\sp\delta_m^{(1)}(\k_2)
% &=\int\mathcal{D}\delta_S\,e^{-S[\delta_S]}
%     \left(2\int_{|\q|<\Lambda} F_2(\q,\k_1-\q)\ssp\delta_L(\q)\ssp
%     \delta_S(\k_1-\q)\right)\delta_S(\k_2) \nonumber\\
% &=2\int_{|\q|<\Lambda} F_2(\q,\k_1-\q)\ssp\delta_L(\q)\ssp
%     \Big\langle\delta_S(\k_1-\q)\ssp\delta_S(\k_2)\Big\rangle_{\delta_L} 
% % &=2\int_{|\q|<\Lambda} F_2(\q,\k_1-\q)\ssp\delta_L(\q)\ssp
% %     (2\pi)^3\delD(\k_1+\k_2-\q)\ssp P_\mrm{lin}(\k_2) \nonumber\\
% =2\sp F_2(\K,-\k_2)\ssp P_\mrm{lin}(\k_2)\ssp\delta_L(\K)
% \eea
\bea
\big\langle
    \delta_m^{(2)}(\k_1)\sp\delta_m^{(1)}(\k_2)
\big\rangle_{\delta_L}
&=\bigg\langle
    \bigg(2\int_{|\q|<\Lambda} F_2(\q,\k_1-\q)\ssp\delta_L(\q)\ssp
    \delta_S(\k_1-\q)\bigg)\,\delta_S(\k_2)
    \bigg\rangle_{\delta_L} \nonumber\\
&=2\int_{|\q|<\Lambda} F_2(\q,\k_1-\q)\ssp\delta_L(\q)\ssp
    \Big\langle\delta_S(\k_1-\q)\ssp\delta_S(\k_2)\Big\rangle 
% &=2\int_{|\q|<\Lambda} F_2(\q,\k_1-\q)\ssp\delta_L(\q)\ssp
%     (2\pi)^3\delD(\k_1+\k_2-\q)\ssp P_\mrm{lin}(\k_2) \nonumber\\
=2\sp F_2(\K,-\k_2)\ssp P_\mrm{lin}(\k_2)\ssp\delta_L(\K)
\eea
where in the first line we have discarded the correlator
$\langle\delta_S\sp\delta_S\sp\delta_S\rangle_{\delta_L}=0$
but kept $\langle\delta_L\sp\delta_S\sp\delta_S\rangle_{\delta_L}=\delta_L\sp\langle\sp\delta_S\sp\delta_S\rangle\neq0$.
In the last equality $\K\equiv \k_1+\k_2$. Note that although we have conditioned on the entire Fourier support of $\delta_L(\x)$, only the mode 
having precisely $\q=\k_1+\k_2=\K$ needs to be considered,
by virtue of statistical homogeneity.
There is also a slight technicality: we need $K$ to be
sufficiently small compared with
$k_1,k_2$ so that in the second line $\delta_S(\k_1-\K)\neq0$.
(This can be assured by picking long and short wavenumbers which 
are well separated, e.g.\ $K\ll\Lambda$ and $k_1,k_2\gg\Lambda$.)
Now, combining these results we have
\be
\langle\delta_m(\k_1)\sp\delta_m(\k_2)\rangle_{\delta_L}
=\langle\delta_S(\k_1)\sp\delta_S(\k_2)\rangle
+f(\k_1,\k_2)\,
    \delta_L(\K)
+\mathcal{O}(\delta_L^2)\,,
\ee
where $f(\k_1,\k_2)=2\sp F_2(\k_1+\k_2,-\k_1)\ssp P_\mrm{lin}(\k_1)+(\k_1\leftrightarrow\k_2)$. This relation
is the basis of quadratic estimation.
We can extract the linear
response by differentiating both sides by $\delta_L(\K')$, then set
$\delta_L=0$ to isolate the linear piece:\footnote{Here we can
either set the Fourier support of $\delta_L$ to zero or just the mode $\delta_L(\K)$;
both yield equivalent results under statistical homogeneity.}
\be\label{eq:2pcf-conditional}
\frac{\partial}{\partial\delta_L(\K)}\,
\big\langle\delta_m(\k_1)\sp\delta_m(\k_2)\big\rangle_{\delta_L}\,\Big|_{\delta_L=0}
=(2\pi)^3\delD(\k_1+\k_2-\K)\ssp f(\k_1,\k_2)\,,
\ee
where we have used that $\partial\delta_L(\K)/\partial\delta_L(\K')
=(2\pi)^3\delD(\K-\K')=(2\pi)^3\delD(\k_1+\k_2-\K')$
(relabelling $\K'\to\K$ above).
Thus we see that $f(\k_1,\k_2)$ corresponds to the
(tree-level) linear response.%
\footnote{A comparison between Eq.~\eqref{eq:2pcf-conditional} and
Eq.~\eqref{eq:fAB-as-deriv} shows that we can either
differentiate then average or average then differentiate;
provided we set $\delta_L=0$ upon differentiation, there is no
difference:
\be\label{eq:response-equivalence}
\frac{\partial}{\partial\delta_L(\K)}
\big\langle{O}_A(\k_1)\sp{O}_B(\k_2)\ssp\big\rangle_{\delta_L}\Big|_{\delta_L=0}
=\bigg\langle\frac{\partial}{\partial\delta_L(\K)}
    {O}_A(\k_1)\sp{O}_B(\k_2)\Big|_{\delta_L=0}\bigg\rangle_{\delta_L}
=\bigg\langle\frac{\partial}{\partial\delta_L(\K)}{O}_A(\k_1)\sp{O}_B(\k_2)\bigg\rangle_{\delta_L}\bigg|_{\delta_L=0}
\ee
where ${O}_A={O}_A[\delta_S,\delta_L]$.
This is valid at all orders.
}

\section{Some explicit calculations}\label{app:response-calculation}

\subsection{Standard response }\label{app:f-matter}
In this appendix we calculate
the leading-order response $f_{AB}$, beginning from the
definition~\eqref{eq:fAB-as-deriv}.
For this we need only go up to second order in
standard perturbation theory (SPT). The tracer overdensity is
$\delta_X(\k)\simeq\delta_X^{(1)}(\k)+\delta_X^{(2)}(\k)$, where
\bea\label{eq:delta-2nd}
\delta_X^{(1)}(\k)
&=b_{1X}\ssp\delta_m^{(1)}(\k),
\qquad
\delta_X^{(2)}(\k)
=\int_\q\, {F}_{2X}(\q,\k-\q)\,
    \delta_m^{(1)}(\q)\sp\delta_m^{(1)}(\k-\q).
% \delta_B(\k)
% &\simeq\delta_B^{(1)}(\k)+\delta_B^{(2)}(\k),
% \quad
% \delta_B^{(1)}(\k)=b_{1B}\ssp\delta_m^{(1)}(\k),
% \quad
% \delta_B^{(2)}(\k)
% =\int\frac{\dif^3\q}{(2\pi)^3}\, {F}^\mrm{mm}_{B}(\q,\k-\q)\,
%     \delta_m^{(1)}(\q)\delta_m^{(1)}(\k-\q).
\eea
Here $\delta_m^{(1)}(\k)$ is a linear matter mode,
$b_{1X}$ linear bias, and the second-order mode-coupling kernel is given by
\bea\label{eq:F2A}
F_{2X}(\k_1,\k_2)
&=b_{1X}F_2(\k_1,\k_2)+b_{2X}+b_{s^2X}
\left(\frac{(\k_1\cdot\k_2)^2}{k_1^2k_2^2}-\frac13\right)\\
&=\left(b_{1X}+\frac{21}{17}b_{2X}\right) F_\mrm{G}(\k_1,\k_2)
    +b_{1X} F_\mrm{S}(\k_1,\k_2)
    +\left(b_{1X}+\frac72b_{s^2X}\right)F_\mrm{T}(\k_1,\k_2)
    \label{eq:F2A-2}
\eea
where $F_2$ is the usual second-order SPT kernel which in the 
second line has been decomposed into G, S, and T kernels (see 
Table~\ref{tab:modecouplings}). In the second line note that 
the shift term is protected from second-order bias 
due to the EP.

Now, the linear response function is given by Eq.~\eqref{eq:fAB-as-deriv}.
The linear response, given by the derivative of the average, can be written as
the average of the derivative, by Eq.~\eqref{eq:response-equivalence}.
The response is then
\bea\nonumber
\frac{\partial}{\partial\delta_L(\K)}
\big\langle\delta_A(\k_1)\sp\delta_B(\k_2)\ssp\big\rangle_0
&=\bigg\langle
    \frac{\partial}{\partial\delta_L(\K)}\,
    \delta_A(\k_1)\sp\delta_B(\k_2)
\bigg\rangle_{0} \nonumber\\
&\simeq
\bigg\langle
    \frac{\partial\delta_A^{(2)}(\k_1)}{\partial\delta_L(\K)}
    \delta_B^{(1)}(\k_2)
\bigg\rangle_0
+
\bigg\langle
    \delta_A^{(1)}(\k_1)
    \frac{\partial\delta_B^{(2)}(\k_2)}{\partial\delta_L(\K)}
\bigg\rangle_0\,,
\label{eq:d2-std}
\eea
where subscript 0 is a shorthand for $\langle\cdots\rangle_{\delta_L}|_{\delta_L=0}$,
and in the second line we substituted the perturbative solutions.
Here we can take $\delta_L(\K)=\delta_m^{(1)}(\K)$ since 
the long mode is in the linear regime.
The derivative is given by
\begin{align}
\frac{\partial\delta_X^{(2)}(\k)}{\partial\delta^{(1)}_m(\K)}
&=\int_\q\,{F}_{2X}(\q,\k-\q)(2\pi)^3\delD(\kl-\q)\sp\delta_m^{(1)}(\k-\q)
+\int_\q\,{F}_{2X}(\q,\k-\q)\sp\delta_m^{(1)}(\q)(2\pi)^3\delD(\kl-\k+\q) \nonumber\\
&=2{F}_{2X}(\kl,\k-\kl)\ssp\delta_m^{(1)}(\k-\kl), \label{eq:d2}
\end{align}
where we have used $\partial\delta_m^{(1)}(\q)/\partial\delta_m^{(1)}(\K)
=(2\pi)^3\delD(\q-\K)$.
Putting everything together in Eq.~\eqref{eq:fAB-as-deriv}
yields the leading-order linear response
\bea\label{eq:response-LO}
f_{AB}(\k_1,\k_2)
% \equiv 2F_2(\kl,\k_1-\kl)\PL(k_2)+(1\leftrightarrow2)
&= 2\sp {F}_{2A}(\k_1+\k_2,-\k_2) P^{(1)}_{mB}(k_2)
+2\sp{F}_{2B}(\k_1+\k_2,-\k_1)P^{(1)}_{Am}(k_1) \\
&= 2\sp\Big[ b_{1B}\sp{F}_{2A}(\K,\k_1-\K)P_L(\K-\k_1)
+\sp b_{1A}\sp{F}_{2B}(\K,-\k_1)P_L(k_1)\Big]
\label{eq:response-LO2}
\eea
where the (linear) power spectra are given by
$\langle\delta_X^{(1)}(\k_1)\sp\delta_m^{(1)}(\k_2)\rangle
=(2\pi)^3\delD(\k_1+\k_2)P_{Xm}^{(1)}(k_1)$ with
$P_{Xm}^{(1)}=b_{1X}P_L$.
Equation~\eqref{eq:response-LO2}, when decomposed in
terms of the G, S, T kernels using Eq.~\eqref{eq:F2A},
yields Eq.~\eqref{eq:f-alpha-AB} in the main text.

\subsection{EP-violating response}
The previous calculation showed in Eq.~\eqref{eq:F2A-2}
that the
mode coupling $F_\mrm{S}$ is affected by second-order
gravitational evolution but \emph{not} by any galaxy bias
operator (besides trivially linear bias).
We will now assume a more general galaxy bias model
which involve shift operators typically forbidden by the EP.
The basic assumption here is that $A$ and $B$ may not be biased tracers
of the matter field only (as required by the equivalence
principle), but may also depend on other fields
such as dark matter $\delta_c$ and baryons $\delta_b$ individually. 
For concreteness, we follow \cite{Bottaro:2023wkd},
hereafter B24,
and take these additional fields to be the
relative density field $\delta_r\equiv\delta_c-\delta_b$
and a relative velocity field $\theta_r$.
Allowing primordial non-Gaussianity, there is also
coupling with the gravitational potential $\Phi$ in the 
bias expansion that we ignore here~\citep{Darwish:2020prn}.
Let us package up these fields into a vector
\be
\bm\phi=\big(\delta_m,\delta_r,\theta_r,\ldots\big)
\ee
The components of this vector will be denoted
$\phi_a$, with $a\in\{m,r,\theta,\ldots\}$.
Previously the tracers were a functional only of the matter field,
$\delta_X=\mathcal{F}_X[\delta_m]$; now
we suppose $\delta_X$ is functional of $\bm\phi$,
\be
\delta_X
=\mathcal{F}_X[\bm\phi]
=\mathcal{F}_X[\delta_m,\delta_r,\theta_r,\cdots].
\ee
At some given time we write
$\delta_X(\k)\simeq\delta_X^{(1)}(\k)+\delta_X^{(2)}(\k)$
with~\citep{Carroll:2013oxa}
\bea
\delta_X^{(1)}(\k)
&=\sum_a J^a_X\,\phi_a^{(1)}(\k), 
\qquad
\delta_X^{(2)}(\k)
=\sum_{a,b}\int_\q\, {F}_X^{ab}(\q,\k-\q)\,
    \phi_a^{(1)}(\q)\phi_b^{(1)}(\k-\q), 
% &=\sum_\mathcal{O}b_\mathcal{O}\ssp \mathcal{O}^{(2)}(\k)
% =\sum_\mathcal{O}b_\mathcal{O}\ssp 
%     \sum_{ab} \int_\q\,{F}^{ab}_\mathcal{O}(\q,\k-\q)\,
%      \phi_a^{(1)}(\q)\phi_b^{(1)}(\k-\q), 
\eea
where  $J^a_X=(b_{1X},b_{rX},b_{\theta X},\ldots)$ is a vector
of linear bias parameters
and $F^{ab}_X$ are arbitrary kernels coupling field $a$
to field $b$.
In other words, we can now have mode coupling between
$\delta_m$ and $\delta_r$, in addition to the previous
self-coupling of $\delta_m$.
We therefore recover Eq.~\eqref{eq:delta-2nd} if we ignore
all but the $F^\mrm{mm}_X$ contribution.
Note however there is an implicit assumption that the second-order
solution is local-in-time.

We will now make the assumption that $\delta_r^{(1)}$,
$\theta_r^{(1)}$, etc, are related to the linear
matter field by a (time-dependent) constant, i.e.\
$\phi_a^{(1)}(\k,z)=G_a(z)\ssp\delta_m^{(1)}(\k,z)$,
where $G_a(z)$ is some scale-independent
function whose precise form will generally depend on the
growth factor of each field or species
(see equations A7 and A8 in B24).
We will work at fixed time so $G_a$ can be treated as 
a constant vector and note that $G_m(z)=1$ by construction.
We then define
$\tilde{b}_{1X}\equiv\sum_a J^a_X\sp G_a$ and
$\tilde{F}_X(\k_1,\k_2)\equiv
\sum_{ab} F^{ab}_X(\k_1,\k_2)\sp G_a\sp G_b$.
We can now repeat the same steps as above to get
[cf.~Eq.~\eqref{eq:response-LO2}]
\bea
f_{AB}(\k,\K-\k)
&= 2\sp\Big[ \tilde{b}_{1B}\sp\tilde{F}_{A}(\K,\k-\K)P_\mrm{lin}(\K-\k)
+\sp \tilde{b}_{1A}\sp\tilde{F}_{B}(\K,-\k)P_\mrm{lin}(k)\Big]
\eea
That is, the response takes the exact same form as 
Eq.~\eqref{eq:response-LO2} but with
$b_{1X}\to \tilde{b}_{1X}$ and
$F_{X}\to \tilde{F}_{X}$. Note that
$\tilde{b}_{1X}=b_{1X}+\cdots$ and
$\tilde{F}_{X}=F_X+\cdots$, where the ellipsis represents terms due to EP violation.

\subsubsection{Model of Bottaro et al.}\label{app:bottaro}
Let us now work out EP-violating kernel $\tilde{F}_X$ for the model
of B24. The bias expansion is
\be
\delta_X=b_{1X}\delta_m+b_{rX}\delta_r+b_{\theta X}\theta_r
+b_{\nabla\delta X}\ssp \v_r\cdot\bm\nabla\delta_m+\cdots
\ee
where we show only the terms relevant for the pole (these are the only operators
which lead to couplings of shift type).
We write $\delta\simeq\delta^{(1)}+\delta^{(2)}$ for each
$\delta_m$, $\delta_r$ and $\theta_r$.
At first order the relative fields are related to
the linear matter field $\delta_m^{(1)}$ by
\be
\delta_r^{(1)}=\frac53\epsilon\ssp\delta_m^{(1)},
\qquad
\theta_r^{(1)}/(-\calH f_r)
=\delta_r^{(1)}=\frac53\epsilon\ssp\delta_m^{(1)},
\qquad
\theta_m^{(1)}/(-\calH f_m)=\delta_m^{(1)},
\ee
see B24.
In our formalism this means $(G_m,G_r,G_\theta)=(1,\frac53\epsilon,-\calH f_r\frac53\epsilon)$ and the
effective linear bias is
\be
\tilde{b}_{1X}=\sum_O b_{OX}\ssp G_O
=b_{1X}+\frac53\epsilon\ssp b_{rX}
    -\calH f_r\frac53\epsilon\ssp b_{\theta X}. 
\ee
At second order (equation A13 and A16 in B24)
\bea
\delta_m^{(2)}(\k)
&=D_m^2\int_\q F_2(\q,\k-\q)\sp
    \delta_{m0}^{(1)}(\q)\sp\delta_{m0}^{(1)}(\k-\q),\\
\theta_m^{(2)}(\k)/(-\calH f_m)
&=D^2_m\int_\q G_{2}(\q,\k-\q)\sp
    \delta_{m0}^{(1)}(\q)\sp\delta_{m0}^{(1)}(\k-\q),\\
\delta_r^{(2)}(\k)
&=\epsilon D_\mrm{cdm}^2\int_\q F_{2r}(\q,\k-\q)\sp
    \delta_{m0}^{(1)}(\q)\sp\delta_{m0}^{(1)}(\k-\q),\\
\theta_r^{(2)}(\k)/(-\calH f_r)
&=\epsilon D_{\mrm{cdm}}^2\int_\q G_{2r}(\q,\k-\q)\sp
    \delta_{m0}^{(1)}(\q)\sp\delta_{m0}^{(1)}(\k-\q),
\eea
where $\delta_{m0}$ is the matter field at some initial time.
At leading order in $\epsilon$ we can take $f_r= 1$ and
replace in the last two expressions $D_\mrm{cdm}$ with $D_m$,
since $D_m\simeq D_\mrm{cdm}(1+ \alpha \epsilon)$, where $\alpha$ is some number, so 
$\epsilon D_\mrm{cdm}^2=\epsilon D_m^2+\mathcal{O}(\epsilon^2)$. From B24 we have: $D_{m}(a,\epsilon)=[1+\frac{6}{5}\epsilon f_{\chi} (\log\frac{a}{a_{\mrm{eq}}}-\frac{181}{90})]D_\mrm{cdm}(a) = (1+\epsilon \alpha)D_\mrm{cdm}(a)$,
where $f_\chi$ is the fraction of interacting dark matter, $a$ is the scale factor and $a_\mrm{eq}$ is the scale factor at matter-radiation equality.
Note that in Appendix~\ref{app:additionalforecasts} we use $\alpha=\frac{6}{5}\epsilon f_{\chi} (\log\frac{a}{a_{\mrm{eq}}}-\frac{181}{90})$ for our linear growth example.

In our formalism we need to compute the effective mode coupling
$\tilde{F}_X=\sum_O b_{OX}\ssp F_O$. Adding up the
kernels above, multiplying each by the appropriate bias
coefficient, we find
\bea
\tilde{F}_X(\k_1,\k_2)
&=\sum_O b_{OX}\ssp F_O \nonumber\\
&=b_{1X}\ssp F_2(\k_1,\k_2)
+\epsilon\ssp b_{rX}\ssp F_{2r}(\k_1,\k_2)
+\epsilon\ssp b_{\theta X}\ssp G_{2r}(\k_1,\k_2)
+\frac53 \epsilon\ssp b_{\nabla\delta X}\ssp F_\mrm{S}(\k_1,\k_2)+\mathcal{O}(\epsilon^2)\,,
\eea
where for the third and fourth terms we have absorbed
$-\calH$ into the definition of the velocity divergence to
make it dimensionless.
Note that the last term has been symmetrized.
Let us focus on the shift terms contained in this expression.
From equation A15 of B24 we have
$F_{2r}=\frac{17}{6}F_\mrm{S}+\cdots$ and from equation A18 we
have $G_{2r}=\frac{7}{3}F_\mrm{S}+\cdots$. Focusing on the
shift terms,
\bea
\tilde{F}_X(\k_1,\k_2)
&=\left(b_{1X}
+\frac{17}{6}\epsilon\ssp b_{rX}
+\frac73\epsilon\ssp b_{\theta X}
+\frac53\epsilon\ssp b_{\nabla\delta X}\right) 
    F_\mrm{S}(\k_1,\k_2)+\text{growth and tidal terms} \nonumber\\
&\equiv b_{\epsilon,\mrm{G}X} F_\mrm{G} + b_{\epsilon,\mrm{S}X} F_\mrm{S} + b_{\epsilon,\mrm{T}X} F_\mrm{T}
\eea
Here the terms proportional to $\epsilon$ lead to
an anti-symmetric shift. Denote the contents of the parentheses
by $\tilde{c}_{\mrm{S},X}$.
Now
\bea
\tilde{b}_{1A}\ssp \tilde{c}_{\mrm{S},B}
&=\left(b_{1A}+\frac53\epsilon\ssp b_{rA}+\frac53\epsilon\ssp b_{\theta A}\right)
\left(b_{1B}
+\frac{17}{6}\epsilon\ssp b_{rB}
+\frac73\epsilon\ssp b_{\theta B}
+\frac53\epsilon\ssp b_{\nabla\delta B}\right) \nonumber\\
&=b_{1A}\ssp b_{1B}
+\epsilon\ssp b_{1A}
\left(\frac{17}{6}b_{rB}+\frac73 b_{\theta B}+\frac53 b_{\nabla\delta B}\right)
+\epsilon\ssp b_{1B}
\left(\frac53b_{rA}+\frac53 b_{\theta A}\right)
+\mathcal{O}(\epsilon^2)
\eea
Clearly without the $\epsilon$ terms this product would
be symmetric. After some algebra we find for the anti-symmetric
part
\be
2\sp\tilde{b}_{1[A}\ssp \tilde{c}_{\mrm{S},B]}
=\epsilon\ssp b_{1A}\frac76
\left(b_{rB}-\calH\frac{10}{7}\bigg(b_{\nabla\delta B}+\frac25b_{\theta B}\bigg)\right)-(A\leftrightarrow B)
\ee
which essentially recovers equation A25 in Bottaro et al.\
(here we have restored the $-\calH$ for the
relative velocity terms, which can be absorbed
into $b_{\nabla\delta}$ and $b_{\theta}$).
From this expression it is easy to extract ${c}_\mrm{S}$.

The full expression including all couplings is
\begin{equation}
 \begin{aligned}
 F_{A}^r 
 &= \left(\frac{203}{90}b_{rA} - \frac{53}{45}\mathcal{H}b_{\theta A} - \frac{2}{3}b_{1A} \frac{6 f_{\chi}}{35} + \frac{5}{3}b_{mrA}-\mathcal{H}\frac{5}{3}b_{\delta \theta A}\right)\frac{21}{17}F_\mrm{G} \\
 &\quad
 +\left(\frac{17}{6}b_{rA} - \frac{7}{3}\mathcal{H}b_{\theta A} - \frac{5}{3}\mathcal{H}b_{\nabla\delta A}\right)F_\mrm{S}
 + \left(\frac{91}{30}b_{rA} - \frac{91}{15}\mathcal{H}b_{\theta A} + b_{1A} \frac{3f_{\chi}}{5}+\frac{35}{6}b_{KrA}-\mathcal{H}\frac{35}{6}b_{KA}\right)F_\mrm{T} \\
& \equiv b_{\epsilon,\mrm{G}A} F_\mrm{G} + b_{\epsilon,\mrm{S}A} F_\mrm{S} + b_{\epsilon,\mrm{T}A} F_\mrm{T}
 \end{aligned}
\end{equation}

\section{Responses in the limit when $K/k$ is small}\label{app:response-expansions}
Here we give expansions for the responses, valid
when $K/k$ is small. This corresponds to a short-leg 
configuration~\citep{Lewis:2011}, with $K\to0$ the squeezed limit.
We can parametrise the response in terms of $K/k$, $k$ and
$\mu=\kl\cdot\k/(Kk)$, and we write
$f^{(\pm)}_\alpha(\k,\kl-\k)=f^{(\pm)}_\alpha(k,K/k,\mu)$.
Substituting into Eq.~\eqref{eq:f_alpha-pm}, the
second-order SPT kernels (see Table~\ref{tab:modecouplings}), we have up to $\mathcal{O}(K/k)$, in terms of Legendre polynomials $\mathcal{L}_{i}(\mu), i=0, 1, 2$:
\bea
f^{(+)}_\mrm{G}(k,K/k,\mu)
&=  P_L(k)\left(\frac{68}{21} \mathcal{L}_{0}(\mu) + \mathcal{O}(K/k)\right), \\
f^{(+)}_\mrm{S}(k,K/k,\mu)
&=P_L(k)
\left[-\mathcal{L}_0(\mu)\left(1+\frac13\frac{\dif\ln P_L}{\dif\ln k}\right)
    + \mathcal{L}_2(\mu)\left(0-\frac23\frac{\dif\ln P_L}{\dif\ln k}\right)
+ \mathcal{O}(K/k)\right], \\
f^{(+)}_\mrm{T}(k,K/k,\mu)
&=P_L(k)
\left[\frac87\times\frac23\sp\mathcal{L}_2(\mu) + \mathcal{O}(K/k)
\right],
\eea
for the symmetric responses, while the anti-symmetric responses are
\bea
f^{(-)}_\mrm{G}(k,K/k,\mu)
&=\mathcal{O}(K/k), \\
f^{(-)}_\mrm{S}(k,K/k,\mu)
&=P_L(k)\left[
    2\sp\mathcal{L}_1(\mu)
    \left(\frac{k}{K}\right) 
    -\left(1+\mu\frac{\dif\ln P_L}{\dif\ln k}\right) +\mathcal{O}(K/k) \right], \\[3pt]
f^{(-)}_\mrm{T}(k,K/k,\mu)
&=\mathcal{O}(K/k).
\eea
Note that $f^{(-)}_\mrm{S}$ contains $\mathcal{L}_1(\mu)(k/K)=\kl\cdot\k/K^2\sim 1/K$ and
that $f^{(-)}_\mrm{G}$ and $f^{(-)}_\mrm{T}$ contain no order $(K/k)^0$ term
(the leading-order contribution for each of these is a dipole).

Thus the total standard matter response~\eqref{eq:standard-response} is
\be
f(k,K/k,\mu)
=\sum_\alpha f_\alpha^{(+)}(k,K/k,\mu)
=P_L(k)\left[\frac{47}{21}-\frac13\frac{\dif\ln P_L}{\dif\ln k}
    +\frac23\mathcal{L}_2(\mu)
    \left(\frac87-\frac{\dif\ln P_L}{\dif\ln k}\right)
    +\mathcal{O}(K/k)\right]
\ee
It is an interesting fact that the leading-order (symmetric)
response is determined by the wavenumber of the short modes only,
i.e.\ does not depend on $K$. This is to be expected because
the response is an intrinsic property of the system.

\section{Quadratic estimators }\label{app:qeestimators}

In this section, we review the quadratic estimator approach to reconstruct the large scale density field $\delta_L(\kl)$. While this has been presented several times in the literature \citep[e.g.][]{Zhu_2016, Foreman:2018gnv, Li:2020uug, Darwish:2020prn, Zhu:2021qmm}, here we re-derive it for the case of mixed fields.

\subsection{Optimal quadratic estimator}
Given tracers $A$ and $B$, a general quadratic estimator 
of the long mode based from the $\alpha$ response is
\be
\widehat{h}^\alpha_{AB}(\kl) = \int_{\k} w_{AB}^{\alpha}(\k, \kl-\k)\delta_A(\k)\delta_B(\kl-\k)\, , \label{eq:generalqe}
\ee
where we assume some general weighting $w_{AB}^{\alpha}(\k, \kl-\k)$ which we will fix below.

To do this, we first require an unbiased cross-correlation of the reconstructed field with the desired field to get the large-scale signal of interest (e.g.\ of an unbiased tracer $C=m$) ():
\begin{equation}
    \big\langle{\widehat{h}^\alpha_{AB}(\kl)\delta_C(-\kl)}\big\rangle'
    =\int_{\k} w_{AB}^{\alpha}(\k, \kl-\k)\big\langle\delta_A(\k)\delta_B(\kl-\k) \delta_C(-\kl)\big\rangle'
    =P_{CC}(\kl)\,.
    \label{eq:crossbispecnormalization}
\end{equation}
Alternatively, we can obtain the same result from a conditional average of Eq.~\eqref{eq:generalqe}, fixing the long-mode of interest $\delta_C$ to get $\langle \widehat{h}^\alpha_{AB}(\kl) \rangle_{\delta_C}=\delta_C(\kl)$.

This constraint on $w^\alpha_{AB}$ can be rearranged into the form
\bea
    I[w_{AB}^{\alpha}] 
    = \int_{\k} w_{AB}^{\alpha}(\k, \kl-\k)\frac{B_{ABC}(\k, \kl-\k, -\kl)}{P_{CC}(\kl)} = \int_{\k} w_{AB}^{\alpha}(\k, \kl-\k) f_{AB}^{\alpha}(\k, \kl-\k)=1 \label{eq:normalizationqe}\,,
\eea
where in the second equality we used our knowledge of the response $f_{AB}$. This will in general contain a bias term, and in this work we assume $f_{AB}^{\alpha}=C^{\alpha}_{(AB)}f_{\alpha}^{(+)}+C^{\alpha}_{[AB]}f_{\alpha}^{(-)}$, i.e.\ the response is decomposed into a tracer dependent and independent terms. While we generally use the $f^\alpha_{AB}$ notation, in practice to build the quadratic estimator the tracer independent component is used.

Another condition is imposed by minimizing a loss function that in our case is defined by the variance of the estimator \eqref{eq:generalqe}. To this end, consider the covariance between $\widehat{h}^\alpha_{AB}(\kl)$ and $\widehat{h}^\beta_{XY}(\kl')$:
\bea
\big\langle\widehat{h}^\alpha_{AB}(\kl) & \widehat{h}^\beta_{XY}(\kl')\big\rangle-\big\langle{\widehat{h}^\alpha_{AB}(\kl)}\big\rangle\big\langle{\widehat{h}^\beta_{XY}(\kl')}\big\rangle \nonumber\\
& =  \int_{\k}\int_{\k'} w_{AB}^{\alpha}(\k, \kl-\k)w_{XY}^{\beta}(\k', \kl'-\k') \nonumber\\
&\qquad\qquad\times\Big[\langle\delta_A(\k)\sp\delta_X(\k')\rangle\langle\delta_B(\kl-\k)\sp\delta_Y(\kl'-\k')\rangle+\langle\delta_A(\k)\delta_Y(\kl'-\k')\rangle\langle\delta_X(\k')\delta_B(\kl-\k)\rangle\Big] \nonumber\\
&= (2\pi)^3 \delD(\kl+\kl')\int_{\k}w_{AB}^{\alpha}(\k, \kl-\k) \nonumber\\
&\qquad\qquad\times\Big[ w_{XY}^{\beta}(-\k, -\kl+\k) P_{AX}(\k)P_{BY}(|\kl-\k|) + w_{XY}^{\beta}(-\kl+\k, -\k) P_{AY}(\k)P_{BX}(|\kl-\k|) \Big] \nonumber\\
&\equiv(2\pi)^3\delD(\kl+\kl') V^{AB,XY}_{\alpha\beta}(\kl)\ .
\eea
Now assuming that the weights are invariant under parity transformation, $w_{XY}^{\beta}(-\k_1, -\k_2) = w_{XY}^{\beta}(\k_1, \k_2)$, and specializing to the case $X=A$, $Y=B$: 
\begin{equation}
\begin{split}
V^{AB}_{\alpha\beta}(\kl) 
\equiv %(2\pi)^3 \delD(\kl+\kl') % LD: don't think this should be here based on the last line above
         V^{AB,AB}_{\alpha\beta}(\K)
= % (2\pi)^3 \delD(\kl+\kl')  % LD: dittot
 \int_{\k}w_{AB}^{\alpha}(\k, \kl-\k)&\Big[ w_{AB}^{\beta}(\k, \kl-\k) \Ptot^{AA}(\k)\Ptot^{BB}(\kl-\k) \\
&
     + w_{AB}^{\beta}(\kl-\k, \k) \Pcross^{AB}(\k)\Pcross^{AB}(\kl-\k) \Big]
\end{split}
\label{eq:variance}
\end{equation}
where $\Ptot^{XX}, X \in \{A,B\}$ is the total power spectrum (signal plus shot noise), and $\Pcross^{XY}, X,Y\in\{A,B\}, X\neq Y$, is the cross-spectrum, which reduces to the total spectrum when $Y=X, X \in \{A,B\}$.
Augmenting constraints using Lagrange multipliers, then looking for stationary points, we get from the equations above the following two constraints (for $\alpha=\beta$, derived by treating $w_{AB}^{\alpha}(\kl-\k, \k)$ and $w_{AB}^{\alpha}(\k, \kl-\k)$ as different, but related by swapping the arguments):
\bea
     0&=2 w_{AB}^{\alpha}(\k, \kl-\k) \Ptot^{AA}(\k)\Ptot^{BB}(\kl-\k) +2 w_{AB}^{\alpha}(\kl-\k, \k)\Pcross^{AB}(\k)\Pcross^{AB}(\kl-\k)-\lambda f_{AB}^{\alpha}(\k, \kl-\k)\,, \nonumber\\
     0&=2 w_{AB}^{\alpha}(\kl-\k, \k) \Ptot^{AA}(\kl-\k)\Ptot^{BB}(\k) +2 w_{AB}^{\alpha}(\k, \kl-\k)\Pcross^{AB}(\k)\Pcross^{AB}(\kl-\k)-\lambda f_{AB}^{\alpha}(\kl-\k,\k)\,. \nonumber 
    % 0&=2 w_{AB}^{\alpha}(\k, \kl-\k)\big[\Ptot^{AA}(\k)\Ptot^{BB}(\kl-\k) +2 \Pcross^{AB}(\k)\Pcross^{AB}(\kl-\k)\big]-\lambda f_{AB}^{\alpha}(\k, \kl-\k)\,, \nonumber\\
    % 0&=2 w_{AB}^{\alpha}(\kl-\k, \k)\big[\Ptot^{AA}(\kl-\k)\Ptot^{BB}(\k) +2 \Pcross^{AB}(\k)\Pcross^{AB}(\kl-\k)\big]-\lambda f_{AB}^{\alpha}(\kl-\k,\k)\,, \nonumber
\eea
Multiplying the first equation by $P^{AA}_\mrm{tot}(\K-\k)P^{BB}_\mrm{tot}(\k)$, 
the second equation by $P^{AB}_\mrm{cross}(\k)P^{AB}_\mrm{cross}(\K-\k)$, then taking the 
difference and rearranging for $w^\alpha_{AB}(\k,\K-\k)$, we find for the optimal weight
\begin{equation}
    w_{AB}^{\alpha}(\k, \kl-\k) = N_{\alpha\alpha}(\kl)\frac{f_{AB}^{\alpha}(\k, \kl-\k)\Ptot^{AA}(\kl-\k)\Ptot^{BB}(\k)-f_{AB}^{\alpha}(\kl-\k, \k)\Pcross^{AB}(\k)\Pcross^{AB}(\kl-\k)}{\Ptot^{AA}(\k)\Ptot^{BB}(\k)\Ptot^{AA}(\kl-\k)\Ptot^{BB}(\kl-\k)-\big[\Pcross^{AB}(\k)\Pcross^{AB}(\kl-\k)\big]^2}\, , \label{eq:fullqeweight}
\end{equation}
with $N_{\alpha\alpha} \equiv {\lambda}/{2}$ derived from the normalization condition \eqref{eq:normalizationqe}. %If $f_{\alpha}$ is symmetric in its arguments, then we can see that this weight is trying to create a minimum variance combination of the two fields $\delta_A, \delta_B$, with 

\subsection{A sub-optimal quadratic estimator}
The weighting~\eqref{eq:fullqeweight} is not separable, hence not well suited for fast evaluation with Fourier transforms, unlike
for typical quadratic estimators. We thus define the weights in the case $A=B$ and $f_\alpha$ symmetric under exchange of its arguments (as it is for the usual $\mrm{G}_+, \mrm{S}_+, \mrm{T}_+$ gravitational kernels):
\begin{equation}
   w_{AA}^{\alpha}(\k, \kl-\k) = N^{AA}
_{\alpha\alpha}(\kl) \frac{f_{AA}^{\alpha}(\k, \kl-\k)}{2 \Ptot^{AA}(\k)\Ptot^{AA}(\kl-\k)}\ .
\end{equation}
Based on this, we use a sub-optimal estimator which is similarly simple form with the following weights
\begin{equation}
w_{AB}^{\alpha}(\k, \K-\k)=N^{AB}_{\alpha\alpha}(\K)\frac{f^\alpha_{AB}(\k,\K-\k)}{2\Ptot^{AA}(\k)\Ptot^{BB}(\K-\k)}\label{eq:symmqeweight}\,,
\end{equation}
%\LDC{This makes the approximation $\Pcross^{AB}=0$ ie. no cross correlations
%$r=0$. Can you (or has someone already?) compute corrections to this to get closer to the optimal weight. Or is it pointless?}
with normalization $N_{\alpha\alpha}$ from Eq.~\eqref{eq:normalizationqe}:
\begin{equation}
    N^{AB}_{\alpha\alpha}(\kl) = \Bigg(\int_{\k} \frac{f_{AB}^{\alpha}(\k,\kl-\k)^2}{2\Ptot^{AA}(\k)\Ptot^{BB}(\kl-\k)}\Bigg)^{-1}\ .  \label{eq:normdef}
\end{equation}

\subsection{Properties of the sub-optimal quadratic estimator}
\begin{figure}[t!]
    \centering
    \includegraphics[width=0.5\linewidth]{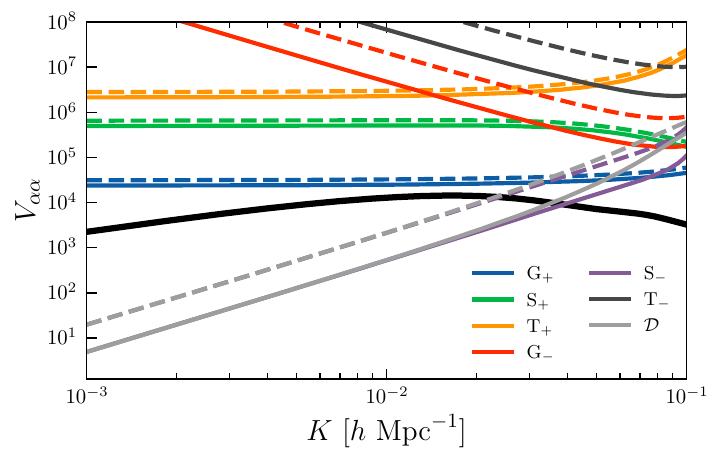}
    \caption{\textbf{QE variance and normalization.} Comparing variance (solid) and normalization (dashed) for different estimators. The displacement estimator (yellow) has a variance comparable to the full anti-symmetric shift estimator (light-blue). It increases more rapidly as we go towards smaller scales, though most of the signature we care about is on large-scales. We over-plot the linear matter power spectrum (black) for reference (though care is required as the true recovered spectrum will be biased in a scale-dependent way). Here we use $b_{1X}=1.6,b_{1Y}=1.2$ and $\bar{n}_A = 3\cdot 10^{-4} h^3 \mrm{Mpc}^{-3},\ \bar{n}_B = 4\cdot 10^{-4} h^3 \mrm{Mpc}^{-3}$.}
    \label{fig:qevariance}
    \vspace{2\baselineskip}
\end{figure}

\paragraph{Reconstruction noise from Gaussian fluctuations.} Using Eq.~\eqref{eq:variance}, the variance of our estimator coming from disconnected Gaussian fluctuations is\footnote{We can improve on the estimator's variance in a couple of ways. The first is by considering a separable GMV-like estimator, with a joint filtering of fields \citep{Maniyar_2021}. The second is by considering $\widehat{h^{\alpha}}_{BA}(\kl)$, and combining the two reconstructions in a minimum variance fashion~\citep{Darwish:2020fwf}:
\begin{equation}
    \widehat{h^{\alpha}}_{\mrm{symm}}(\kl) = x_{AB}\widehat{h^{\alpha}}_{AB}(\kl)+x_{BA}\widehat{h^{\alpha}}_{BA}(\kl)\ ,
\end{equation}
where the weights $x_{AB}, x_{BA}$ are derived by minimizing the variance of this estimator. The basic idea is that the quadratic estimator presented here is essentially the product of a Wiener-filtered (WF) field and an inverse-variance filtered (IVF) field. This implies an asymmetry in the modes we capture from $A$ and $B$. From $A$ we have more large scales, while from $B$ we have smaller scales.} 
\bea
    V^{AB}_{\alpha\beta}(\kl) &= \int_{\k}w_{AB}^{\alpha}(\k, \kl-\k)  \Big[w_{AB}^{\beta}(\k, \kl-\k) \Ptot^{AA}(\k)\Ptot^{BB}(\kl-\k) + w_{AB}^{\beta}(\kl-\k, \k)\Pcross^{AB}(\k)\Pcross^{AB}(\kl-\k) \Big] 
    \nonumber\\
&= N_{\alpha\alpha}^{AB}(\kl)N_{\beta\beta}^{AB}(\kl) \int_{\k} 
\frac{f_\alpha(\k,\kl-\k)}{4 \Ptot^{AA}(\k)\Ptot^{BB}(\kl-\k)} 
\Big[ f_\beta(\k,\kl-\k) + f_\beta(\kl-\k,\k) \frac{\Pcross^{AB}(\k)\Pcross^{AB}(|\kl-\k|)}{\Ptot^{AA}(\kl-\k)\Ptot^{BB}(\k)} \Big] 
\label{eq:variancedef}\ .
\eea
As a check, if $A=B$ and $\alpha=\beta$, we get back the normalization~\eqref{eq:normdef}, as it is the usual for the standard QEs. Figure \ref{fig:qevariance} shows the variances and normalization for different gravitational couplings. In particular, we see that the deflection estimator used in the main text (solid yellow) recovers the variance of the full anti-symmetric shift term (solid light-blue). On the other hand, the normalization is different with respect to the variance, by construction. The matching is recovered once we use the optimal estimator, as shown in Figure \ref{fig:qevarianceoptimal}.

\begin{figure}
    \centering
    \includegraphics[width=0.5\linewidth]{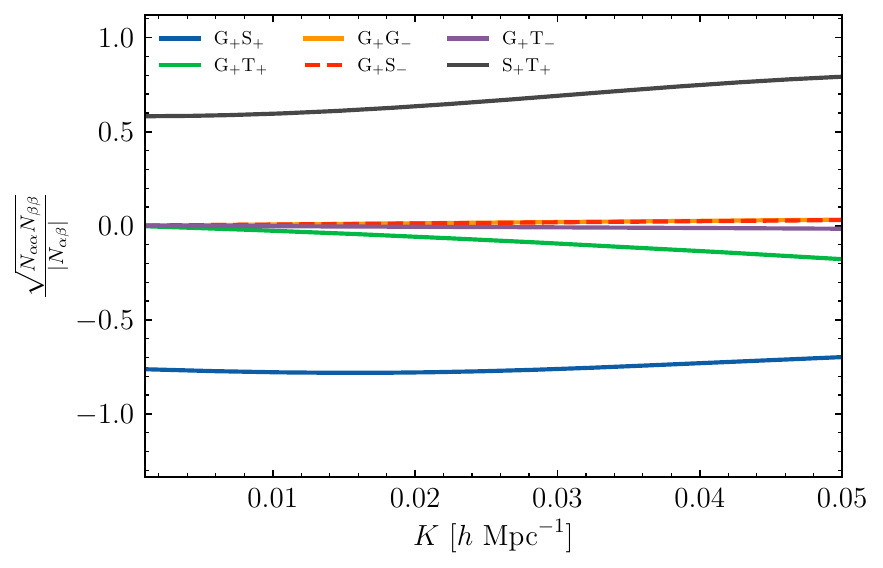}
    \caption{\textbf{Variance correlation between QEs.} Cross-correlation coefficient of the variances between some QE estimators, defined as $r = {\sqrt{N_{\alpha\alpha} N_{\beta\beta}}}/{N_{\alpha\beta}}$. We can understand the amount of correlation between estimators by looking at the corresponding squeezed limit responses in Section \ref{app:response-expansions}.\footnote{To be more precise, we actually use normalization for this plot, not the variances, even though we obtain the same results. Due to some Monte Carlo noise in our integration of variances, the curves obtained using normalization are the same as the ones from variances, but less noisy.}
    }
    \label{fig:qecorrs}
\end{figure}

\begin{figure}
    \centering
    \includegraphics[width=0.7\linewidth]{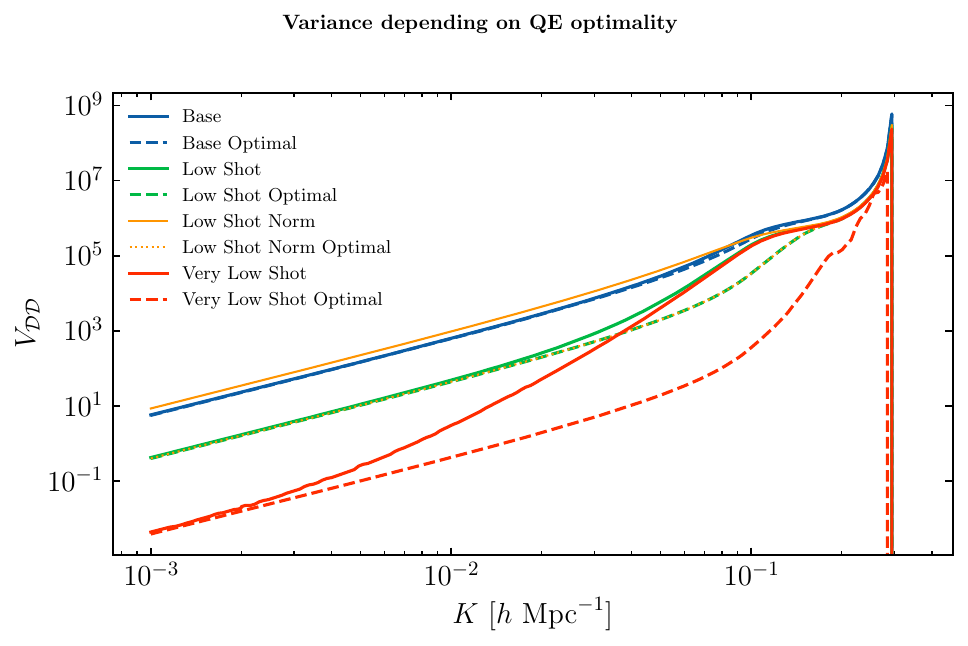}
    \caption{\textbf{Variance depending on QE optimality.} Comparing variance and normalization for our sub-optimal and optimal displacement estimator. We consider two cases with equal linear biases. The one already present in Figure \ref{fig:qevariance} (Base) ($\bar{n}_A = 3\cdot 10^{-4} h^3 \mrm{Mpc}^{-3},\ \bar{n}_B = 4\cdot 10^{-4} h^3 \mrm{Mpc}^{-3}$), a low-shot noise configuration ($\bar{n}_A = 3\cdot 10^{-3} h^3 \mrm{Mpc}^{-3},\ \bar{n}_B = 5\cdot 10^{-3} h^3 \mrm{Mpc}^{-3}$), and a very low-shot noise one ($\bar{n}_A = 3\cdot 10^{-1} h^3 \mrm{Mpc}^{-3},\ \bar{n}_B = 5\cdot 10^{-1} h^3 \mrm{Mpc}^{-3}$). The first two could represent cases where we access a fraction of Euclid or LSST galaxies to study EP violations, while the third can be seen as a limiting case.
    In both cases, our sub-optimal estimator is able to capture the optimal noise on large-scales. We also plot the norm for the low-shot noise case (brown): the sub-optimal norm (solid) does not match the sub-optimal variance (green). We need the optimal estimator for this matching to happen (dashed-green vs dotted-brown). But for the very low shot-noise configuration our sub-optimal estimator is not able to match the optimal case (pink).}
    \label{fig:qevarianceoptimal}
\end{figure}

Finally, Figure \ref{fig:qecorrs} shows the correlation of the variances among some QE estimators. In particular, we see that the anti-symmetric shift $\mrm{S}_-$ has very little correlation with the usual growth estimator (red line). This can be understood by recalling that on large scales $\mrm{S}_-$ probes a dipole signature in the gravitational response, while $\mrm{G}_+$ probes a monopole.\footnote{Note that in Figure \ref{fig:qecorrs} the correlation between the tidal $\mrm{T}_+$ and growth $\mrm{G}_+$ estimators is also small, as the tidal is mainly a quadrupole, in the squeezed limit. This is different to figure 16 of \cite{Darwish:2020prn}, where a large correlation between growth and tidal estimators was found. This was due to a bug in how the tidal estimator was implemented in the theory calculation (see \href{https://github.com/Saladino93/quadraticrecforlss/commit/be503b297e5a77074bbc7bb8a7a16c4cb79f237f}{here}). The main arguments and results of that work are unaffected.}

\paragraph*{Bias.} In practice the bispectrum appearing in Eq.~\eqref{eq:crossbispecnormalization} contains multiple mode couplings, and the condition~\eqref{eq:normalizationqe} will not perfectly hold. Hence, the true estimator acquires a bias. Taking the conditional average of the estimator, with fixed $\delta_L(\kl)$, we get
\bea
\wideaverage{\widehat{h^{\alpha}}_{AB}}_{\delta_L(\kl)}
&= \int_{\k}w^{AB}_{\alpha}(\k, \kl-\k)f_{AB}(\k, \kl-\k)\delta^{(1)}_m(\kl)
% &=\bigg(\int_{\k}\sum_{\beta} 
%     w^{AB}_{\alpha}(\k, \kl-\k) f_{AB}^{\beta}(\k, \kl-\k)\bigg) \delta^{(1)}_m(\kl) 
\eea
where $f_{AB}=\sum_\beta f^{\beta}_{AB}= \sum_\beta (C^{\beta}_{(AB)}f_{\beta}^{(+)}+C^{\beta}_{[AB]}f_{\beta}^{(-)})$ is the full response and we swapped the sum with the integral. This can be compared with Eq.~\eqref{eq:biasedexpectation}. While we could bias-harden (`deproject' from contaminants) our estimator \citep[e.g.][for LSS]{Namikawa_2013,Darwish:2020prn}, we choose not to explore this avenue in a first application of the QEs for the EP.

\subsubsection{QE Cross-spectrum}
The cross-correlation of the QE with an external matter tracer $\delta_C$ is
\begin{equation}
    \langle\widehat{h}^\alpha_{AB}(\kl) \delta_C(\kl') \rangle  = \int_{\k} w^{AB}_{\alpha}(\k, \kl-\k)\langle \delta_A(\k)\delta_B(\kl-\k) \delta_C(-\kl) \rangle\ . \label{eq:reccrossfull}
\end{equation}
We can see that this is effectively a bispectrum. By construction, it contains signal and shot-noise contributions \citep[e.g.][]{Jeong2010, Chan_2017, Ginzburg_2017, Darwish:2020prn}.

The QE shot-noise component is
\begin{equation}
    P_{C\alpha,\mrm{shot}}(\kl) = \int_\k w^{AB}_{\alpha}(\k, \kl-\k) B^{ABC}_{\mrm{shot}}(\k,\kl-\k,-\kl)\ . \label{eq:bishotnoise}
\end{equation}
And this is derived from the shot-noise bispectrum:
\begin{equation}
    B^{ABC}_{\mrm{shot}}(\k_1,\k_2,\k_3)=\frac{1}{\bar{n}^2_A}\delta^K_{AB}\delta^K_{AC}+\frac{1}{\bar{n}_C}\delta^{K}_{AC}\Pcross^{AB}(\k_2)+\frac{1}{\bar{n}_B}\delta_{BC}^{K}\Pcross^{CA}(\k_1)+\frac{1}{\bar{n}_A}\delta^{K}_{AB}\Pcross^{BC}(\k_3)\ ,
\end{equation}
where $\delta^{K}_{ij}$ is the Kronecker delta, $\k_1=\k,\ \k_2=\kl-\k,\ \k_3=-\kl$, and the cross-power spectrum becomes a total auto-power spectrum when two fields are the same. As an example, if $A=B$ and $C$ is matter, then we just get $\Pcross^{Am}(-\kl)/{\bar{n}_A}$ as shot-noise bispectrum. If $A \neq B$ but $A=C$, then we get $\Pcross^{AB}(\k_2)/{\bar{n}_A}$. This is the case depicted in Figure \ref{fig:simsgalaxydouble} where we cross-correlate our LRG-like galaxy with the QE obtained from LRG and ELG.

\subsubsection{QE Auto-spectrum}

In the same way, we write the total auto-spectrum as
\bea
    &\langle  \widehat{h^{\alpha}}_{AB}(\kl) \widehat{h^{\beta}}_{CD}(\kl') \rangle - \langle  \widehat{h^{\alpha}}_{AB}(\kl) \rangle \langle \widehat{h^{\beta}}_{CD}(\kl') \rangle \nonumber\\
    &\qquad\qquad
    = \int_{\k} \int_{\k'} w^{AB}_{\alpha}(\k, \kl-\k)w^{CD}_{\beta}(\k', \kl'-\k') \langle \delta_A(\k)\delta_B(\kl-\k) \delta_C(\k')\delta_D(-\kl'-\k') \rangle  \label{eq:recautofull}
\eea

The QE shot-noise component is
\begin{equation}
    P_{\alpha\beta,\mrm{shot}}(\kl) = \int_\k \int_{\k'} w^{AB}_{\alpha}(\k, \kl-\k)w^{CD}_{\beta}(\k', -\kl+\k') T^{ABCD}_{\mrm{shot}}(\k,\kl-\k,\k',-\kl-\k')\ . \label{eq:trishotnoise}
\end{equation}

By generalizing the calculation of appendix B in \cite{Darwish:2020prn}, the shot-noise trispectrum for four different discrete tracers is
\begin{multline}
    T^{ABCD}_{\mrm{shot}}(\k_1,\k_2,\k_3, \k_4)=\frac{1}{\bar{n}^3_A}\delta^K_{AB}\delta^K_{AC}\delta^K_{CD} \\+  \delta^K_{AB}\delta^K_{AC}\frac{1}{\bar{n}_A^2}\Pcross^{AD}(\k_4)+\delta^K_{AB}\delta^K_{AD}\frac{1}{\bar{n}_A^2}\Pcross^{AC}(\k_3)+\delta^K_{AC}\delta^K_{AD}\frac{1}{\bar{n}_A^2}\Pcross^{AB}(\k_2)+\delta^K_{BC}\delta^K_{BD}\frac{1}{\bar{n}_B^2}\Pcross^{AB}(\k_1) \\
    + \delta^K_{AB}\delta^K_{CD}\frac{1}{\bar{n}_A}\frac{1}{\bar{n}_C}\Pcross^{AC}(\k_1+\k_2)+\delta^K_{AC}\delta^K_{BD}\frac{1}{\bar{n}_A}\frac{1}{\bar{n}_B}\Pcross^{AB}(\k_1+\k_3)+\delta^K_{AD}\delta^K_{BC}\frac{1}{\bar{n}_A}\frac{1}{\bar{n}_B}\Pcross^{AC}(\k_1+\k_4)\\
    +\delta^K_{AB}\frac{1}{\bar{n}_A}B^{ACD}(\k_1+\k_2,\k_3,\k_4)+\delta^K_{AC}\frac{1}{\bar{n}_A}B^{ABD}(\k_1+\k_3,\k_2,\k_4)+\delta^K_{AD}\frac{1}{\bar{n}_A}B^{ABC}(\k_1+\k_4,\k_2,\k_3)\\+\delta^K_{BC}\frac{1}{\bar{n}_B}B^{ABD}(\k_2+\k_4,\k_1,\k_3)+\delta^K_{BD}\frac{1}{\bar{n}_B}B^{BAC}(\k_2+\k_4,\k_1,\k_3)+\delta^K_{CD}\frac{1}{\bar{n}_C}B^{CAB}(\k_3+\k_4,\k_1,\k_2)
    \,,
\end{multline}
where $\k_4=-(\k_1+\k_2+\k_3)$.
When $A=C$, $B=D$, $A \neq B$, the case relevant for us, this becomes
\begin{multline}
    T^{AB}_{\mrm{shot}}(\k_1,\k_2,\k_3, \k_4)=\frac{1}{\bar{n}_A}\frac{1}{\bar{n}_B}\Pcross^{AB}(\k_1+\k_3)
    +\frac{1}{\bar{n}_A}B^{ABB}(\k_1+\k_3,\k_2,\k_4)+\frac{1}{\bar{n}_B}B^{BAA}(\k_2+\k_4,\k_1,\k_3)
    \,
\end{multline}

\subsection{Application to simulations }\label{sec:sims}
We apply our estimator to cosmological simulations. We do not aim for a rigorous analysis, rather show how we can get reasonable predictions for the QE with simple tools. 

We use four simulations from the \textsc{AbacusSummit} suite of cosmological $N$-body simulations, run with the high-accuracy \textsc{Abacus} code \citep{Garrison}. The simulations cover a $2\  h^{-1} \rm{Gpc}$ box, containing $6912^3$ particles each with a mass of $M_{\rm{part}}=2.1\times 10^{9} h^{-1}M_{\odot}$. The simulations exceed “Cosmological Simulation Requirements” for the DESI survey \citep{Maksimova}, and hence are well suited for a first demonstration of our estimator, though we do not expect any signal; they are based on a {\it Planck} 2018 $\Lambda$CDM cosmology and the bias from our estimator acts just as a consistency test. 

As we want to showcase the application of the QE on mixed matter tracers, the specific choice of objects does not matter here, even if not realistic. Using the \textsc{Abacusutils} code we populate each realization with LRG- and ELG-like DESI samples at $z = 0.5$ \citep{Hadzhiyska}.\footnote{\url{https://abacusutils.readthedocs.io/}}
We implement the estimator of Eq.~\eqref{eq:estimatorfourier} through FFTs from \textsc{scipy} to get our estimated field \citep{2020SciPy-NMeth}. This can also be compactly written in real space as a usual shift estimator%\footnote{We get the unnormalized reconstruction through the convolution theorem $\mathcal{F}[\mathcal{F}^{-1}[\bard_X]\mathcal{F}^{-1}[\barl_Y]]$.}
\begin{equation}
    \frac{N_{XY}^{\DD}(\kl)}{2}\mathcal{F}\left\{\frac{\mathbf{\nabla}}{\mathbf{\nabla}^2}\left[ \mathbf{\nabla}\barl_X(\mathbf{x}) \bard_Y(\mathbf{x})\right]\right\}
\end{equation}
where $\barl_X(\mathbf{x})=\mathcal{F}^{-1}[\barl_X(\k)]$
and $\bard_Y(\mathbf{x}) = \mathcal{F}^{-1}[\bard_Y(\k)]$. For each simulation we calculate the filtering total power spectrum as $\Ptot^{XX}=b_{1X}^{2}\PL+\bar{n}_X^{-1}$, where $\bar{n}_X= {N_{X}}/{V}$ is the number density of objects $X$ in volume $V$. To get a first estimate of the $b_1$ bias we use the \textsc{ZCV} module, that employs the technique of control variates to reduce the variance on measured simulations' bias parameters with the analytical Zel'dovich approximation \citep{Kokron_2022, DeRose_2023, Hadzhiyska_2023}.%\footnote{We thank Borayana Hadzhiyska for offering support in measuring additional bias parameters from the \textsc{Abacus} simulations.}

To obtain the predictions, Eqs.~\eqref{eq:forecast_autorspec} and \eqref{eq:forecast_crossspec}, we need $C_{(XY)}^{\alpha}, \alpha \in \{\mrm{G_+},\mrm{S_+},\mrm{T_+}\}$ (note that we do not need the asymmetric biases as we are working in a standard cosmology). For our purposes we do not have high accuracy requirements, so we do the following: first, we build quadratic estimators for $\alpha \in \{\mrm{G_+},\mrm{S_+},\mrm{T_+}\}$. These are sensitive to $C_{(XY)}^\mrm{G}=b_{1X}b_{1Y}+\frac{21}{17}\frac{1}{2}(b_{2X}b_{1Y}+b_{2Y}b_{1X})$,  $C_{(XY)}^\mrm{S}=b_{1X}b_{1Y}$, $C_{(XY)}^\mrm{T}=b_{1X}b_{1Y}+\frac{7}{2}\frac{1}{2}(b_{s^2X}b_{1Y}+b_{s^2Y}b_{1X})$ respectively, as shown in Table \ref{tab:modecouplings_effective_coefficients}. First, the $b_{1}$ parameter is measured by taking the combination  $\langle \hat{P}_{\mathrm{gm}}\rangle_{\mrm{sims}}/\langle \hat{P}_{\mathrm{mm}}\rangle_{\mrm{sims}}$, where $g$ is the discrete tracer, and $m$ the corresponding (rescaled initial) matter density field used in the simulation (to make the measurements we use \textsc{nbodykit}; \citealt{Hand:2017pqn}). We use scale cuts $K_{\mrm{min}}=0.003 h\mrm{Mpc}^{-1}$ and $K_{\mrm{max}}=0.02 h\mrm{Mpc}^{-1}$, to ensure linear bias and linear matter. This allows us to secure a reference estimate $\widehat{C_{(XY)}^{S}}_{\mrm{first}}$ from $\widehat{b_{1X}}$ and $\widehat{b_{1Y}}$ estimates of the linear bias parameters. We then measure $\langle \hat{P}_{m\widehat{h}^{\beta}_{g_Xg_Y} }\rangle_{\mrm{sim}}/\langle \hat{P}_{\mathrm{mm}}\rangle_{\mrm{sims}},\beta \in \{\mrm{G_+},\mrm{S_+},\mrm{T_+}\}$ to get a cosmic variance free estimate of the QE biases. We jointly fit these measurements, without accounting for any covariances, using scale cuts of measurement $K_{\mrm{min}}=0.005 h\mrm{Mpc}^{-1}$ and $K_{\mrm{max}}=0.05 h\mrm{Mpc}^{-1}$. In doing all of this, we employ the \texttt{curve\_fit} function from \textsc{Scipy}, with weights calculated from the square of the linear matter power spectrum. We check that using $\widehat{C_{(XY)}^{S}}_{\mrm{first}}$ in the QE fits, instead of freeing it, gives similar results. This allows us to get the prediction shown in Figure \ref{fig:simsgalaxydouble}.
%\footnote{In the beginning we tried using }
%use the \textsc{AbacusSummit} \textsc{ZCV} module, that employs the technique of control variates to reduce the variance on measured simulations' bias parameters with the analytical Zeldovich's approximation \cite{Kokron_2022, DeRose_2023, Hadzhiyska_2023}. Specifically, the code measures Lagrangian biases that we translate into Eulerian biases using the following relations: $b_{1,E}=1+b_{1,L}$; $b_{2,E}=(2(a_1+a_2)b_{1,L}+a_2b_{2,L})/2, a_1=1, a_2=-\frac{17}{21}$; $b_{s^2,E}=-\frac{2}{7}b_{1,L}+b_{s^2,L}$ \cite{bias_review, Abidi_2018}.\ODC{I need to divide like this $b_{2,E}/2$ to get correct biases.}
%Figure \ref{fig:simsmatteronly} shows the auto and cross-spectra of the normalized reconstruction with the linear density field. For comparison, Figure \ref{fig:simsgalaxydouble} shows the same, but doubles the maximum mode of reconstruction, and Figure \ref{fig:simsmatteronly} shows an application on a matter only simulation rescaled by some linear bias. 

In addition to Figure \ref{fig:simsgalaxydouble}, we also show additional comparisons for the QE applied on a matter only simulation, and the cross-correlation coefficients with matter of various QEs.

While our simple predictions are not perfect, due to simple theory modelling and bias treatment, they are able to match reasonably well the measured spectra. This allows us to get an idea of the inner workings of the quadratic estimator, though we defer to future work more detailed studies on simulations and data applications.

\begin{figure}
    \centering
    \includegraphics[width=\linewidth]{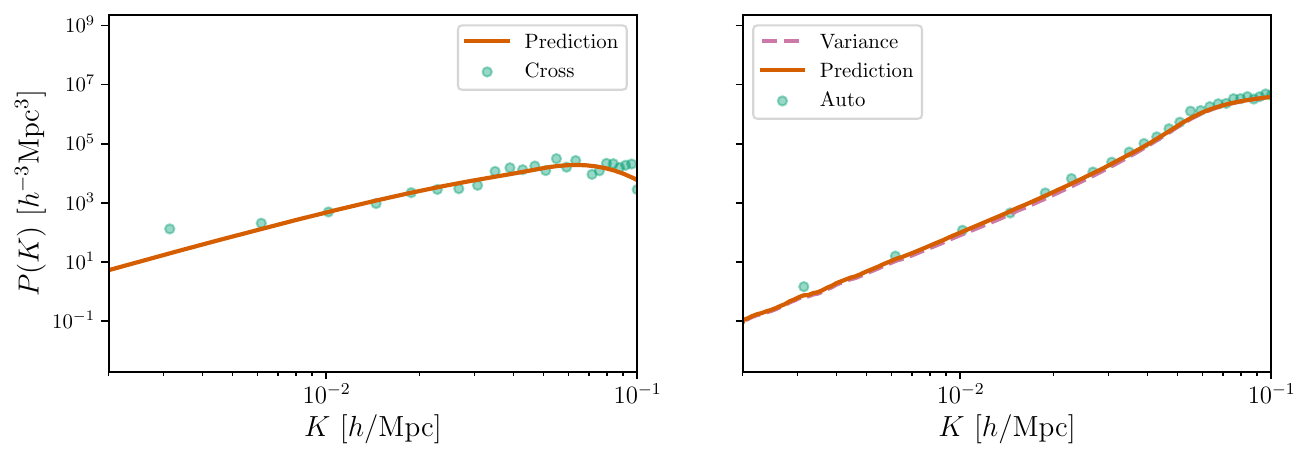}
    \caption{Our estimator applied on a matter-only simulation (rescaled by a linear bias $b=1.6$). Scattered points are measurements (green) and the predictions are shown in orange, accounting for $b_1,b_2,b_{s^2}$ biases. We use modes of reconstruction within $k_{\mrm{min,rec}}=0.05 h\mathrm{Mpc}^{-1}$ and $k_{\mrm{max,rec}}=0.1 h\mathrm{Mpc}^{-1}$. {\textit{Left panel:}}
    The cross-correlation between the reconstruction and the input linear matter. {\textit{Right panel:}} The auto-correlation of the reconstruction. The simulation is mainly explained by a Gaussian variance contribution (dashed-pink).}
    \label{fig:simsmatteronly}
\end{figure}

\begin{figure}
    \centering
    \includegraphics[width=\linewidth]{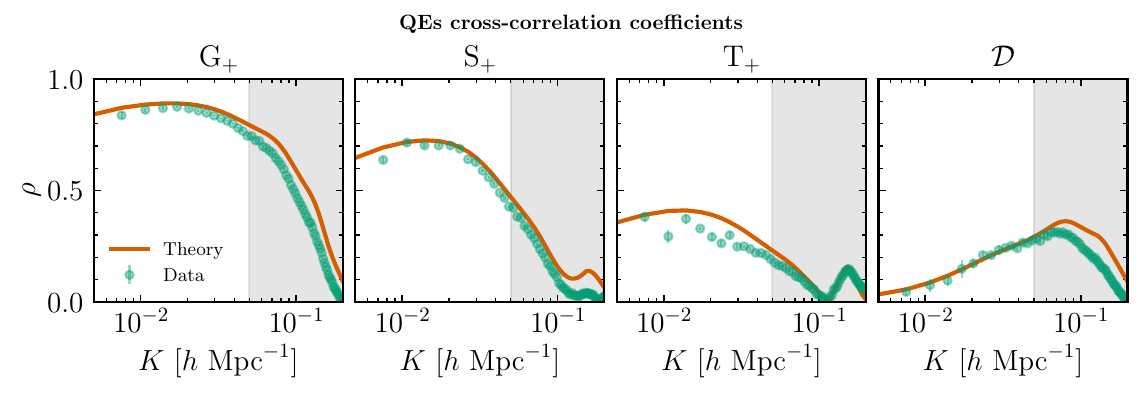}
    \caption{Cross-correlation coefficient $\rho={\Pcross^{\widehat{h}m}}/{(\Ptot^{\widehat{h}\widehat{h}}\Ptot^{mm})^{1/2}}$ with the input linear matter field for different estimators for the following galaxy combinations: mixing LRG and ELG (green). We use four simulations to get the mean measurements, and the error bars are the standard deviation of the mean. Our theoretical predictions are based on rough fits. We use the same reconstruction modes as in Figure \ref{fig:simsgalaxydouble}}
    \label{fig:crosscorrabacus}
\end{figure}

Finally, Figure \ref{fig:crosscorrabacus} presents cross-correlation coefficients of reconstructions with the input linear field. With our simple modelling and for a high maximum reconstruction mode of $k_{\mrm{max,rec}}=0.2 h\mathrm{Mpc}^{-1}$ we can get good fits to the simulations. The standard growth estimator ($\mathcal{G_{+}}$) has the highest cross-correlation coefficient with the input. We note that all the estimators except the deflection $\mathcal{D}$ are symmetrized in the input fields. When there is only one input field, the deflection estimator is basically the usual shift $\mathcal{S_+}$.

\section{A heuristic CMB lensing-like derivation of the displacement estimator }\label{app:cmblensingder}
In this appendix we give a simple derivation of the anti-symmetric shift response, a la CMB lensing. Suppose the density field $\delta_X$ experiences a shift, in such a way that at the observed position $\tilde{\x}$ we see
\begin{equation}
    \tilde\delta_{X}(\tilde{\mathbf{x}}) =  \delta_X(\mathbf{x}+\Delta \mathbf{x}_X)\ ,
\end{equation}
where we ignore galaxy bias for simplicity (it can be reabsorbed in the left-hand side). This expression tells us that the observed field is just a displaced linear field. This is similar to CMB lensing, where the temperature $T_{\mathrm{CMB}}$ is remapped due to matter fluctuations through some deflection field $\mathbf{d}$, $T_{\mathrm{CMB}}(\mathbf{n}) \rightarrow T_{\mathrm{CMB}}(\mathbf{n}+\mathbf{d})$ \citep{LEWIS_2006}.
% \footnote{Note what is happening for galaxies is not a remapping, but we will keep this simple picture in mind here. To model galaxies one can rigorously employ Lagrangian Perturbation Theory (LPT).  The Lagrangian picture treats tracer bias and displacement separately, and so it is very convenient. One could start from $1+\delta_A(\x)=\int d^3 \q F^{A}(\q) \delD(\x-\q-\mathbf{\Psi}_A(\q,\tau))$, where $\q$ represents Lagrangian positions. Then, we would need to compare galaxies at the same Lagrangian position. But here we make a naive modeling by comparing densities at the same Eulerian positions.} 

Provided $\Delta\x\ll\x$, we may Taylor expand in the displacement field:
\begin{equation}
    \tilde\delta_{X}(\k_1) = \delta_{X}(\k_1)+i\int_{\q_1} (\k_1-\q_1)\cdot \Delta\x_X(\q_1) \delta_X(\k_1-\q_1)\ ,
\end{equation}
We now take the expectation value, for fixed displacement (assumed to be independent of the small-scale observables):
\begin{equation}
    \langle \tilde\delta_{A}(\k_1)\tilde\delta_{B}(\k_2) \rangle_{\Delta \x} = - i\big[\Delta\x_A(\k_1+\k_2)    \cdot \k_2 P_{AB}(\k_2) + \Delta\x_B(\k_1+\k_2)  \cdot \k_1 P_{AB}(\k_1)\big]  \ ,
\end{equation}
where $\k_1 \neq \k_2$ and subscript $\Delta\x$ means we do not average over the displacement. We need only consider modes that respect the triangle equality $\k_1+\k_2 = \kl$, with $K\ll k_1,k_2$. Furthermore, we assume $\Delta\mathbf{x}_X = \alpha_X \Delta\mathbf{x}$, where $\alpha_X$ is some model-dependent parameter determining the acceleration of object $X$ (with $\alpha_A=\alpha_B$ if the EP holds). Neglecting terms second order in $\Delta\x_X$, we have
\begin{equation}
    \langle \tilde\delta_{A}(\k_1)\tilde\delta_{B}(\k_1) \rangle_{\Delta \x} \simeq  i \Delta\x(\kl)\cdot \k_1 \big[ \alpha_A P_{AB}(\k_1) - \alpha_B  P_{AB}(\k_1) + \alpha_A \nabla_{\k}P_{AB}(\k_1)\cdot \kl \big]  \ .
\end{equation}
In terms of the long-wavelength potential $\phi_L(\mbf{x})$, the displacement is given by $\Delta\mathbf{x} = - \bm{\nabla}\phi_L$ (ignoring the unimportant growth factor), and so $ \Delta\mathbf{x}(\kl) = -i {\kl}/{K^2} \delta_L(\kl)$. Substituting this into the foregoing equation yields
\begin{equation}\label{eq:2pcf-lensinglike}
    \langle \tilde\delta_{A}(\k_1)\tilde\delta_{B}(\k_1) \rangle_{\Delta \x} 
    \simeq ( \alpha_A  - \alpha_B )  \frac{\kl\cdot \k_1 }{K^2} \delta_L(\kl) P_{AB}(\k_1)\ .
\end{equation}
This expression can be compared with those from CMB lensing~ \citep{Hu:2001kj, LEWIS_2006}.
Note that by differentiating Eq.~\eqref{eq:2pcf-lensinglike} by $\delta_L$ we obtain the
response~\eqref{eq:fAB-as-deriv}, shown here in the limit $K\ll k_1$.

% See Refs.~\cite{Baldauf:2012, Marinucci:2024add} for alternative derivations. 
Alternatively, we can obtain Eq.~\eqref{eq:2pcf-lensinglike} by taking a Lagrangian perturbation theory style approach. In particular, by transforming to a frame that removes the uniform acceleration of object $A$, we can isolate the relative displacement of object $B$, $\Delta\mathbf{x}_{BA}=\Delta\mathbf{x}_B-\Delta\mathbf{x}_A= -(\alpha_B-\alpha_A)\bm{\nabla}\phi_L$. Thus, if the EP holds we expect $\Delta\mbf{x}_{BA}=0$, i.e. the shift gravitational term does not depend on the object. But suppose this is not the case. From the point of view of object $A$ (undisplaced in its frame),
object $B$ is transported from its initial position as $\delta_B(\mathbf{x}+\Delta \mathbf{x}_{BA}) \approx \delta_B(\mathbf{x})-(\alpha_B-\alpha_A)\bm{\nabla}\phi_L\cdot \bm{\nabla}\delta_B(\mathbf{x})$. Cross-correlating $A$ and $B$, keeping the long-wavelength mode
$\phi_L$ fixed, we have 
$\langle \delta_{A}(\mathbf{x})\sp\delta_B(\mathbf{x}+\Delta \mathbf{x}_{BA}) \rangle_{\phi_L}
\simeq
\Delta\mbf{x}_{BA}\cdot \langle \mathbf{\nabla}\delta_B(\mathbf{x})\sp \delta_{A}(\mathbf{x})\rangle$,
which is Eq.~\eqref{eq:2pcf-lensinglike} upon rewriting $\Delta\x_{BA}$ in terms of
$\delta_L$.
% Another way to derive this expression is to consider directly the two-point function $\langle \delta_A \delta_B\rangle$, and expand at linear order in the modulating field in the mid-point, assuming it is slowly varying.

\section{Forecasts }\label{app:forecasts}
Our forecasts are based on the Fisher matrix formalism. In this Section we give details about the formalism, present additional combinations that can be used to bound EP violations and alternative results. While in the main text we focused on constraining $C_{[AB]}^\mrm{S}$, in Section \ref{app:additionalforecasts} we will focus on constraining $\epsilon$. This means that we model the additional bias parameters $b_{\epsilon,\alpha X}$. Finally, we discuss a comparison with a simple bispectrum estimator in Section \ref{app:forecastsbispectrum}.

\subsection{Fisher matrix formalism}

The Fisher matrix provides an estimator of the inverse covariance of the maximum likelihood estimate under the assumption of Gaussianity around the peak of the likelihood  \citep{Dodelson:2020bqr}. For an observable $\mathcal{O}$, the Fisher information per mode $\mathbf{K}$ is given in general by \citep[e.g.][]{Tegmark_1997, heavens2010statisticaltechniquescosmology, Dodelson:2020bqr}
\begin{equation}
  \tilde{F}_{mn}(\kl)=\int_{\q_1}\cdots\int_{\q_{n-1}}\frac{\partial\mathcal{O}}{\partial\theta_m}\mrm{Cov}^{-1}(\mathcal{O})\frac{\partial\mathcal{O}}{\partial\theta_n}\label{eq:fisherstandard}
\end{equation}
where {$\theta_m, \theta_n$} are parameters of interest, $\mrm{Cov}(\mathcal{O})$ is the covariance matrix of the observable $\mathcal{O}$, and we impose the constraint $\kl+\sum_i \q_i = 0$ for the $n$-point correlation function being measured. In our case, the internal wavevectors $\q_i,i \in \{1,...,n-1\}$ used for the quadratic estimator reconstruction range from $k_{\mrm{rec,min}}$ to $k_{\mrm{rec,max}}$.

The total Fisher information is obtained by integrating over the long modes:
\begin{equation}
    F_{mn} = V\int_{K_{\rm{min}}}^{K_{\rm{max}}} \frac{ K^2 dK}{2\pi^2}\tilde{F}_{mn}(\kl)\ ,
\end{equation}
where we implicitly set no directional dependence in the Fisher matrix per mode (e.g. we do not consider redshift-space distortions). The integration limits span from a survey fundamental mode $K_{\mathrm{min}} = 2\pi/V^{1/3}$, where $V$ is the volume, to some $K_{\mathrm{max}}$ that we choose $K_{\mathrm{max}}\leq k_{\mrm{rec,max}}$.
%in units $h\ \mrm{Mpc}^{-1}$.

When considering Fisher matrices of a survey spanning several redshift bins $i$, we calculate the Fisher matrix $F^i_{mn}$ for that redshift bin, with $K_{\mrm{min}}$ depending on its volume $V_i$. Finally, the overall Fisher matrix, assuming independence of redshift bins, is $F^{\mrm{sum}}_{mn} = \sum F^{i}_{mn}$.

\subsubsection{Power spectra definitions }\label{sec:powerspectradef}

%We use power spectra from all possible combinations of available fields. Our dataset includes auto-power spectra $P_{\text{tot}}^{XX}$ and cross-power spectra $P_{\text{cross}}^{XY}$, where the fields are defined as $\text{Data} = \big\{P_{\text{tot}}^{XX}, P_{\text{cross}}^{XY} : X,Y \in \{A,B,\mathcal{D},\mrm{G_+}\}, X \neq Y\big\}$. Here $\mathcal{D}$ represents the reconstructed displacement field and $\mathrm{G_+} = \mrm{G}$ is the standard growth estimator. When not specified, from now on $X,Y$ are abstract labels for $A,B$.

Our analysis considers two galaxy tracers $A$ and $B$ to perform QE reconstruction. We combine the reconstructed field with the original galaxy fields to measure the EP violations through the $\epsilon$-scale-dependent bias $b_{\DD}(\kl)$.\footnote{Note that in principle we could use a third galaxy tracer, another reconstruction, or the matter distribution itself too.}

We consider the following observables $\mathcal{O}$ in our Fisher matrix analysis:
%\begin{itemize}
%    \item Large-scale galaxy power spectra follow the standard form:
%\begin{equation}
%    \Ptot^{XY}(\mathbf{K}) = b_{1X}b_{1Y} P_\mrm{lin}(\mathbf{K}) + \delta^\mrm{K}_{XY} N_{XX,\text{shot}}\ , \label{eq:forecast_autogspec}\,
%\end{equation}
%where $b_{1X}, b_{1Y}$ are linear bias parameters and %$N_{XX,\text{shot}}=1/\bar{n}_X$ is the Poissonian shot-noise contribution with number density $\bar{n}_X$.
%\LDC{Since this $N$ can be confused with the $N$
%in the normalization of the QE, can we just write skip writing $N$ here and just put $1/n$ in Eq.~\eqref{eq:forecast_autogspec}? Or is $N$ here needed elsewhere.}
\begin{itemize}
\item Correlations among the displacement estimator $\widehat{h}^{\mathcal{D}}_{AB}(\kl)$ and growth estimator $\widehat{h}^{\mathrm{G_+}}_{AB}(\kl)$:
\begin{equation}
\Ptot^{\alpha\beta}(\mathbf{K})
\equiv
\big\langle \widehat{h}^\alpha_{AB}(\kl)
\widehat{h}^\beta_{AB}(\kl') \big\rangle'
=b_{\alpha}(\kl)b_{\beta}(\kl) P_L(\mathbf{K}) + V_{\alpha\beta}(\kl) + P_{\alpha\beta,\text{shot}}(\kl)\ ,\label{eq:forecast_autorspec}
\end{equation}
where  $V_{\alpha\beta}$ captures reconstruction noise from Gaussian fluctuations, $P_{\alpha\beta,\mrm{shot}}$ is the auto-spectrum reconstruction shot-noise, as this correlation probes a four-point function of discrete tracers. The effective biases $b_{\alpha},b_{\beta}$ for $\alpha,\beta \in \{\DD,\mrm{G}_+\}$ are calculated in a similar way to Eq.~\eqref{eq:biasedexpectation} (we omit the $AB$ superscript indices for brevity).
\item Cross power spectra between reconstruction $\widehat{h}^\alpha_{AB}(\kl)$ and tracer $X\in \{A,B\}$:
\begin{equation}
    \Pcross^{X\alpha}(\mathbf{K}) \equiv
\big\langle \widehat{h}^\alpha_{AB}(\kl)
\delta_X(\kl')\big\rangle'= b_{1X} b_{\alpha}(\mathbf{K}) P_L(\mathbf{K}) + P_{X\alpha,\text{shot}}(\mathbf{K})\ , \label{eq:forecast_crossspec} 
\end{equation}
where $P_{X\alpha,\text{shot}}$ is the cross-shot noise component, as the cross-spectrum probes a three-point function of discrete tracers.
\item Large-scale galaxy cross-power spectrum:
\begin{equation}
    \Pcross^{AB}(\mathbf{K}) 
    = \big\langle\delta_A(\K)\sp\delta_B(\K')\big\rangle'
    = b_{1A}b_{1B} P_L(\mathbf{K})\ , \label{eq:forecast_crossgspec}\,
\end{equation}
where $b_{1A}, b_{1B}$ are linear bias parameters and we assume no common shot-noise between the two galaxy populations.
\end{itemize}

We deliberately exclude the auto-power spectra $\Ptot^{AA}$ and $\Ptot^{BB}$. While including these spectra could lead to more robust marginalized constraints, we consider scenarios where they are not usable due to systematic effects. We include these auto-power spectra only in covariance calculations.

Table~\ref{tab:datasetsused} summarizes different possible data combinations. The most conservative approach relies entirely on the reconstruction cross-correlation $P_{\mathrm{cross}}^{\mathcal{D}\mathrm{G_+}}$, which contains cosmological information even when standard galaxy cross-correlations are unavailable, albeit with higher noise levels.

\begin{table}[t!]
\centering
\caption{
Summary of different combinations of spectra considered in the Fisher forecast, and their shorthands. 
These spectra depend on $\epsilon$ and the bias parameters $b_{1X}$, $b_{2X}$, and $ b_{s^2X}$, $X \in \{A,B\}$.
Note that $\mathcal{D} \otimes \mathrm{G_+}$
is based purely on spectra from reconstructed fields,
while the rest of the combinations also consider spectra from cross-correlations
with galaxy fields (on large scales).
}
\label{tab:datasetsused}
\centering
\begin{tabular}{ll}
Combination & Spectra Included \\
\hline
$\mathcal{D} \otimes \mathrm{G_+}$ & $P_{\mathrm{cross}}^{\mathcal{D}\mathrm{G}_+}$ only \\[1pt]
$\mathcal{D} \otimes \mathrm{Galaxies}$ & $P_{\mathrm{cross}}^{XY}$ for $X,Y \in \{A,B,\mathcal{D}\}$, $X\neq Y$ \\[1pt]
$\mathrm{G_+} \oplus \mathrm{Galaxies}$ & $P_{\mathrm{cross}}^{XY}$ for $X,Y \in \{A,B,\mathrm{G}_+\}$, $X\neq Y$; and $P^{\mathrm{G_+}\mathrm{G_+}}_{\mathrm{tot}}$ \\[1pt]
$\mathcal{D} \oplus \mathrm{Galaxies}$ & As in $\mathcal{D} \otimes \mathrm{Galaxies}$ but adding $P_{\mathrm{tot}}^{\mathcal{D}\mathcal{D}}$ \\[1pt]
$\mathcal{D} \oplus \mathrm{G}_+ \oplus \mathrm{Galaxies}$ & As in $\mathcal{D} \oplus \mathrm{Galaxies}$ but adding $P_{\mathrm{cross}}^{X\mathrm{G}_+}, P^{\mathrm{G_+}\mathrm{G_+}}_{\mathrm{tot}}$ \\
\end{tabular}
\end{table}

\subsubsection{Analytical expressions}

Although the literature has extensive discussions about Fisher matrix calculations, here we briefly mention some analytical results for convinience of the reader. We will focus on two fields only, $\DD$ and $X$. We assume tracer $X$ insensitive to $\epsilon$. Calculations involving more than two fields can be performed using dedicated symbolic computation codes such as \textsc{Sympy} \citep{10.7717/peerj-cs.103}. While expressions become more involved for three or more fields, the underlying ideas remain the same.

\paragraph{Fisher information from cross-correlations only (one power spectrum only).} We begin by examining the cross-spectrum between our reconstructed field $\widehat{h}^{\mathcal{D}}_{AB}$ and a matter tracer $\delta_X$ (such as weak lensing or galaxy distributions) on large-scales:
\begin{equation}
    \langle \delta_X(\mathbf{K}') \widehat{h}^{\mathcal{D}}_{AB}(\mathbf{K}) \rangle'  = \Pcross^{X\DD}(\mathbf{K}) =  b_{1X}b_{\DD}(\epsilon,\mathbf{K})\PL(\mathbf{K}) + N_{X\DD,\rm{shot}}(\mathbf{K}Z\, , \label{eq:reccross}
\end{equation}
where $N_{X\DD,\rm{shot}}(\mathbf{K})$ represents the bispectrum cross-shot-noise contribution.

Using Eq.~\eqref{eq:fisherstandard} and Gaussian error bars for the cross-spectrum \eqref{eq:reccross}, the Fisher information matrix for the $\epsilon$ parameter is
\begin{equation}
    \tilde{F}_{\epsilon\epsilon}[{\Pcross^{X\DD}}] = \tilde{F}_{\epsilon\epsilon}^{\mrm{rec}}(\kl) = \frac{(\partial_{\epsilon}\Pcross^{X\DD}(\mathbf{K}))^2}{\Ptot^{XX}(\kl) P_{\DD\DD}(\mathbf{K})+\Pcross^{X\DD}(\mathbf{K})^2}\ . \label{eq:fishercrossanalytical}
\end{equation}

For $X$, we have $\Ptot^{XX}=b_{1X}^2\PL+N_X$, where $N_X$ is a shot-noise component. On the other hand, the reconstructed field has:
\begin{equation}
    \Ptot^{\DD\DD}(\mathbf{K}) = b_{\DD}^2(\kl)\PL(\kl)+V_{\DD\DD}(\kl)+P_{\DD\DD,\mrm{shot}}(\kl)\ ,
\end{equation}
where $V_{\DD\DD}$ is the Gaussian variance, and $N_{\DD\DD,\mrm{shot}}$ is the trispectrum auto-shot-noise induced component.

\paragraph{Fisher information from cross and auto-correlations.} We now combine the auto-spectra of $X$ and $\DD$ with their cross-correlation. One can use again Eq.~\eqref{eq:fisherstandard} by packaging together auto- and cross-correlations in a data vector $\mathbf{d}$, then considering their covariance matrix $C$. Alternatively, a useful formula to have is \citep[e.g.][]{Tegmark_1997}:
\begin{equation}
    \tilde{F}_{\epsilon\epsilon}[{P^{A\DD}_{\mrm{cross+auto}}}](\kl) = \tilde{F}^{\mrm{joint}}_{mn}(\kl) = \frac{1}{2}\mathrm{Tr}\Big[ \partial_mC(\kl)C^{-1}(\kl)\partial_nC(\kl)C^{-1}(\kl)\Big]\ , \label{eq:fisherjointanalytical}
\end{equation}
where $C$ is the total covariance matrix of a zero-mean data vector $\mathbf{d}$. This can be then specialized for example for $m=n=\epsilon$. For a more explicit formulation of this expression see equation 52 in \cite{Darwish:2020prn}.

\subsubsection{Implementation }\label{sec:numericalforecasts}

The Fisher matrices are computed by numerically evaluating Eq.~\eqref{eq:fisherstandard} and Eq.~\eqref{eq:fisherjointanalytical}. We evaluate integrals and derivatives using Monte Carlo integration via the \textsc{torchquad} and \textsc{jax} libraries \citep{jax2018github, Gomez_torchquad_Numerical_Integration_2021}, which provide significant computational speed-up on GPUs compared to CPUs. Automatic differentiation in \textsc{jax} is used for derivative calculations.

Unless otherwise specified, our baseline reconstruction employs $k_{\mathrm{min,rec}}=0.051\,h\mrm{Mpc}^{-1}$ and $k_{\mathrm{max,rec}}=0.15\,h \mrm{Mpc}^{-1}$, a regime where standard perturbation theory is reliable. Our analysis integrates Fisher matrix information over scales from some $K_{\mrm{min}}$ to $K_{\mrm{max}} \sim 0.05\, h \mathrm{Mpc}^{-1}$. We choose $K_{\mrm{max}}< k_{\mathrm{max,rec}}$ to establish a separation between the modulating large-scale mode and the modulated local spectra.

\begin{figure}
    \centering
    \includegraphics[width=0.5\linewidth]{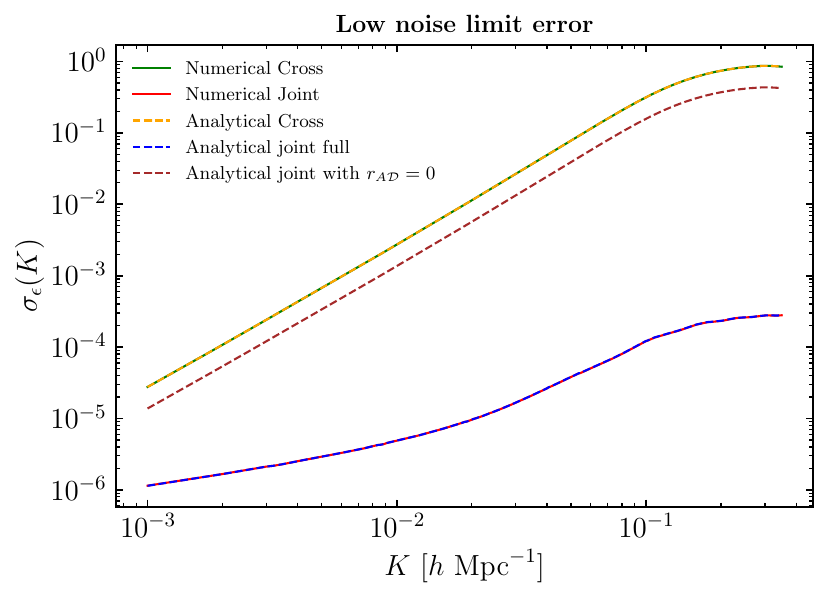}
    \caption{Forecast comparing numerical and analytical results. We assume a very small noise here (shot and reconstruction). We compare the cross-spectrum only error bars from Eq.~\eqref{eq:fishercross} (dashed orange) and the code (solid green). We do the same for the joint analysis numerically (solid red) and analytically (dashed red). Finally, we show the joint analysis but assuming a zero cross-correlation between reconstruction and the galaxy tracer (dashed brown). We can see that the error bars are boosted, as we do not any more reap the benefits of cosmic variance cancellation.}
    \label{fig:forecast_analytical}
\end{figure}

As a cross-check of our numerical code, we compare the analytical and numerical predictions of the unmarginalized per mode Fisher matrix results, in the low noise limit. We show this comparison in Figure~\ref{fig:forecast_analytical} where we plot $\sigma_\epsilon = \sqrt{F_{\epsilon\epsilon}^{-1}}$, with $F$ a Fisher matrix from Eq.~\eqref{eq:fishercrossanalytical} or \eqref{eq:fisherjointanalytical}. 
% We also made full tests using \textsc{Sympy} to compare \textsc{JAX} based results, for the standard noise case.

\section{More forecasts }\label{app:additionalforecasts}

This section forecasts the constraining power of the displacement estimator, which stems from the unique $1/K$ dipole generated in the squeezed bispectrum by EP violations (see Section \ref{sec:poleonly}). Our analysis also reveals that, for the specific models and configurations considered here, the auto-spectrum of the displacement estimator contributes very little to improving the constraints on $\epsilon$. Furthermore, we emphasize the crucial benefit of the QE formalism to allow for easy combinations with other tracers. As we will show, this allows for a powerful synergy between the displacement and growth estimators, leading to more robust results.

In the main text we focussed on the dressed quantity
$\epsilon \times C_{[AB]}^\mrm{S}$ to facilitate comparison with the
literature, in particular the work of \cite{Graham:2025fdt}.
Here we focus on $\epsilon$ itself, which connects more directly with theory.

In the main text we made the assumption that $C_{[AB]}^{\alpha}$ is an independent bias parameter, but in practice it depends on other bias parameters, including the linear bias parameter $b_{1X}$. While $b_{1X}$ will be constrained to high accuracy from the galaxy power spectrum, $b_{\epsilon,\alpha X}$ is poorly constrained and not well understood. As the combination of all of these parameters is anti-symmetric, marginalizing over them will blow up any attempt to constrain $\epsilon$. Hence, in the forecasts below we will fix $b_{\epsilon,\alpha X}$ to some fiducial value.

\subsection{Simplified Setup}

We construct various Fisher matrices $F_{ab}$ using the power spectra listed in Table~\ref{tab:datasetsused}. For all retained parameters, we impose flat priors. Our parameter set includes the standard bias parameters \(\{b_{1X}, b_{2X}, b_{s^2X}\}\) for \(X \in \{A,B\}\), but excludes the \(\epsilon\)-related bias parameters \(\{b_{\epsilon,GX}, b_{\epsilon,SX}, b_{\epsilon,TX}\}\), which are poorly constrained due to limited knowledge (the precise details of which are model dependent). %This leads to two approaches: (1) directly constraining the combinations $\epsilon \times C_{(AB)},C_{[AB]}^{\alpha},\alpha \in \{G,S,T\}$, or (2) constraining $\epsilon$ while fixing the poorly constrained parameters. 

For the forecasts presented here we use a simple setup. We consider a DESI-like survey \citep{desicollaboration2016desiexperimentisciencetargeting}, with a number density $\bar{n} \sim 5\times 10^{-4}\,h^{3}\mathrm{Mpc}^{-3}$ and volume $V \sim 30 h^{-3}\mathrm{Gpc}^{3}$ at $z = 0.5$; specifically we have $\bar{n}_A=\bar{n}/3$, $\bar{n}_B =\bar{n}/4$, $b_{1A}=1.6$, and $b_{1B}=1.3$. We use the fitting formula of \cite{Lazeyras:2015lgp} to get $b_{2A},b_{2B}$; for the tidal biases we use the co-evolution prediction $b_{s^2A}=-2/7(b_{1A}-1)$, $b_{s^2B}=-2/7(b_{B}-1)$, assuming zero tidal bias in Lagrangian space \citep{Abidi_2018}. 
For a survey of volume $V$, we assume only a fraction of the surveyed objects to be suitable for analysis.

\paragraph{Specifying an EP violating model.} We finally specify fiducial values for the additional biases $b_{\epsilon,\alpha X},\alpha \in \{\mrm{G},\mrm{S},\mrm{T}\}$. We use a simplified version of the fifth-force model of B24, presented in a in basic form in Section \ref{sec:epviolationmodel}, and specified in Appendix~\ref{app:bottaro}. To summarize, B24 considers a dark fifth force (affecting dark matter but not baryons), resulting in a nonzero velocity bias (between matter and galaxies) and an enhanced growth of structure (that we will ignore for now). The simplified version we consider has a new bias $b_r$ quantifying the response of galaxy densities due to changes in the local baryon-CDM ratio. Unlike the well determined $b_1$, the relative bias $b_r$ is not as well determined, though some preliminary work has been carried out \citep{Schmidt:2016coo, Chen:2019cfu, Barreira_2020, Khoraminezhad:2020zqe}. Following \cite{Schmidt:2016coo} we simply assume $b_r$ is order unity, with $b_{rA}=b_{rB}=b_r=1$ (generally these parameters will take on different values for $A$ and $B$). However, the most important parameter combination for a detection of EP violation is $C_{[AB]}^\mrm{S} \neq 0$. Provided this is nonzero, setting $b_{rA}=b_{rB}$ is fine for a fiducial model (remember that we care about $(b_{1B}b_{rA}-b_{1A}b_{rB})/2$). In making this choice, we ensure that the marginalization is robust to varying values of the linear bias parameters. We then have $b_{\epsilon,\mrm{S}X}=17/6b_r$,  $b_{\epsilon,GX}=203/90 b_r$, and $b_{\epsilon,\mrm{T}X}=  91/30 b_r$ from the model in Appendix~\ref{app:bottaro}.

When marginalizing over galaxy bias parameters we consider only up to second-order bias: $b_{1X}, b_{2X}, b_{s^2X}$, with $X \in \{A,B\}$. % \st{and $\epsilon$}.
We do not marginalize over the poorly constrained parameters $b_{\epsilon,\alpha X},\alpha \in \{\mrm{G},\mrm{S},\mrm{T}\}$.

\subsection{Constraints from the anti-symmetric shift response }\label{sec:poleonly}

As a first study of the estimator's performance, we forecast $\epsilon$ constraints expected solely from the shift $S$ (both symmetric and anti-symmetric). That is, we consider a scenario where EP violation leads to $b_{\epsilon,\mrm{S}X}\neq 0$, but $b_{\epsilon,\mrm{G}X}=0$ and $b_{\epsilon,\mrm{T}X}=0$ (changes to the shift only). We will find that the constraints are solely due to the anti-symmetric shift ($\mrm{S}_{-}$), 
parametrized by $C^\mrm{S}_{[AB]}$, with the symmetric shift $C^\mrm{S}_{(AB)}$ carrying very little weight. The fiducial model assumes $\epsilon=0$, $b_{\epsilon,\mrm{S} X} \sim 1$, and $b_{\epsilon,\mrm{G} X}=b_{\epsilon,\mrm{T} X} = 0$ for both tracers $A$ and $B$.

%\LDC{This doesn't make sense. In this section C, do you consider only anti-symmetric shift? Or do you consider other responses too? Because in this section you show plots with other responses.
%So it's not true that we `forecast $\epsilon$ constraints expected solely from the anti-symmetric shift'.
%} \ODC{Let me clarify in the text. Overall, the shift term is used. But basically what gives constraints is its anti-symmetric part.}

\subsubsection{Analytical estimates}

%\LDC{this section feels a bit top heavy (analytical stuff has lots more discussion than the numerical results). My suggestion is: start with the numerical results (section~\ref{sec:practical-performance}) then justify it by saying `we can understand these numerical results by looking at a simple analytical model...' etc. That way we won't need to have subsections and the most important results comes first}
%\LDC{if you agree I will change it}

Before presenting numerical results, we analytically examine the estimator's expected behaviour. This builds on previous work with amplitude parameters \citep[e.g.][]{Schmittfull_2015, dePutter:2018jqk, Darwish:2020prn}. We focus on constraining $\epsilon$ using $\widehat{h}^\mathcal{D}_{AB}$ in combination with field $A$, assumed insensitive to $\epsilon$. We ignore marginalization over nuisance parameters for simplicity.

In the low-noise limit, Eq.~\eqref{eq:fishercrossanalytical} gives the
following Fisher information on $\epsilon$, from the cross-correlation
$\Pcross^{A\DD}$ only:
\begin{equation}\label{eq:F-single}
    \tilde{F}_{\epsilon\epsilon}[{\Pcross^{A\DD}}](\kl) = \frac{1}{2}\left(\frac{\partial_\epsilon b_\DD(\kl)}{b_\DD(\kl)}\right)^2 \,,
\end{equation}
where $b_\DD$ is the effective bias given by
Eq.~\eqref{eq:effective-bias}. This clearly reaches a ceiling in the constraining power of $\epsilon$, in the sense that lower noise will not improve this figure.

When including the $\mathcal{D}\mathcal{D}$ auto-spectrum, using Eq.~\eqref{eq:fisherjointanalytical} we get the following Fisher information
\begin{equation}
     \tilde{F}_{\epsilon\epsilon}[{P^{A\DD}_{\mrm{cross+auto}}}](\kl) = \frac{2-r_{A\DD}^2}{1-r_{A\DD}^2}\left(\frac{\partial_\epsilon b_\DD(\kl)}{b_\DD(\kl)}\right)^2
     =2\,\frac{2-r_{A\DD}^2}{1-r_{A\DD}^2}\,F_{\epsilon\epsilon}[{\Pcross^{A\DD}}](\kl) \,,
     \label{eq:F-joint}
\end{equation}
where $r_{A\DD}={\Pcross^{A\DD}}/{(\Ptot^{\DD\DD}\Ptot^{AA})^{1/2}}$ is a field-level correlation coefficient.

\begin{figure*}[t!]
\centering
\includegraphics[width=.49\textwidth]{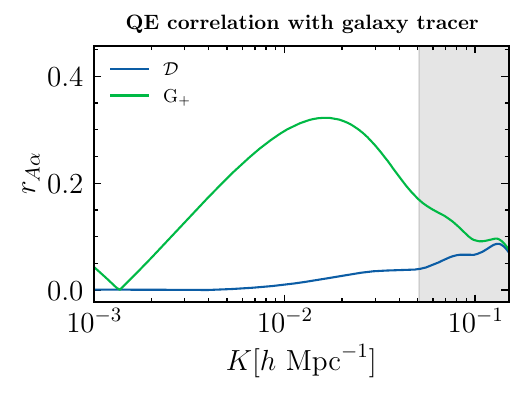}%
\includegraphics[width=.5\textwidth]{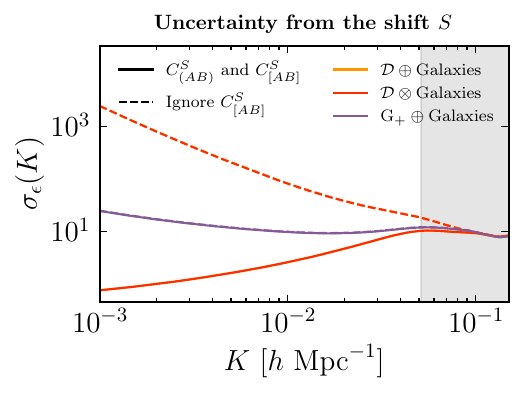}
\caption{{\textit{Left panel:}} Cross-correlation coefficient $r_{A\alpha}$ between $\delta_A$ and the reconstructed matter mode $\widehat{h}^{\alpha}_{AB}$, for $\alpha=\DD$ and $\mrm{G}_+$. Compared to the displacement estimator, the growth estimator has higher fidelity to the galaxy field. As in Figure~\ref{fig:simsgalaxydouble}, grey bands represent the
range of wavenumbers used in the reconstruction.
{\textit{Right panel:}} $\epsilon$ uncertainty per wavenumber $K$ with (solid curves) and without (dashed curves) the anti-symmetric shift. Note that even with a very low $r_{A\alpha}$, the displacement estimator ($\DD \otimes \mrm{Galaxies}$ and $\DD \oplus \mrm{Galaxies}$, red and orange respectively, though they are identical in this case) has a much higher sensitivity to $\epsilon$ with respect to the growth estimator (purple). The improvement in $\sigma_\epsilon$ from the anti-symmetric shift ($\mrm{S}_-$) is clear when comparing the solid and dashed red curves.
}
\label{fig:ignoreshift}
\end{figure*}

In the limit $r_{A\DD}\to1$, this expression leads to the expected cosmic variance cancellation (same mode measured multiple times)~\citep{seljakPhysRevLett.102.021302, McDonald_2009}. Unfortunately, in our case cosmic variance cancellation is difficult to realize in practice,
as the displacement estimator~\eqref{eq:estimatorfourier} turns out to be a poor tracer of matter, and hence of the galaxy field $A$. The left panel of Figure \ref{fig:ignoreshift} illustrates this, where for comparison we also show the higher fidelity of the growth estimator to $A$, as measured from the cross-correlation coefficient.

Despite this, the displacement estimator manages to capture a $1/K$ scaling from the response function, and is thus sensitive to changes in $\epsilon$. The right panel of Figure~\ref{fig:ignoreshift} shows the predicted Gaussian errors per mode on $\epsilon$, calculated from taking the reciprocal of the Fisher matrix per mode. The displacement estimator, see $\mathcal{D} \otimes \mathrm{Galaxies}$ (solid red) and $\mathcal{D} \oplus \mathrm{Galaxies}$ (solid orange), although in this case both are practically identical, outperforms the growth estimator (solid purple curve) by a significant margin. The better performance is due to the displacement estimator's sensitivity to the anti-symmetric shift component. This interpretation is supported by setting $C^\mrm{S}_{[AB]} = 0$ which removes the anti-symmetric shift component and degrades the displacement estimator forecasts by several orders of magnitude, leaving only information from the symmetric component $C^\mrm{S}_{(AB)}$. In contrast, the growth estimator remains virtually insensitive to these anti-symmetric effects. (The picture changes however if we include, say, $C^\mrm{G}_{[AB]}$ from $\epsilon$ through $b_{\epsilon,\mrm{G}X} \neq 0$ due to EP violation.)

\subsubsection{Numerical results}\label{sec:practical-performance}
\begin{figure}
    \centering
\includegraphics[width=0.9\linewidth]{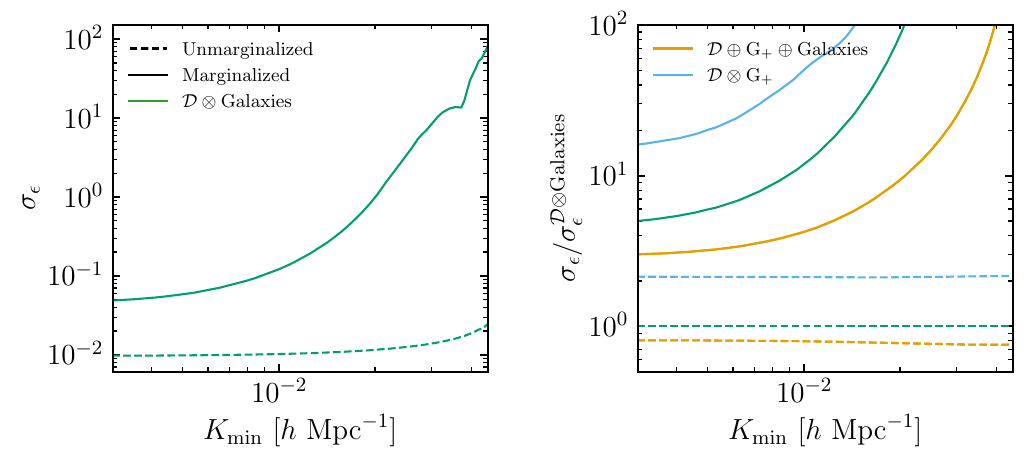}
    \caption{\textbf{Overall constraints on $\epsilon$ from the displacement estimator using $C_{[AB]}^\mrm{S}$.} {\textit{Left panel:}} Unmarginalized (dashed) and marginalized (solid) constraints from the $\DD \otimes \mrm{Galaxies}$ (green) combination. Note that we do not show $\DD \oplus \mrm{Galaxies}$ as it gives identical results. {\textit{Right panel:}} Comparison between uncertainties from different data combinations. Here we use the baseline configuration and
    normalize $\sigma_\epsilon$ to the unmarginalized error from $\DD \otimes \mrm{Galaxies}$ (dashed green in the left panel).
    }
    \label{fig:forecast_base}
\end{figure}

We now verify the arguments above through numerical forecasts, as implemented following Section \ref{sec:numericalforecasts}. Figure \ref{fig:forecast_base} shows the overall uncertainty on $\epsilon$  as a function of the minimum wavenumber $K_{\mrm{min}}$. The left panel focuses on cross-correlations only, $\mathcal{D} \otimes \mathrm{Galaxies}$ (green). Including the auto-spectrum of $\widehat{h}^{\DD}_{AB}$ yields negligible improvement, and is hence omitted in the plot. Marginalization over nuisance parameters increases uncertainties by a factor of about three compared with unmarginalized constraints (compare solid and dashed lines).

The right panel compares two alternative combinations relative to our unmarginalized $\mathcal{D} \otimes \mathrm{Galaxies}$ case. The reconstruction-only cross-correlations $\mathcal{D} \otimes \mathrm{G_+}$ combination (light blue) uses information from large scales without requiring external tracers, but suffers significant degradation when nuisance parameters are marginalized (solid vs.\ dashed lines). The combined set, $\mathcal{D} \oplus \mathrm{G_+} \oplus \mathrm{Galaxies}$,  incorporates both reconstruction and galaxies cross-correlation and reconstruction auto-correlations, and proves to be more robust to marginalization. Indeed, the marginalized constraints on $\epsilon$ improve by $40\%$ with respect to $\mathcal{D} \otimes \mathrm{Galaxies}$ marginalized constraints.

These results indicate that anti-symmetric shift alone does not yet yield competitive constraints on $\epsilon$, consistent with the bispectrum analysis of B24 (see their appendix where they consider unmarginalized constraints from the bispectrum pole). This limitation motivates exploring additional observational strategies to enhance sensitivity to $\epsilon$.

\begin{figure}
    \centering
    \includegraphics[width=0.5\linewidth]{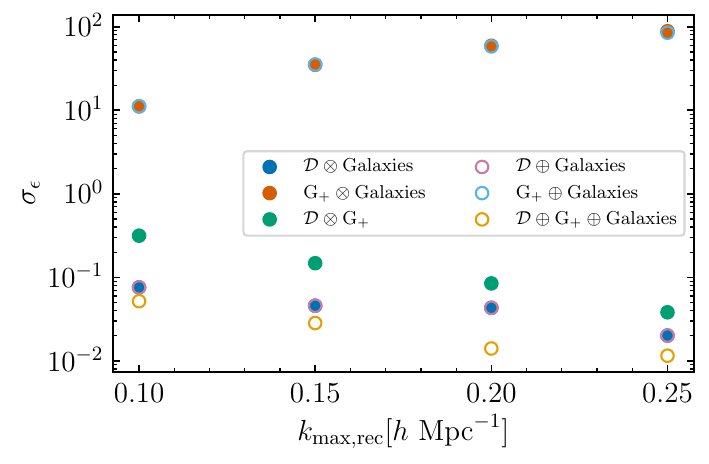}
\caption{\textbf{Marginalized error bars in function of maximum mode of reconstruction} $k_{\mrm{max,rec}}$. We show the constraints on $\epsilon$ as we vary the maximum mode of reconstruction used for QEs, and data combinations. By far, the $\DD \oplus \mrm{G_+} \oplus \mrm{Galaxies}$ gives the best and most robust constraints, showing the importance of synergies of QEs.}
    \label{fig:kmaxfunction}
\end{figure}

\subsection{Constraints from other signatures}\label{sec:generalfore}
A violation of the EP can also lead to an enhanced growth of structure. In particular, in the fifth force model of B24,
it was shown that even for a single tracer
competitive constraints on $\epsilon$ can be obtained through an enhanced linear growth factor (specifically, B24 considers a model with a large fraction of self-interacting dark matter where this is true). 

Here we impose a tracer-scale independent growth modification $D_G(z,\epsilon)=D_G(z)(1+\alpha\epsilon)+\mathcal{O}(\epsilon^2)$, where $D_G$ is the growth function in $\Lambda-$CDM, and $\alpha$ is some expansion parameter capturing the modification of growth (that we keep fixed, see Appendix \ref{app:bottaro}). Effectively, this can be absorbed inside the linear bias $b_{1X} \rightarrow \tilde{b}_{1X}= b_{1X}D_G(z)(1+\alpha\epsilon)$. This is reflected in modification of large scale power spectra, see Eqs.~\eqref{eq:forecast_autorspec}, \eqref{eq:forecast_crossspec}, and \eqref{eq:forecast_crossgspec}.\footnote{Note that when we build the quadratic estimator, we have an inverse variance filtering step: in this we assume fiducial $\Lambda$-CDM cosmology, with $\epsilon=0$.} We use the $\alpha$ as obtained from the enhanced log-term in B24.

In case a model does not predict this, one can alternatively look at potential $\epsilon$ signatures in the growth $C^\mrm{G}_{AB}$ and tidal $C^\mrm{T}_{AB}$ bias responses, in addition to the $1/K$ anti-symmetric shift bias. We implement this by considering $b_{\epsilon,\alpha X} \neq 0, \alpha \in \{\mrm{G},\mrm{S},\mrm{T}\}$.

%\LDC{I'm confused whether the enhanced growth factor has anything to do with the $G_+$ estimator. Presumably you are including the
%`log-enhanced' growth factor in the response of $G_+$? But it seems though that you do not need all the QE formalism to get the benefit
%from the `log-enhanced' growth factor - you can just do some standard kind of power spectrum analysis (don't need the responses at all). Is that right?}

\begin{figure}
    \centering
    \includegraphics[width=0.9\linewidth]{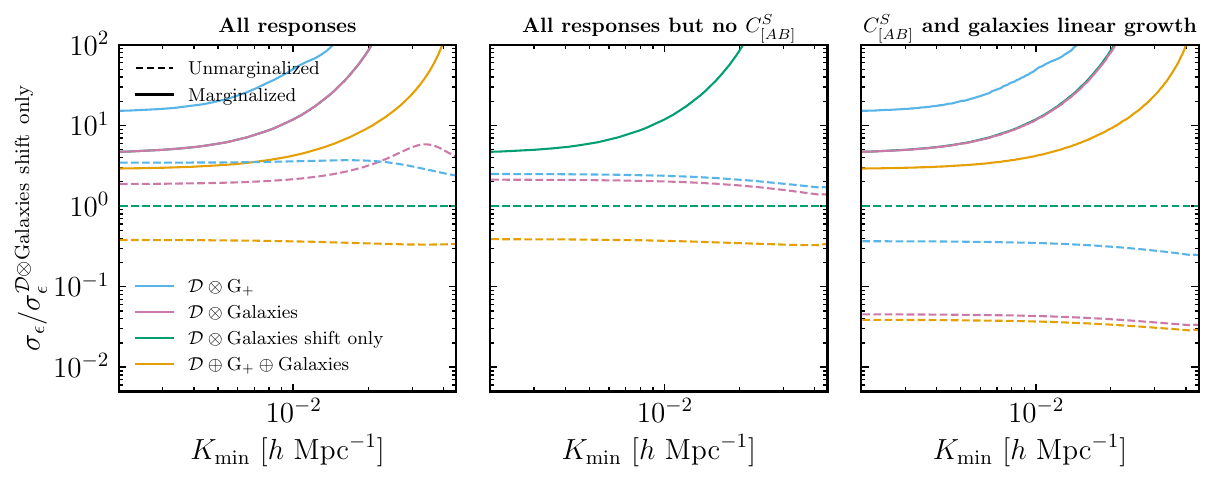}
    \caption{\textbf{Relative constraints on $\epsilon$ from additional signatures beyond $C_{[AB]}^\mrm{S}$.} {\textit{Left panel:}} $\epsilon$ constraints when including all responses, symmetric and anti-symmetric, $C_{AB}^{\alpha}\neq0,\alpha\in\{\mrm{G},\mrm{S},\mrm{T}\}$. %\LDC{You mean when including all *anti-symmetric* response, right? Because $C_{[AB]}^{\alpha}$, with the square brackets, is for the anti-symmetric part only. } 
    All curves are relative to the baseline case in the left panel of Figure \ref{fig:forecast_base} (green dashed curve).
    {\textit{Centre panel:}} constraints when using all responses except the
    anti-symmetric shift, $C_{[AB]}^\mrm{S}=0$.
    {\textit{Right panel:}} $\epsilon$ constraints when using the anti-symmetric shift response only with enhanced linear growth.}
    \label{fig:forecast_variations}
\end{figure} 

Figure \ref{fig:forecast_variations} demonstrates the impact of exploiting different EP violation signatures to constrain $\epsilon$. Using the combined $\mathcal{D} \oplus \mathrm{G_+} \oplus \mathrm{Galaxies}$, the left panel shows that including all three signatures can improve unmarginalized constraints by around an order of magnitude relative to the shift-only baseline (green dashed). This would potentially give $\sigma_\epsilon \sim 10^{-2}$. On the other hand, we see that the unmarginalized constraints of the cross-information $\DD \otimes \mrm{Galaxies}$ (pink dashed) yields \textit{worse} constraints, by a factor of two. This is somewhat surprising, as one would expect including all responses to give better results in the displacement estimator. However, these unmarginalized constraints depend on the choice of our fiducial parameters, and in this case there are partial cancellations among different signatures in the $\DD$ epsilon dependent bias. But we can see that once we marginalize over bias parameters, we achieve similar constrains (solid lines, note that the pink solid line overlaps with the green solid line). 

The middle panel of Figure \ref{fig:forecast_variations} shows that with our fiducial parameters the unmarginalized constraints, without the anti-symmetric shift, can be improved by up to an order of magnitude with respect to our unmarginalized baseline (green dashed). However, marginalization washes out these improvements, showing the importance of the anti-symmetric shift.

Finally, the right panel shows what happens if we combine $C_{[AB]}^\mrm{S}$ with a log-enhanced growth from B24. In this case we can achieve about a hundred-fold improvement on the unmarginalized constraints. However, marginalization degrades constraints to levels similar to those seen in Figure \ref{fig:forecast_base}.
This consistent post-marginalization behaviour across different cases shows that the anti-symmetric shift is indeed 
a robust signature of EP violation.

%When we include $C^\mrm{S}_{[AB]}$ and galaxies linear growth (which incorporates growth effects), the constraints remain competitive even without the full suite of signatures. This indicates that growth-related EP violations provide the primary source of statistical information.

\subsection{Marginalized constraints}

\begin{figure}
    \centering
    \includegraphics[width=\linewidth]{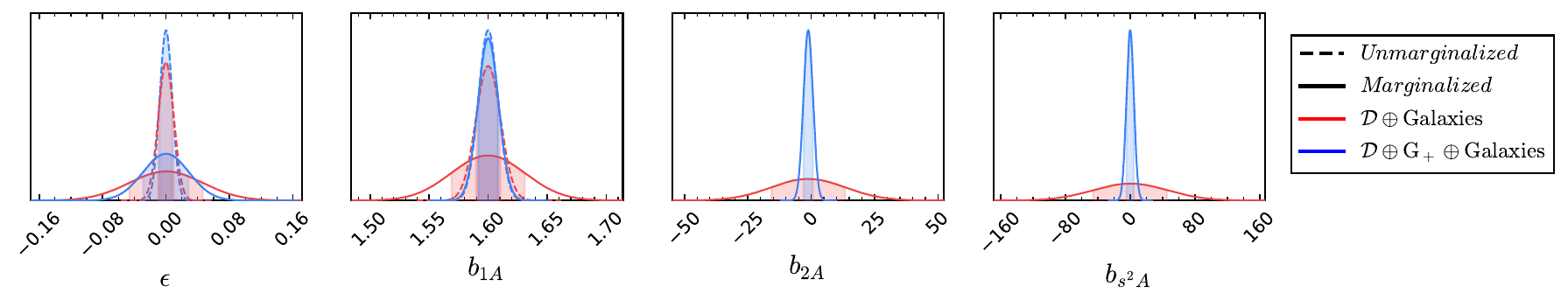}
    \caption{\textbf{Marginalization over standard bias parameters.} We show the effect of marginalization over bias parameters of tracer $A$.  In particular, note the benefit of adding the growth estimator in constraining $b_2$ and $b_{s^2}$. We highlight the flexibility of the QE approach in combining a variety information to improve parameter constraints.}
    \label{fig:distributions_plot}
\end{figure}

Figure \ref{fig:distributions_plot} shows marginalized vs unmarginalized constraints on bias parameters depending on whether we include (red) or do not include (blue) the nonlinear growth estimator $\mrm{G}_+$. We clearly see how biases are well constrained when including this (compare blue vs red). Roughly speaking, the growth estimator is taking a squared (filtered) galaxy field. Similar applications have been already explored and applied to data in the context of projected bispectra in CMB lensing cross-correlations \citep[e.g.][]{Farren:2023yna}.
% and it will be an important topic for future cosmological analyses.

\begin{figure}[h!]
    \centering
    \includegraphics[width=0.7\linewidth]{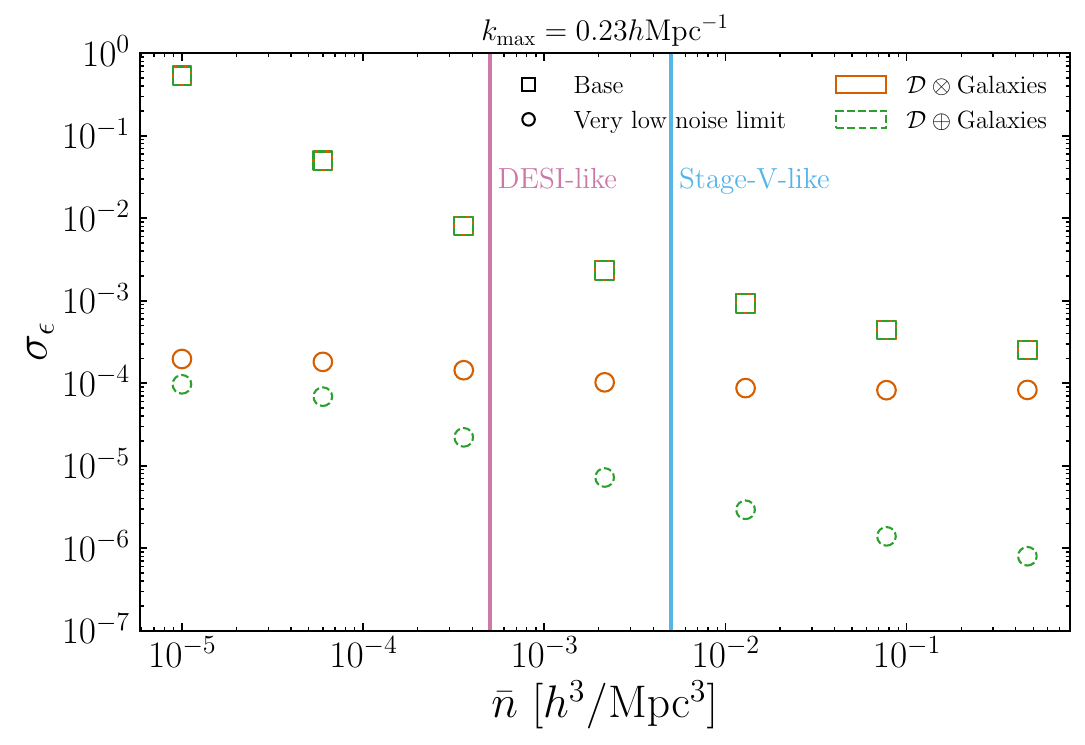}
    \caption{Expected marginalized constraints on $\epsilon$ for varying $\bar{n}$ (from which $\bar{n}_A=\bar{n}/3,\bar{n}_B=\bar{n}/4$). For this example, assume a fixed volume of $V=30 h^{-3}\mrm{Gpc^{3}}$. Compared to the rest of the text, we consider a higher maximum mode of reconstruction $k_{\mrm{max,rec}}=0.23 h\,\mrm{Mpc}^{-1}$. Bias parameters are fixed to be the same as the ones used in Figure \ref{fig:forecast_base}. We show the outcome of constraining $\epsilon$ using our displacement estimator (square), a more optimal one (diamond), or if we consider the very low noise limit (circle). Each one of these uses data from just the cross $\Pcross^{A\DD}$ (orange) or the joint $P^{A\DD}_{\mrm{cross+auto}}$. We note that even having a very high $\bar{n}$ does not dramatically improve constraints (see squares and diamonds) as opposed to the very low noise case, where cosmic variance cancellation enters into play. From here, we see that given that cosmic variance cancellation is difficult to achieve, it may be better to focus on a survey with larger volume rather than one with a smaller volume and very high number density.}
    \label{fig:nbareps}
\end{figure}

\subsection{Varying number densities}

Figure \ref{fig:nbareps} shows the dependence of constraints on number density $\bar{n}$, assuming $\bar{n}_A = 1/3 \bar{n}$ and $\bar{n}_B = 1/4\bar{n}$. The numerical results, cross+auto (green) and cross (orange), are in excellent agreement even for high number densities. This is because the displacement estimator is still very noisy, 
even in low shot-noise regimes. For comparison, in the case when there is both
low shot noise and reconstruction noise (circle), we see that the
cross+auto (green) overcomes the cross only case (orange).

\subsection{Information comparison with a simple bispectrum estimator }\label{app:forecastsbispectrum}
Here we make a quick comparison between the information content in cross-correlations, and that in a simple bispectrum estimator. We take as our starting point Eq.~\eqref{eq:fishercrossanalytical} and continue our calculations from there. 

Using Eq. \eqref{eq:biasedexpectation}, the derivative with respect to $\epsilon$ is
\bea
\partial_\epsilon P_{X\mathcal{D}}(\K)
&= b_{1X}\partial_{\epsilon}b_{\mathcal{D}}(\K) \PL(\K)  \nonumber\\
&= b_{1X}\PL(\K) N_{\mathcal{DD}}(\K) 
 \sum_{\beta \in \{\mathrm{G,S,T}\}}  \Big[\partial_{\epsilon}C^{\beta}_{(AB)} R^{\mathcal{D}}_{+\beta}(\K) + \partial_{\epsilon}C^{\beta}_{[AB]}R^{\mathcal{D}}_{-\beta}(\K)\Big] \ .
\eea

Since EP violations primarily manifest through the anti-symmetric shift response, we focus on\footnote{The $\epsilon$ term in the symmetric coefficients $C^{\alpha}_{(XY)}$ is suppressed by $\epsilon$, that can be of order of $\epsilon \sim10^{-3}$ \citep{Bottaro:2023wkd}. For growth and tidal this suppression is true also for the anti-symmetric part $C^{\alpha}_{[XY]}$. On the other hand, the anti-symmetric shift term is the cleanest one. Coupled to its scale behavior, this makes it the strongest.}
\bea
\partial_\epsilon \Pcross^{X\DD}(\kl) 
&= b_{1X}\PL(\kl) N_{\DD\DD}(\kl) \partial_{\epsilon}C^{S}_{[AB]}R_{-S}(\kl) \nonumber\\
&= b_{1X}\PL(\kl)\frac{\left(b_{1B} b_{\epsilon,SA}-b_{1A} b_{\epsilon,SB}\right)}{2}
\equiv b_{1X}\sp b_{\epsilon}^{AB}\PL(\kl)\ ,
\eea
where we use the fact that $N_{\DD\DD}(\kl)=1/R_{-S}(\kl)$.
Inserting these expressions back into the Fisher matrix per long mode [Eq.~\eqref{eq:fishercrossanalytical}], we have
\bea
\tilde{F}_{\epsilon\epsilon}^{\mrm{rec}}(\kl) 
&= \frac{(b^{AB}_{\epsilon})^2 b_{1X}^2  \PL^2}{b_{\DD}^2 b_{1X}^2\PL^2 \big\{ [1+N_X/(\PL b_{1X}^2)][1+(V_{\DD\DD}+N_{\DD\DD,\mrm{shot}})/(\PL b_{\DD}^2)]+[1+N_{X\DD,\mrm{shot}}/(b_{1X}b_{\DD}\PL)]^2\big\}} 
% &= \frac{(b^{AB}_{\epsilon})^2}{b_{\DD}^2(\kl)} 
%    \frac{1}{\left[ [1+N_X/(\PL(\kl)b_{1X}^2)][1+(V_{\DD\DD}(\kl)+N_{\DD\DD,\mrm{shot}}(\kl))/(\PL(\kl)b_{\DD}^2(\kl))]+[1+N_{X\DD,\mrm{shot}}(\kl)/(b_{1X}b_{\DD}(\kl)\PL(\kl))]^2\right]}
   \label{eq:fishercross}
\eea
{where for brevity we have suppressed the $\K$ dependence in 
$b_\DD$, $P_L$, $V_{\DD\DD}$, $N_{\DD\DD,\mrm{shot}}$, and $N_{X\DD,\mrm{shot}}$.}
The denominator in this expression accounts for cosmic variance contributions and noise components, limited by the number density of objects we have access to, and the reconstruction modes used to recover the large scales. On the other hand, past forecasting work to constrain $\epsilon$ usually ignores the cosmic variance contribution.

\subsubsection{Comparison with a sub-optimal bispectrum estimator}
We would like to compare the Fisher information \eqref{eq:fishercross} with some form of a separable bispectrum estimator. Our cross-correlation \eqref{eq:reccross} is an indirect probe of the squeezed bispectrum
\begin{align}
\langle \delta_X(-\mathbf{K})\delta_A(\mathbf{k}_1)\delta_B(\mathbf{k}_2) \rangle\ = (2\pi)^3\delta_D(-\kl+\k_1+\k_2)B^{XAB}(-\kl, \k_1, \k_2), \label{eq:bispectrumcross}
\end{align}
where $\mathbf{k}_1, \mathbf{k}_2$ are the wavevectors of the short modes. This cross-bispectrum can be written as a fiducial bispectrum times the amplitude parameter of the EP violation: $\epsilon B^{\mrm{theo},\epsilon=1}(\K, \k_1,\k_2)$, if we assume no other effects arise in the bispectrum. From this we can estimate $\epsilon$, following a similar approach to constraining local primordial non-Gaussianity parameter $f_\mrm{NL}$, or any other bispectrum amplitude parameter.

%To get the information content for $\epsilon$ from this bispectrum we can recalculate the Fisher matrix as done above. 
%Nevertheless, to connect with existing literature we take a slightly different presentation. We follow the context of constraining the local primordial non-Gaussianity amplitude $f_{NL}$, as this argument can be general for amplitude parameters of the bispectrum. 
Let us consider a naive separable estimator for $\epsilon$, motivated by a maximum-likelihood estimator, assuming weak non-Gaussianity \citep{Babich:2005en, Fergusson_2012, Regan_2012}:\footnote{Note that the factor of $2$ here comes because we are using the weight in the case of $A=B$, as $\mrm{Var}(B^{XAB}) \approx \Ptot^{XX}(\kl)[\Ptot^{AA}(\k_1)\Ptot^{BB}(\k_2)+P_{\mrm{cross}}^{AB}(\k_1)P_{\mrm{cross}}^{AB}(\k_2)]$.}
\begin{equation}
    \widehat{\epsilon} = N_{\widehat{\epsilon}} \int_{\K}\int_{\k} \frac{B^{\mrm{theo},\epsilon=1}(\k, \K-\k, -\K)}{2\Ptot^{XX}(\K)\Ptot^{BB}(\K-\k)\Ptot^{AA}(\k)}[\delta_A(\k)\delta_B(\K-\k)\delta_X(-\K)]\ ,
\end{equation}
where $N_{\widehat{\epsilon}}$ is a normalization term, the power spectra in the denominator 
are to be understood as inverse-variance filters, and the integrand can be seen as a Wiener filter of a bispectrum.\footnote{We ignore the linear terms in the density field $\langle \delta_A(\k)\delta_B(\K-\k) \rangle \delta_C(-\K)+\mathrm{perms}$, as they will average to zero, assuming statistical homogeneity.} Note that the integral over $\k$ runs from $k_{\mrm{min,rec}}$ to $k_{\mrm{max,rec}}$.
Let us write a per mode estimator $\kl$ and use the response expression for the bispectrum:
\bea
    \widehat{\epsilon}(\K) &= \bard_X(-\K)\frac{N_{\widehat{\epsilon}}}{2} \int_{\k} B^{\mrm{theo},\epsilon=1}(\k, \K-\k, -\K)[\bard_A(\k)\bard_B(\K-\k)]  \nonumber \\
    &= \bard_X(-\K)\frac{N_{\widehat{\epsilon}}}{2}b_{1X}\PL(\K)\int_{\k} f_{AB}^{\epsilon=1}(\k, \K-\k)[\bard_A(\k)\bard_B(\K-\k)]\ .
\eea
This clearly shows that the per mode bispectrum estimator $\widehat{\epsilon}(\K)$ is simply given by a QE cross-correlated with a (Wiener-filtered) external field. Indeed, we can write this expression as:
\begin{equation}
  \widehat{\epsilon}(\K) = \frac{N_{\widehat{\epsilon}}  b_{1X}\PL(\K)}{\Ptot^{XX}(\kl)}N^{-1}_{\DD\DD}(\kl) \delta_X(-\kl)\widehat{h}^\DD_{AB}(\kl)= N_{\widehat{\epsilon}} W(\kl)\delta_X(-\kl)\widehat{h}^\DD_{AB}(\kl)\ .
\end{equation}
It is clear that, modulo an $\epsilon$ independent additional `weight' $W(\kl)$, we get a similar result to what we have previously obtained; this implies equality between $\widehat{\epsilon}(\kl)$ Fisher information matrix and Eq.~\eqref{eq:fishercross}. The important point is that in this case we can re-use existing technology developed for the QE, including for the higher-order shot-noise contributions.

Finally, to get the overall amplitude we calculate a weighted relative cross-spectrum sum:
\begin{equation}
    \widehat{\epsilon} = N_{\widehat{\epsilon}}\int_{\kl}  \Pcross^{X\mrm{matter}}(\kl) W(\kl) \frac{\delta_X(-\kl)\widehat{h}^\DD_{AB}(\kl)}{(\Pcross^{X\mrm{matter}}(\kl))}\ ,
\end{equation}
whose variance is
\begin{equation}
    \sigma^2_{\epsilon} = N_{\widehat{\epsilon}}^2\int_{\kl}  (\Pcross^{X\mrm{matter}}(\kl) W(\kl))^2 \frac{\sigma^2(\Pcross^{X\DD})}{(\Pcross^{X\mrm{matter}}(\kl))^2}\ ,
\end{equation}
with $\sigma^2(\Pcross^{X\DD})$ the variance already calculated in the previous section.
The normalization is simply
\begin{equation}
    N_{\widehat{\epsilon}} = \Big(\int_{\kl} W(\kl) \Pcross^{X\DD}(\kl)|_{\epsilon=1} \Big)^{-1}\ .
\end{equation}
Using the weights previously described, we get the following constraint (signal-to-noise) from $\widehat{\epsilon}$:
\begin{equation}
    \sigma_{\epsilon}^{-2} 
    %=\Big(\int_{\kl} \frac{\Pcross^{X\mrm{matter}}(\kl)\Pcross^{X\DD}(\kl)|_{\epsilon=1}}{\Ptot^{XX}(\kl) N_{\DD\DD}(\kl)}  \Big)^{2} \Big(\int_{\kl} \frac{(\Pcross^{X\mrm{matter}}(\kl))^2}{(\Ptot^{XX}(\kl) N_{\DD\DD}(\kl))^2}\frac{\sigma^2(\Pcross^{X\DD})(\kl)}{(\Pcross^{X\mrm{matter}}(\kl))}\Big)^{-1} = \\
= \left(\int_{\kl} \frac{(\Pcross^{X\DD}(\kl)|_{\epsilon=1})^2}{\Ptot^{XX}(\kl) N_{\DD\DD}(\kl)}  \right)^{2} \left(\int_{\kl} \frac{\Pcross^{X\DD}(\kl)|_{\epsilon=1})^2}{(\Ptot^{XX}(\kl) N_{\DD\DD}(\kl))^2}\sigma^2(\Pcross^{X\DD})(\kl)\right)^{-1}\ .
\end{equation}
This is different from what we get from the integrated Fisher information \eqref{eq:fishercross}:
\begin{equation}
    \int_{\kl}\tilde{F}^{\mrm{rec}}_{\epsilon\epsilon} = \int_{\kl} \frac{(\Pcross^{X\DD}(\kl)|_{\epsilon=1})^2}{\sigma^2(\Pcross^{X\DD})(\kl)} \label{eq:integratedfishercross}\ .
\end{equation}
The same result can be obtained
if we assume the variance is $\sigma^2(\Pcross^{X\DD})(\kl)\sim\Ptot^{XX}(\kl) N_{\DD\DD}(\kl)$ and assume $A=B$ (so that variance and normalization are the same, $V_{\DD}=N_{\DD}$).

\paragraph{Improving on the naive sub-optimal bispectrum estimator.}
While the two expressions should lead to similar results for $A=B$, the case $A \neq B$ leads to a discrepancy. We can restore agreement by considering a general weight $W$:
\begin{equation}
    \widehat{\epsilon} = N_{\widehat{\epsilon}} \int_{\kl} W(\kl) \widehat{\epsilon}(\kl)\ ,
\end{equation}
and minimizing the variance of $\widehat{\epsilon}$ asking $N_{\widehat{\epsilon}} \int_{\kl} W(\kl) \Pcross^{X\DD} = 1$ to recover $\epsilon$. 
Indeed, now we get a proper inverse variance weighting
\begin{equation}
   W(\kl) = \frac{\Pcross^{X\DD}(\kl)|_{\epsilon=1}}{\sigma^2(\Pcross^{X\DD})(\kl)}\ ,\ \ N_{\widehat{\epsilon}} = \Big(\int_{\kl} \frac{(\Pcross^{X\DD}(\kl)|_{\epsilon=1})^2}{\sigma^2(\Pcross^{X\DD})(\kl)}\Big)^{-1}\ ,
\end{equation}
and the constraint is simply:
\begin{equation}
    \sigma^{-2}_{\epsilon} = \frac{1}{N_{\widehat{\epsilon}}} = \int_{\kl} \frac{(\Pcross^{X\DD}(\kl)|_{\epsilon=1})^2}{\sigma^2(\Pcross^{X\DD})(\kl)}\ ,
\end{equation}
in agreement with Eq.~\eqref{eq:integratedfishercross}.

\end{document}